\newif\iftr
\trtrue  %

\iftr
\documentclass[10pt]{article}  %
\else
\documentclass[prodmode]{acmsmall}
\fi

\newif\iflncs\lncsfalse

\iftr 
\usepackage[round, sort, numbers, authoryear]{natbib}
\bibpunct[, ]{[}{]}{;}{a}{}{,}
\let\citeN\cite
\let\cite\citep
\usepackage[left=3cm,right=3cm]{geometry}
 
\usepackage{authblk} %

\usepackage{titlesec}
\titleformat{\paragraph}[runin]{\em}{}{}{}[.] %

\fi

\PassOptionsToPackage{svgnames,x11names,dvipsnames}{xcolor}
\usepackage{xcolor}
\usepackage{tikz}
\usetikzlibrary{patterns}

\usepackage{array,tabularx}
\usepackage{paralist}
\usepackage{enumerate}
\usepackage[algoruled,vlined,english]{algorithm2e}
\usepackage[color=green!20]{todonotes}
\usepackage[breaklinks]{hyperref}
\usepackage{url}
\usepackage{xspace}
\usepackage[draft]{fixme}
\usepackage{verbatim}
\usepackage{graphicx}
\usepackage{amssymb}
\usepackage{amsmath} 
\usepackage{listings}
\iftr %
\usepackage{amsthm}
\newtheorem{theorem}{THEOREM}[section]

\newtheorem{lemma}[theorem]{LEMMA}

\theoremstyle{definition}
\newtheorem{definition}[theorem]{Definition}

\newtheorem{example}[theorem]{EXAMPLE}
\fi

\newtheorem{obs}[theorem]{Observation}
\newtheorem{corol}[theorem]{Corollary}

\let\com=\newcommand
\com{\bthm}{\begin{theorem}}
\com{\ethm}{\end{theorem}}
\com{\bdfn}{\begin{definition}}
\com{\edfn}{\end{definition}}
\com{\blem}{\begin{lemma}}
\com{\elem}{\end{lemma}}
\com{\bcor}{\begin{corol}}
\com{\ecor}{\end{corol}}
\com{\bexm}{\begin{example}}
\com{\eexm}{\end{example}}
\com{\bobs}{\begin{obs}}
\com{\eobs}{\end{obs}}

\com{\bprf}{\begin{proof}}
\iflncs
\com{\eprf}{\qed\end{proof}}
\else
\com{\eprf}{\end{proof}}
\fi

\newenvironment{indexedequations}[1]{\begingroup\def\theequation{\arabic{equation}$_{#1}$}}{\endgroup}
\newcommand{\myref}[2]{\hyperref[#1]{\arabic{#1}$_{#2}$}}
\newcommand{\myeqref}[2]{(\myref{#1}{#2})}

\newcommand{\cdont}{{\cdot}}   %
\newcommand{\tev}[2]{\begin{pmatrix} #1, & \ #2 \end{pmatrix}}  %

\let\vect=\vec
\renewcommand{\vec}[1]{\mathbf{#1}}
\newcommand{\tuple}[1]{\langle #1 \rangle}
\newcommand{\set}[1]{\{ #1 \}}

\newcommand{\ints}{\ensuremath{\mathbb Z}\xspace}
\newcommand{\nats}{\ensuremath{\mathbb N}\xspace}
\newcommand{\rats}{\ensuremath{\mathbb Q}\xspace}
\newcommand{\ratsp}{\ensuremath{\mathbb{Q}_+}\xspace}
\newcommand{\reals}{\ensuremath{\mathbb R}\xspace}

\newcommand{\size}[1]{\Vert #1 \Vert}
\newcommand{\vtxsize}[1]{\Vert #1 \Vert_v}
\newcommand{\bitsize}[1]{\Vert #1 \Vert_b}
\newcommand{\fctsize}[1]{\Vert #1 \Vert_f}
\newcommand{\convhull}[0]{\mathrm{convhull}}
\newcommand{\cone}[0]{\mathrm{cone}}
\newcommand{\recess}[1]{{\mathcal R_{#1}}}
\newcommand{\states}{\poly{C}}
\newcommand{\transitions}{\poly{Q}}

\newcommand{\trcv}[2]{\ensuremath{\bigl(\begin{smallmatrix}{#1}\hfill\\{#2}\end{smallmatrix}\bigr)}}
\newcommand{\trdisp}[2]{\ensuremath{\begin{pmatrix}{#1}\hfill\\{#2}\end{pmatrix}}}
\let\tr=\trcv
\newcommand{\rfp}[2]{(#1{,}#2)}
\newcommand{\rfcoeff}[0]{\lambda}
\newcommand{\prcoeffa}[0]{\mu}
\newcommand{\prcoeffb}[0]{\eta}
\newcommand{\prcs}[4]{\Gamma(#1,#2,#3,#4)}

\newcommand{\poly}[1]{{\mathcal #1}}
\newcommand{\inthull}[1]{{#1}_I}
\newcommand{\intpoly}[1]{I({#1})}

\newcommand{\affhull}[1]{\ensuremath{\mathtt{aff.hull}(#1)}}
\newcommand{\rint}[1]{\ensuremath{\mathtt{ri}(#1)}}

\newcommand{\eqdef}[0]{\,{{\stackrel{\mathrm{def}}{=}}}\,}

\newcommand{\linrfq}[0]{\ensuremath{\textsc{LinRF}(\ensuremath{\rats})}\xspace}
\newcommand{\linrfz}[0]{\ensuremath{\textsc{LinRF}(\ensuremath{\ints})}\xspace}

\newcommand{\llinrfq}[0]{\ensuremath{\textsc{LexLinRF}(\ensuremath{\rats})}\xspace}
\newcommand{\llinrfz}[0]{\ensuremath{\textsc{LexLinRF}(\ensuremath{\ints})}\xspace}

\newcommand{\mlc}[0]{\ensuremath{\mathit{MLC}}\xspace}
\newcommand{\slc}[0]{\ensuremath{\mathit{SLC}}\xspace}

\newcommand{\lrf}[0]{\ensuremath{\mathit{LRF}}\xspace}
\newcommand{\lrfs}[0]{\ensuremath{\mathit{LRFs}}\xspace}
\newcommand{\llrf}[0]{\ensuremath{\mathit{LLRF}}\xspace}
\newcommand{\llrfs}[0]{\ensuremath{\mathit{LLRFs}}\xspace}
\newcommand{\witness}[0]{\ensuremath{W}\xspace}
\newcommand{\hwitness}[0]{\ensuremath{W_{\!\!H}}\xspace}
\newcommand{\wspace}[2]{\ensuremath{\mathit{WS}(#1{,}#2)}\xspace}
\newcommand{\wspacecons}[2]{\ensuremath{\mathit{\Psi_{WS}}(#1{,}#2)}\xspace}
\newcommand{\trans}[0]{{\mbox{\tiny T}}}
\newcommand{\pointinS}[1]{(\rfcoeff_0#1,\vect{\rfcoeff}#1,\vect{\mu}#1,\vect{\eta} #1)}

\newcommand{\lp}[0]{\ensuremath{\mathit{LP}}\xspace}

\newcommand{\tvpi}[0]{\ensuremath{\mathit{TVPI}}\xspace}
\newcommand{\ptvpi}[0]{\ensuremath{\mathit{PTVPI}}\xspace}
\newcommand{\mlcwhile}[0]{\ensuremath{\mathit{loop}}\xspace}
\newcommand{\while}[0]{\ensuremath{\mathit{while}}\xspace}
\newcommand{\wdo}[0]{\ensuremath{\mathit{do}}\xspace}
\newcommand{\llrfsym}[0]{\ensuremath{\tau}\xspace}
\newcommand{\nollrf}[0]{\mbox{\textsc{None}}\xspace}
\newcommand{\nollrfwitness}[2]{\Phi(#1,#2)}

\newcommand{\ranked}[0]{\ensuremath{\mathbf{R}}}
\newcommand{\closure}[1]{\ensuremath{\overline{U}_{#1}}}
\newcommand{\diff}[1]{\ensuremath{\Delta #1}}

\newcommand{\irankfinder}[0]{i\textsc{RankFinder}\xspace}

\newcommand{\shorteq}[0]{
 \rule[.3ex]{3pt}{0.2pt}\llap{\rule[.5ex]{3pt}{0.2pt}}}

\begin{document}

\iftr %

\title{Ranking Functions for Linear-Constraint Loops}
\author[1]{Amir M. Ben-Amram\thanks{amirben@mta.ac.il}}
\author[2]{Samir Genaim\thanks{samir.genaim@fdi.ucm.es}}
\affil[1]{The Academic College of Tel-Aviv Yaffo}
\affil[2]{Complutense University of Madrid}

\maketitle

\begin{abstract}
  In this paper we study the complexity of the problems: given a loop,
  described by linear constraints over a finite set of variables, is
  there a linear or lexicographical-linear ranking function for this
  loop?
  While existence of such functions implies termination, these
  problems are not equivalent to termination. When the variables range
  over the rationals (or reals), it is known that both problems are
  PTIME decidable.
  However, when they range over the integers, whether for single-path
  or multipath loops, the complexity has not yet been determined. We
  show that both problems are coNP-complete.  However, we point out
  some special cases of importance of PTIME complexity.
  We also present complete algorithms for synthesizing linear and
  lexicographical-linear ranking functions, both for the general case
  and the special PTIME cases. 
  Moreover, in the rational setting, our algorithm for synthesizing
  lexicographical-linear ranking functions extends existing ones,
  because our definition for such functions is more general, yet it
  has PTIME complexity.
\end{abstract}

\fi

\section{Introduction}
\label{sec:introduction}

Termination analysis has received considerable attention and
nowadays several powerful tools for the automatic termination analysis
of different programming languages and computational models
exist~\cite{DBLP:conf/rta/GieslTSF04,CPR06,DBLP:conf/fmco/AlbertAGPZ07,DBLP:journals/toplas/SpotoMP10,%
kstw2010-cav,harris2011alternation}.
Much of the recent development in termination analysis has benefited
from techniques that deal with one  loop at a time, where a
loop is specified by a  
loop guard and a (non-iterative) loop body.

Very often, these loops are abstracted so that
the state of the program during the loop is represented by a finite
set of integer variables, the loop guard is
a conjunction of linear inequalities, and the body  modifies the variables in an affine
linear way, as in the following example:
\begin{equation}
\label{eq:intro:loop1}
\begin{array}{l}
\while~( x_2-x_1 \le 0, x_1+x_2 \ge 1 ) ~\wdo~ x_2' = x_2-2x_1+1, x_1'=x_1
\end{array}
\end{equation}
where primed variables represent the values at the completion of an iteration. 
When the variables are modified in the loop body so that they are not affine linear functions of the
old ones, 
the effect is sometimes captured (or approximated)
 using \emph{linear constraints}. For example, the C loop
``\lstinline!while (4*x1>=x2 && x2>=1)$~$x1=(2*x1+1)/5;!'', which involves integer division, can be represented by linear constraints as follows (since \lstinline!2*x1+1! is always positive)
\begin{equation}
\label{eq:intro:loop2}
\begin{array}{l}
\while~(4x_1 \ge x_2, x_2 \ge 1) ~\wdo~ 5x_1' \le 2x_1+1, 5x_1' \ge 2x_1-3, x_2'=x_2
\end{array}
\end{equation}
Linear constraints might also be used to model changes to
data structures, the variables representing a
size abstraction such as length of
lists, depth of trees, etc.~\cite{LS:97,DBLP:conf/popl/LeeJB01,DBLP:journals/toplas/BruynoogheCGGV07,DBLP:journals/toplas/SpotoMP10,DBLP:conf/popl/MagillTLT10}.
For a precise definition of the loop representations we consider, see Section~\ref{sec:prelim};
they also include \emph{multipath loops} where alternative paths in the loop body are represented.

A standard technique to prove the termination of a loop is to find a ranking function. 
Such a function maps a program state (a valuation of the variables) into an element of some
well-founded ordered set, such that the value descends (in the appropriate order) whenever
the loop completes an iteration.  
Since descent in a well-founded set cannot be infinite, this proves that the loop must terminate.
This definition of ``ranking function" is very general; in practice, researchers have often
limited themselves to a convenient and tractable form of ranking function, so that an algorithm
to find the function---if there is one---might be found.

A frequently used class of ranking functions is based on
\emph{affine linear functions}. In this case, we seek a function
$\rho(x_1,\dots,x_n) = a_1x_1+\dots+a_n x_n + a_0$, with the rationals
as a co-domain, such that 
\begin{enumerate}[(i)]
\item\label{intro:lrf1} $\rho(\bar{x}) \ge 0$ for any valuation $\bar{x}$ that satisfies
  the loop guard; and
\item\label{intro:lrf2} $\rho(\bar{x})-\rho(\bar{x}') \ge 1$ for any transition (single
  execution of the loop body) that starts in $\bar{x}$ and leads to
  $\bar{x}'$.
\end{enumerate}
This automatically induces the piecewise-linear ranking function:
$f(\bar{x}) = \rho(\bar{x})+1$ if $\bar{x}$ satisfies the loop guard and
$0$ otherwise, with the non-negative rationals as a co-domain but
ordered w.r.t. $a \succeq b$ if and only if $a\ge b+1$ (which is
well-founded).
For simplicity, we call $\rho$ itself a \emph{linear ranking function}
instead of referring to $f$.

An algorithm to find a linear ranking function 
using linear programming (\lp)
was found by multiple researchers in different places and times and
in some alternative versions~\cite{Feautrier92.1,DBLP:conf/pods/SohnG91,DBLP:conf/tacas/ColonS01,DBLP:conf/vmcai/PodelskiR04,DBLP:journals/tplp/MesnardS08,ADFG:2010}.
Since \lp has a polynomial-time
complexity, most of these methods yield polynomial-time algorithms.
Generally speaking, they are based on the fact that
\lp can precisely decide whether a given inequality is implied by
a set of other inequalities, and 
can even be used to generate any implied inequality.
After all, conditions ~\eqref{intro:lrf1} and \eqref{intro:lrf2} above are inequalities that should be implied by the constraints that define the loop guard and body. 
This approach can,
in a certain sense, be \emph{sound and complete}.

Soundness means that it produces a correct
linear ranking function, if it succeeds; completeness means that if a linear ranking function exists, it will succeed.
In other words, there are no \emph{false negatives}.  
A completeness claim appears in some of the references,
and we found it cited several times.
In our opinion, it has created a false impression that the Linear Ranking problem for linear-constraint
loops with \emph{integer variables} was completely solved (and happily classified as polynomial time).

The fly in the ointment is the fact that these solutions are only complete when the variables range \emph{over the rationals},
which means that the linear ranking function has to fulfill its requirements for any rational valuation of the
variables that  satisfies the loop guard.
But this may lead to a false negative if the variables are,
in fact, integers. The reader may turn to the two loops above and note that both do not 
terminate over the rationals at all (for the first, consider $x_1=x_2=\frac{1}{2}$; for the second,
$x_1=\frac{1}{4}$ and $x_2=1$). But they have linear ranking functions valid for all integer valuations,
which we derive in Section~\ref{sec:rf-synthesis}.

This observation has led us to 
investigate the Linear Ranking problem for single-path
and multipath linear constraint loops.  
We present several fundamental new results on this problem.
We have confirmed that this problem is indeed harder in the integer
setting, proving it to be coNP-complete (as a decision problem),
even for loops that only manipulate integers in a finite range.
On a positive note, this shows that there \emph{is} a complete solution, even if exponential-time.
We give such a solution both to the decision problem and to the synthesis problem.
The synthesis algorithm is based on first computing the
integer hull of the transition polyhedron defined by the loop
constraints, which may require exponential time, and then applying an
\lp{}-based solution (one which is complete over the rationals).
The crux of the coNP-completeness proof is that we rely on the
\emph{generator representation} of the (integer hull of) the
transition polyhedron. We provide sufficient and necessary conditions
for the existence of a linear ranking function that use the vertices
and rays of this representation. This also leads to an alternative
synthesis algorithm.

Another positive aspect of our results, for the practically-minded
reader, is that some special cases of importance do have a PTIME
solution, because they reduce (with no effort, or with a
polynomial-time computation) to the rational case.
We present several such cases, which include, among others, loops in
which the body is a sequence of linear affine updates with integer
coefficients, as in loop~\eqref{eq:intro:loop1} above, and the
condition is defined by either an extended form of \emph{difference
  constraints}, a restricted form of \emph{Two Variables Per
  Inequality} constraints, or a cone (constraints where the free
constant is zero).  Some cases in which the body involves linear
constraints are also presented.

But linear ranking functions do not suffice for all loops, and, in particular for multipath loops,
\emph{lexicographic-linear ranking functions} %
are a natural extension.  Such functions are a tuple
of affine functions, such that in every iteration of the loop, the value of the tuple decreases lexicographically.
Such a function will work, for example, 
for the following multipath loop
\begin{equation}
\label{intro:ex:llrf}
\begin{array}{rll}
\mlcwhile: 
 \set{x_1 \ge 0,x_2 \ge 0, x_1'=x_1-1} \vee 
 \set{x_1 \ge 0,x_2 \ge 0, x_2'=x_2-1, x_1'=x_1}
\end{array}
\end{equation}
where in the first path $x_1$ decreases towards zero and $x_2$ is
changed unpredictably,
since there is no constraint on $x_2'$; this could arise, for instance, from $x_2$ being set to the result
of an input from the environment, or a function call for which we have no invariants.
In the second path $x_2$ decreases
towards zero and $x_1$ is unchanged.
Clearly,  $\tuple{x_1,x_2}$ always decreases lexicographically, but there can be no single
linear ranking function for this loop.

In Section~\ref{sec:llinrf} we analyze the complexity of the decision problem: is there a lexicographic-linear ranking function for a given loop?
We also give a complete synthesis algorithm.  Our point of departure (corresponding to the case of linear ranking functions)
is the known polynomial-time algorithm of~\citeN{ADFG:2010}, based on \lp, that is claimed to be complete---%
and as explained above, is only complete when one extends the domain of the variables to the rationals.
We show that the corresponding decision problem %
is, like the case of
linear ranking function, coNP-complete when the variables are restricted to hold integers. We also give a novel complete
synthesis algorithm.  The algorithm is of exponential-time
complexity, but becomes polynomial-time in special cases corresponding to those identified in the context
of linear ranking functions.

We also consider the application of the algorithm to the setting of rational data; in this setting it
has polynomial-time complexity and extends the one of
\citeN{ADFG:2010}, because our class of ranking functions is more general. The algorithm produces
a function that descends lexicographically in the rationals; for example, if it produces $\tuple{x_1,x_2}$,
it ensures that in every possible transition either $x_1>x_1'$ and $x_1\ge 0$ 
or $x_1=x_1'$ and $x_2>x_2'$ and $x_2\ge 0$.
If one is only interested in integer data, such a function proves termination, and this relaxation to the
rationals is therefore sound. Over the rationals, however,
this lexicographic order is not
well-founded --- simply because the order $(>)$ on $\ratsp$ is not
(consider the sequence $x_1= \frac{1}{2}, \frac{1}{3}, \frac{1}{4}, \dots$).
Interestingly,  we prove that a function that descends in the lexicographic extension of the order $(>)$
can always be turned into one
that descends in the lexicographic extension of the order $a\succeq b$ (defined as
$b\ge a+1$), and therefore implies termination.

We prove some properties of our synthesis algorithm, for example that the dimension (the length of the tuple)
of the functions it produces is always the smallest possible.

Our results should be of interest to all users of ranking functions, and in fact their use
goes beyond termination proofs. For example, they
have been used to provide an upper bound on the number of iterations
of a loop in \emph{program complexity analysis}~\cite{DBLP:journals/jar/AlbertAGP11,ADFG:2010}
and to automatically parallelize computations~\cite{Feautrier92.1,Darte2010}.
We remark that in termination analysis, 
the distinction between integers
and rationals has already been considered, 
both regarding ranking-function generation~\cite{Feautrier92.1,DBLP:conf/concur/BradleyMS05,CookKRW10} 
and the very decidability of the termination problem~\cite{Ben-AmramGM:toplas2012,Tiwari:04,Braverman06}. 
All these works left the integer case open.
Interestingly, our results provide an insight on how to make the
solution proposed by~\citeN{DBLP:conf/concur/BradleyMS05}, for
synthesizing linear ranking functions, complete (see
Section~\ref{sec:related-work}).

Our tool \irankfinder implements 
the algorithms mentioned above (and more) and
can be tried out online (see Section~\ref{sec:implementation}).

This paper is organized as follows. Section~\ref{sec:prelim} gives
definitions and background information regarding linear-constraint
loops, linear and lexicographic-linear ranking functions, and the
mathematical notions involved.
Section~\ref{sec:coNP} proves that the decision problem ``is there a
linear ranking function for an integer loop'', is coNP-complete, and
also presents an exponential-time ranking-function synthesis
algorithm. Section~\ref{sec:ptime} discusses PTIME-solvable cases.
Section~\ref{sec:llinrf} studies the complexity of the decision
problem ``is there a lexicographic-linear ranking function for a given
loop'', both for integer and rational data, and proves that it is
coNP-complete and PTIME respectively. It also develops corresponding
complete synthesis algorithms.
Section~\ref{sec:implementation} describes a prototype implementation.
Section~\ref{sec:related-work} surveys related previous work.
Section~\ref{sec:conclusions} concludes.  A conference version of this
paper, including the results on linear ranking functions (but not
lexicographic-linear ranking functions), has been presented at
POPL 2013~\cite{Ben-AmramG13popl}.

\section{Preliminaries}
\label{sec:prelim}

In this section we recall some results on (integer) polyhedra on which
we will rely along the paper, define the kind of loops we are
interested in, and define the \emph{linear} and
\emph{lexicographic-linear ranking function} problems for such loops.

\begin{figure}

\begin{center}
\begin{minipage}{5cm}
\begin{tikzpicture}[scale=0.5]

  \draw [->,thick] (0,0) node  {} -- (9,0) node (xaxis) [right] {$x_1$};
  \draw [->,thick] (0,0) node  {} -- (0,8) node (yaxis) [above] {$x_2$};

  \coordinate (a) at (0.5,3.5);
  \coordinate (b) at (4.5,7.5);
  \coordinate (c) at (3.333333333,0.6666666);
  \coordinate (e) at (8.5,3.25);

  \fill[fill=black!20] (a) -- (b) -- (e) -- (c) -- cycle;

  \fill[] (a) circle (3pt);
  \fill[] (c) circle (3pt);

  \draw [thick]  (a) -- (b) {};
  \node[rotate=45] () at (2.5,6) {\small \textcolor{DarkBlue}{$x_2{-}x_1{\le}3$}}; 

  \draw [thick]  (a) -- (c) {};
  \node[rotate=-45] () at (1.7,1.8) {\small \textcolor{DarkBlue}{${-}x_1{-}x_2{\le}{-}4$}}; 

  \draw [thick]  (c) -- (e) {};
  \node[rotate=27] () at (6.5,1.7) {\small \textcolor{DarkBlue}{$\frac{1}{2}x_1{-}x_2{\le}1$}}; 

  \draw[dotted] (yaxis |- a) node[left] {\scriptsize $\mathbf{\frac{7}{2}}$}
            -| (xaxis -| a) node[below] {\scriptsize $\mathbf{\frac{1}{2}}$};

  \draw[dotted] (yaxis |- c) node[left] {\scriptsize $\mathbf{\frac{2}{3}}$}
            -| (xaxis -| c) node[below] {\scriptsize $\mathbf{\frac{10}{3}}$};

  \node[] at (8,7) {\textcolor{purple}{$\poly{P}$}};

  \draw [opacity=0.8,red,very thick,dashed,->]  (3,3) -- (5.5,5.5) {};
  \draw [opacity=0.8,red,very thick,dashed,->]  (3,3) -- (6.3,4.4) {};

\end{tikzpicture}
\end{minipage}
~~~~~~~~~~~~~~~~
\begin{minipage}{5cm}
\begin{tikzpicture}[scale=0.5]

  \draw [->,thick] (0,0) node  {} -- (9,0) node (xaxis) [right] {$x_1$};
  \draw [->,thick] (0,0) node  {} -- (0,8) node (yaxis) [above] {$x_2$};

  \coordinate (a) at (0.5,3.5);
  \coordinate (b) at (4.5,7.5);
  \coordinate (c) at (3.333333333,0.6666666);
  \coordinate (e) at (8.5,3.25);
  \coordinate (f) at (3,1);
  \coordinate (g) at (1,3);
  \coordinate (h) at (1,4);
  \coordinate (i) at (4,1);

  \fill[pattern=dots]  (b) -- (h) -- (g) -- (f) -- (i) -- (e) -- cycle;
  \fill[fill=black!20] (a) -- (h) -- (g) -- cycle;
  \fill[fill=black!20] (f) -- (c) -- (i) -- cycle;

  \fill[] (f) circle (3pt);
  \fill[] (g) circle (3pt);
  \fill[] (h) circle (3pt);
  \fill[] (i) circle (3pt);

  \draw [thick]  (f) -- (i) {};
  \node[fill=white, inner sep =0] () at (3.55,1.45) {\small \textcolor{blue}{$x_2{\ge}1$}}; 

  \draw [thick]  (g) -- (h) {};
  \node[rotate=90,fill=white, inner sep =0] () at (1.5,3.45) {\small \textcolor{blue}{$x_1{\ge}1$}}; 

  \draw [thick]  (h) -- (b) {};
  \node[rotate=45] () at (2.5,6) {\small \textcolor{DarkBlue}{$x_2{-}x_1{\le}3$}}; 

  \draw [thick]  (g) -- (f) {};
  \node[rotate=-45] () at (1.7,1.8) {\small \textcolor{DarkBlue}{${-}x_1{-}x_2{\le}{-}4$}}; 

  \draw [thick]  (i) -- (e) {};
  \node[rotate=27] () at (6.5,1.7) {\small \textcolor{DarkBlue}{$\frac{1}{2}x_1{-}x_2{\le}1$}};

  \draw[dotted] (yaxis |- f) node[left] {\scriptsize $\mathbf{1}$}
            -| (xaxis -| f) node[below] {\scriptsize $\mathbf{3}$};

  \draw[dotted] (yaxis |- g) node[left] {\scriptsize $\mathbf{3}$}
            -| (xaxis -| g) node[below] {\scriptsize $\mathbf{1}$};

  \draw[dotted] (yaxis |- h) node[left] {\scriptsize $\mathbf{4}$}
            -| (xaxis -| h) node[below] {\scriptsize $\mathbf{1}$};

  \draw[dotted] (yaxis |- i) node[left] {\scriptsize $\mathbf{1}$}
            -| (xaxis -| i) node[anchor=west,below] {\scriptsize $\mathbf{4}$};

  \phantom{\draw[dotted] (yaxis |- a) node[left] {\scriptsize $\mathbf{\frac{7}{2}}$}
            -| (xaxis -| a) node[below] {\scriptsize $\mathbf{\frac{1}{2}}$};}

  \phantom{\draw[dotted] (yaxis |- c) node[left] {\scriptsize $\mathbf{\frac{2}{3}}$}
            -| (xaxis -| c) node[below] {\scriptsize $\mathbf{\frac{10}{3}}$};}

  \node[] at (8,7) {\textcolor{purple}{$\inthull{\poly{P}}$}};
  \draw [opacity=0.8,red,very thick,dashed,->]  (3,3) -- (5.5,5.5) {};
  \draw [opacity=0.8,red,very thick,dashed,->]  (3,3) -- (6.3,4.4) {};

\end{tikzpicture}
\end{minipage}
\end{center}
 
\caption{A polyhedron $\poly{P}$ and its integer hull $\inthull{\poly{P}}$.}
\label{fig:poly}

\end{figure}

\subsection{Integer Polyhedra}
\label{sec:prelim:polyhedra}

We recall some useful definitions and properties, all
following~\citeN{Schrijver86}.

\paragraph{Polyhedra}
A \emph{rational convex polyhedron} $\poly{P} \subseteq \rats^n$
(\emph{polyhedron} for short) is the set of solutions of a set of
inequalities $A\vec{x} \le \vec{b}$, namely $\poly{P}=\{
\vec{x}\in\rats^n \mid A\vec x \le \vec b \}$, where $A \in \rats^{m
  \times n}$ is a rational matrix of $n$ columns and $m$ rows, $\vec
x\in\rats^n$ and $\vec b \in \rats^m$ are column vectors of $n$ and
$m$ rational values respectively.
We say that $\poly{P}$ is specified by $A\vec{x} \le \vec{b}$.
We use calligraphic letters, such as $\poly{P}$ and $\poly{Q}$ to
denote polyhedra.
The set of \emph{recession directions} of a polyhedron $\poly{P}$
specified by $A\vec{x} \le \vec b$ is the set $\recess{\poly{P}} = \{
\vec{y}\in\rats^n \mid A\vec{y} \le \vec{0}\}$. 

\bexm
\label{ex:poly:polyhedron}
Consider the polyhedron $\poly{P}$ of Figure~\ref{fig:poly} (on the
left). The points defined by the gray area, and the black borders, are
solutions to the system of linear inequalities $x_2-x_1\le 3 \,\land\,
-x_1-x_2\le -4 \,\land\, \frac{1}{2}x_1-x_2\le 1$. 
\eexm

\paragraph{Integer Polyhedra}
For a given polyhedron $\poly{P} \subseteq \rats^n$ we let
$\intpoly{\poly{P}}$ be $\poly{P} \cap \ints^n$, i.e., the set of
integer points of $\poly{P}$. The \emph{integer hull} of $\poly{P}$,
commonly denoted by $\inthull{\poly{P}}$, is defined as the convex
hull of $\intpoly{\poly{P}}$, i.e., every rational point of
$\inthull{\poly{P}}$ is a convex combination of integer points. This
property is fundamental to our results.
It is known that $\inthull{\poly{P}}$ is also a polyhedron.  An
\emph{integer polyhedron} is a polyhedron $\poly{P}$ such that
$\poly{P} = \inthull{\poly{P}}$. We also say that $\poly{P}$ is \emph{integral}.

\bexm
\label{ex:poly:inthull}
The integer hull $\inthull{\poly{P}}$ of polyhedron $\poly{P}$ of
Figure~\ref{fig:poly} (on the left) is given in the same figure (on
the right). It is defined by the dotted area and the black border, and
is obtained by adding the inequalities $x_1 \ge 1$ and $x_2 \ge 1$ to
$\poly{P}$. The two gray triangles next to the edges of  $\inthull{\poly{P}}$ are subsets of
$\poly{P}$ that were eliminated when computing $\inthull{\poly{P}}$.
\eexm

\paragraph{Generator representation} 
Polyhedra also have a \emph{generator representation} in terms of
vertices and rays%
\footnote{Technically, the $\vec x_1,\ldots,\vec x_n$ are only
  vertices if the polyhedron is \emph{pointed}.}%
, written as
\[
\poly{P} = \convhull\{\vec x_1,\dots,\vec x_m\} + \cone\{\vec
y_1,\dots,\vec y_t\} \,.
\]
This means that $\vec x\in \poly{P}$ if and only if $\vec x =
\sum_{i=1}^m a_i\cdot \vec x_i + \sum_{j=1}^t b_j\cdot \vec y_j$ for some
rationals $a_i,b_j\ge 0$, where $\sum_{i=1}^m a_i = 1$.  Note that $\vec
y_1,\dots,\vec y_t$ are the recession directions of $\poly{P}$, i.e.,
 $\vec{y}\in\recess{\poly{P}}$ if and only if
$\vec{y}=\sum_{j=1}^t b_j \cdot \vec{y}_j$ for some rationals $b_j\ge 0$.
If $\poly{P}$ is integral, then there is a generator representation in
which all $\vec{x}_i$ and $\vec{y}_j$ are integer.
An empty polyhedron is represented by an empty set of vertices and rays.

\bexm
\label{ex:poly:genrep}
The generator representations of $\poly{P}$ and $\inthull{\poly{P}}$ of
Figure~\ref{fig:poly} are
\[
\begin{array}{rl}
\poly{P} = & \convhull\{(\frac{1}{2},\frac{7}{2}),(\frac{10}{3},\frac{2}{3})\}+ \cone\{(1,1),(7,3)\}\\[1ex]
\inthull{\poly{P}} = & \convhull\{(1,3),(1,4),(3,1),(4,1)\}+ \cone\{(1,1),(7,3)\}\\
\end{array}
\]
The points in $\convhull$ are vertices, they correspond to the points
marked with $\bullet$ in Figure~\ref{fig:poly}. 
The rays are the vectors $(1,1),(7,3)$;  they describe a direction, rather than a specific point, and are therefore
represented in the figure as arrows.
Note that the
vertices of $\inthull{\poly{P}}$ are integer points, while those of
$\poly{P}$ are not.
The point $(3,2)$, for example, is defined as
$\frac{5}{17}\cdot(\frac{1}{2},\frac{7}{2}) +
\frac{12}{17}\cdot(\frac{10}{3},\frac{2}{3})\}+
\frac{1}{2}\cdot(1,1)+0\cdot(7,3)$ in $\poly{P}$, and as
$0\cdot(1,3)+\frac{1}{3}\cdot(1,4)+0\cdot(3,1)+\frac{2}{3}\cdot(4,1)+
0\cdot(1,1)+0\cdot(7,3)$ in $\inthull{\poly{P}}$.
\eexm

\paragraph{Faces} 
If $\vec{c}$ is a nonzero vector and $a=\max\{\vec{c}\cdot\vec{x} \mid
\vec{x}\in \poly{P}\}$, then $\poly{H}=\{\vec{x}\in\rats^n \mid
\vec{c}\cdot\vec{x}=a\}$ is called a supporting hyperplane for
$\poly{P}$.
A \emph{non-empty} subset $\poly{F}\subseteq\poly{P}$ is called a
\emph{face} if $\poly{F}=\poly{P}$ or $\poly{F}$ is an intersection of
$\poly{P}$ with a supporting hyperplane~\cite[p.~101]{Schrijver86}. In
the latter case we say that $\poly{F}$ is a \emph{proper} face of
$\poly{P}$.
Alternatively, $\poly{F}$ is face of $\poly{P}$ if and only if it can
be obtained by turning some inequalities of $A\vec{x}\le b$ to
equalities~\cite[Sec.~16.3, p.~231]{Schrijver86}.
It is known that a polyhedron $\poly{P}$ is integral if and only if
every face of $\poly{P}$ includes an integer point~\cite[Sec.~16.3,
p.~231]{Schrijver86}. This implies that the faces of an integral
polyhedron $\poly{P}$ are integral.

\bexm
\label{ex:poly:faces}
Polyhedron $\poly{P}$ of Figure~\ref{fig:poly} has $5$ \emph{proper}
faces, each corresponds to either a black segment or a vertex (a point
marked with $\bullet$). For example, the segment between
$(\frac{1}{2},\frac{7}{2})$ and $(\frac{10}{3},\frac{2}{3})$ is a
proper face, and it can be obtained by turning the inequality
$-x_1-x_2\le -4$ to $-x_1-x_2 = -4$ in $\poly{P}$.
Similarly, polyhedron $\inthull{\poly{P}}$ of Figure~\ref{fig:poly}
has $9$ \emph{proper} faces, in this case each includes an integer
point.
\eexm

\paragraph{Dimension of polyhedra}
Let $A^{\shorteq}\vec{x}\le \vec{b}^{\shorteq}$ be the set of all implicit equalities in
$A\vec{x}\le \vec{b}$ ($\vec{a}_i\cdot\vec{x}\le \vec{b}_i$ is an
implicit inequality if $\vec{a}_i\cdot\vec{x}= \vec{b}_i$ holds for
any $\vec{x}\in\poly{P}$). The \emph{affine hull} of $\poly{P}$ is
defined as $\affhull{\poly{P}}=\{ \vec{x}\in\rats^n \mid A^{\shorteq}\vec{x}=
\vec{b}^{\shorteq} \}$.
The dimension of the affine hull is the dimension of the linear
subspace $\{ \vec{x} \mid A^{\shorteq}\vec{x} = \vec{0}\}$ (i.e, the
cardinality of the bases). Alternatively, it is equal to $n$ minus the
rank of the matrix $A^{\shorteq}$. The \emph{dimension} of a polyhedron
$\poly{P}\subseteq\rats^n$, denoted by $\dim(\poly{P})$, is equal to
the dimension of its affine hull.
The dimension of the empty polyhedron, by convention, is $-1$. The
dimension of a proper face of $\poly{P}$ is at least $1$ less than
that of $\poly{P}$. Note that when $\dim(\poly{P})=0$ then $\poly{P}$
is a single point.

\bexm
\label{ex:poly:dim}
Both $\poly{P}$ and $\inthull{\poly{P}}$ of Figure~\ref{fig:poly} have
dimension $2$. Their proper faces that are defined by segments
(resp. vertices) have dimension $1$ (resp. $0$).
\eexm

\paragraph{Relative interior} 
The \emph{relative interior} of $\poly{P}$ is defined as
$\rint{\poly{P}}=\{\vec{x} \mid \exists \epsilon>0\ .\
B(\vec{x},\epsilon)\cap\affhull{\poly{P}} \subseteq \poly{P} \}$ where
$B(\vec{x},\epsilon)$ is a ball of radius $\epsilon$ centered on
$\vec{x}$. Intuitively, it is the set of all points which are not on
the ``edge'' of $\poly{P}$.
Note that $\vec{x}\in\rint{\poly{P}}$ if and only if
$\vec{x}\in\poly{P}$ and $\vec{x}$ does not belong to any proper face
of $\poly{P}$.
When $\dim(\poly{P})=0$, the single point of $\poly{P}$ is in the
relative interior (since $\poly{P}$ does not have any proper face).

\bexm
\label{ex:poly:rint}
Consider the polyhedra of Figure~\ref{fig:poly}. The relative interior
of $\poly{P}$ is defined by the gray area, and that of
$\inthull{\poly{P}}$ by the dotted area, i.e., we exclude the points
on the black segments of each polyhedron (which are proper faces as
explained in Example~\ref{ex:poly:rint}).
\eexm

\paragraph{Size of polyhedra} 
Complexity of algorithms on polyhedra is measured in this paper by
running time, on a conventional computational model (polynomially
equivalent to a Turing machine), as a function of the \emph{bit-size}
of the input.  Following~\citeN[Sec. 2.1]{Schrijver86}, we define the bit-size of
an integer $x$ as $\size{x} = 1 + \lceil \log (|x|+1)\rceil$; the bit-size of
an $n$-dimensional vector $\vec a$ as $\size{\vec a} = n+\sum_{i=1}^n \size{a_i}$;
and the bit-size of  an inequality
$\vec{a}\cdot \vec{x} \le c$ as $1+\size{c}+\size{\vec{a}}$.
For a polyhedron $\poly{P} \subseteq \rats^n$ defined by $A\vec x \le \vec b$,
we let $\bitsize{\poly{P}}$ be the bit-size of 
$A\vec x \le \vec b$, which we can take as the sum of the sizes of the inequalities.
The \emph{facet size}, denoted by $\fctsize{\poly{P}}$, is the
smallest number $\phi \ge n$ %
 such that $\poly{P}$ may be described by
\emph{some} $A\vec x \le \vec b$ where 
each inequality in $A\vec x \le \vec b$ fits in $\phi$
bits. Clearly, $\fctsize{\poly{P}} \le \bitsize{\poly{P}}$.
The \emph{vertex size}, denoted by $\vtxsize{\poly{P}}$, is the smallest
number $\psi \ge n$ such that $\poly{P}$ has a generator representation in
which each of $\vec x_i$ and $\vec y_j$ fits in $\psi$ bits (the size of a vector is calculated
as above). For integer polyhedra, we restrict the generators to be integer.
The following theorems state some relations between the different
bit-sizes defined above, they are later used to polynomially bound the
bit-size of some set of integer points of $\inthull{\poly{P}}$. They 
are from \citeN{Schrijver86} (Th.~10.2, p.~121, and Cor.~17.1a,17.1b,
p.~238), who cites~\citeN{DBLP:conf/focs/KarpP80}.

\bthm
\label{thm:size}
Let $\poly{P}$ be a rational polyhedron in $\rats^n$; then
$\vtxsize{\poly{P}} \le 4n^2\fctsize{\poly{P}}$ and $\fctsize{\poly{P}} \le
4n^2\vtxsize{\poly{P}}$.  
\ethm

\bthm 
\label{thm:PIsize}
Let $\poly{P}$ be a rational polyhedron in $\rats^n$;
then $\vtxsize{\inthull{\poly{P}}} \le 6n^3\fctsize{\poly{P}}$ 
and $\fctsize{\inthull{\poly{P}}} \le 24n^5\fctsize{\poly{P}}$.
\ethm

\subsection{Multipath Linear-Constraint Loops}
\label{sec:prelim:loops}

A \emph{single-path} linear-constraint loop (\slc for short) over
$n$ variables $x_1,\ldots,x_n$ has the form
\begin{equation} \label{eq:ilc-loop} 
  \mathit{while}~(B\vec{x} \le
  \vec{b})~\mathit{do}~ A\begin{pmatrix}\vec{x}\phantom{'}\\
    \vec{x}'\end{pmatrix} \le \vec{c}
\end{equation}
where $\vec{x}=(x_1,\ldots,x_n)^\trans$ and
$\vec{x}'=(x_1',\ldots,x_n')^\trans$ are column vectors, and for some
$p,q>0$, $B \in {\rats}^{p\times n}$, $A\in {\rats}^{q\times 2n}$,
$\vec{b}\in {\rats}^p$, $\vec{c}\in {\rats}^q$.
The constraint $B\vec{x} \le \vec{b}$ is called \emph{the loop
  condition} (a.k.a. the loop guard) and the other constraint is
called \emph{the update}. 
The update is called \emph{deterministic} if, for a given $\vec x$
(satisfying the loop condition) there is at most one $\vec{x}'$
satisfying the update constraint. The update is called \emph{affine linear}
if it can be rewritten as
\begin{equation}
\vec{x}' = A'\vec{x} + \vec{c}'
\end{equation}
for a matrix $A'$ and vector $\vec{c}'$ of appropriate dimensions.
We say that the loop is a \emph{rational loop} if $\vec{x}$ and
$\vec{x}'$ range over $\rats^n$, and that it is an \emph{integer loop}
if they range over $\ints^n$.

We say that there is a transition from a state $\vec{x}\in\rats^n$ to
a state $\vec{x}'\in\rats^n$, if $\vec{x}$ satisfies the condition and
$\vec{x}$ and $\vec{x}'$ satisfy the update.
A transition can be seen as a point $\trcv{\vec{x}}{\vec{x}'} \in \rats^{2n}$, where its first $n$
components correspond to $\vec{x}$ and its last $n$ components to
$\vec{x}'$. For ease of notation, we denote $\tr{\vec{x}}{\vec{x}'}$ by
$\vec{x}''$.
The set of all transitions $\vec{x}''\in \rats^{2n}$ will be denoted,
as a rule, by $\transitions$.
The transition polyhedron $\transitions$ is specified by the set of inequalities 
$A''
\vec{x}'' \le \vec{c}''$ where
\begin{align*}
A''  & = \begin{pmatrix} B & 0 \\ \multicolumn{2}{c}{A} \end{pmatrix}  &
\vec c'' & = \begin{pmatrix} \vec b \\ \vec c \end{pmatrix}
\end{align*}
Note that we may assume that $\transitions$ does not include the
origin, for if it includes it, the loop is clearly non-terminating 
(this condition is easy to check). Hence, $\transitions$ is not
a cone (i.e., $m \ge 1$ in the generator representation).
The polyhedron defined by the loop condition $B\vec{x}\le\vec{b}$ will
be denoted by $\states$ and referred to as the condition polyhedron.

A \emph{multipath} linear-constraint loop (\mlc for short)
differs by having alternative loop conditions and updates, which are,
in principle, chosen non-deterministically (though the constraints may
enforce a deterministic choice):
\begin{equation} \label{eq:multipath}
\mlcwhile~\bigvee_{i=1}^k \left[ B_i\vec{x} \le  \vec{b}_i \,\land\, A_i\begin{pmatrix}\vec{x}\phantom{'}\\ \vec{x}'\end{pmatrix} \le \vec{c}_i \right]
\end{equation}
This means that the $i$-th update can be applied if the $i$-th
condition is satisfied. 
Following the notation of \slc loops, the transitions of an \mlc loop
are specified by the transition polyhedra $\transitions_1,\ldots,\transitions_k$,
where each $\transitions_i$ is specified by $A''_i\vec{x}''\le
\vec{c}''_i$. The polyhedron defined by the condition
$B_i\vec{x}\le\vec{b}_i$ is denoted by $\states_i$.

For simplifying the presentation, often we write loops with
explicit equalities and inequalities %
instead of the matrix representation. We also might refer to loops by
their corresponding transition polyhedra, or the sets of inequalities
that define these polyhedra.

\subsection{Linear Ranking Functions}
\label{sec:prelim:lrf}

An affine linear function $\rho: \rats^n \to \rats$ is %
of the form
$\rho(\vec{x}) = \vect{\rfcoeff}\cdot\vec{x} + \rfcoeff_0$ where
$\vect{\rfcoeff}\in\rats^n$ is a row vector and $\rfcoeff_0\in\rats$.
For ease of notation we sometimes refer to an affine linear function 
using the row vector $\rfp{\rfcoeff_0}{\vect{\rfcoeff}} \in \rats^{n+1}$.
For a given function $\rho$, we define the function
$\diff{\rho}:\rats^{2n}\mapsto\rats$ as
$\diff{\rho}(\vec{x}'')=\rho(\vec{x})-\rho(\vec{x}')$.
Next we define when an affine linear function is a \emph{linear ranking
  function} (\lrf for short) for a given rational or integer 
\mlc loop.

\bdfn
\label{def:linearrf}
Given a set $T\subseteq \rats^{2n}$, representing transitions,
we say that $\rho$ is a \lrf for $T$
if the following
hold for every $\vec{x}'' \in T$:

\begin{align}
 \rho(\vec{x})  \ge 0  \,, \label{eq:lrf1}\\
 \diff{\rho}(\vec{x}'')  \ge 1 \,. \label{eq:lrf2} 
\end{align}
\edfn

\noindent
We say that $\rho$ is a \lrf for a rational loop, specified by $\transitions_1,\ldots,\transitions_k$, when
it is a \lrf for all of $\transitions_1,\ldots,\transitions_k$
(equivalently, it is a \lrf for $\bigcup_{i=1}^k \transitions_i$).
We say that $\rho$ is a \lrf for an integer loop, specified by $\transitions_1,\ldots,\transitions_k$
polyhedra, when it is a \lrf for all of $\intpoly{\transitions_1},{\ldots},\intpoly{\transitions_k}$.

Clearly, the existence of a \lrf implies
termination of the loop.
Note that in~\eqref{eq:lrf2} we require $\rho$ to decrease at least by
$1$, whereas in the literature~\cite{DBLP:conf/vmcai/PodelskiR04} this
$1$ is sometimes replaced by $\delta>0$.  It is easy to verify that
these definitions are equivalent as far as the existence of a \lrf is
concerned.

\bdfn 
The decision problem \emph{Existence of a \lrf} is
defined by

\smallskip
\noindent
\begin{tabular}{ll}
 \emph{Instance}:& an \mlc loop. \\
 \emph{Question}:& does there exist a \lrf for
this loop? 
\end{tabular}

\smallskip
\noindent
The decision problem is denoted by \linrfq and \linrfz for rational
and integer loops respectively. 
\edfn

\noindent
It is known that \linrfq is
PTIME-decidable~\cite{DBLP:conf/vmcai/PodelskiR04,DBLP:journals/tplp/MesnardS08}.
In this paper, we focus on  \linrfz.

\subsection{Ranking Functions}
\label{sec:prelim:llrf}

A \emph{$d$-dimensional affine function} $\llrfsym: \rats^n \to \rats^d$ is a
function of the form $\llrfsym = \tuple{\rho_1,\dots,\rho_d}$,
where each component $\rho_i :\rats^{n} \to \rats$ is an affine function.
The number $d$ is informally called the \emph{dimension} of the
function (technically, it is the dimension of the co-domain).
Next we define when a $d$-dimensional affine function is a
\emph{lexicographic-linear ranking function} (\llrf for short) for a
given rational or integer \mlc loop.

\bdfn 
\label{def:lexlinearrf}
Let $T\subseteq \rats^{2n}$ be a given set, representing transitions,
and $\llrfsym=\tuple{\rho_1,\dots,\rho_d}$ a $d$-dimensional affine
function.
We say that $\llrfsym$ is a \llrf for $T$
 if and only if for every
$\vec{x}'' \in T$ there
exists $i\le d$ such that the following hold
\begin{alignat}{ 2 }
 \forall j < i \ .\   && \diff{\rho_j}(\vec{x}'') &\ge 0 \,, \label{eq:llrf1}\\
 \forall j \le i \ .\ && \rho_j(\vec{x}) &\ge 0          \,, \label{eq:llrf2}\\
                      && \diff{\rho_i}(\vec{x}'') &\geq 1\,. \label{eq:llrf3} 
\end{alignat}
We say that $\vec{x}''$ is ranked by $\rho_i$.
\edfn

As for \lrfs, we say that $\llrfsym$ is a \llrf for a rational loop $\transitions_1,\dots,\transitions_k$ when it is
a \llrf for $\bigcup_{i=1}^k \transitions_i$, and that it is a \llrf for the corresponding integer loop if it is a \llrf for
$\bigcup_{i=1}^k \intpoly{\transitions_i}$.

Note that in~\eqref{eq:llrf3} we require $\rho_i$ to decrease at
least by $1$. As for the case of \lrfs, this $1$ can be replaced by any
$\delta_i>0$. It is easy to verify that these definitions are
equivalent as far as the existence of a \llrf is concerned.
The existence of a \llrf implies termination of the loop.
This may be justified by converting the function into one that decreases in a well-founded set;
such a function is 
$$\hat\llrfsym(\vec{x}) = \tuple{\max(0, \rho_1(\vec{x})), \dots, \max(0, \rho_d(\vec{x}))},$$
whose co-domain is $\tuple{ \rats_+^d, \preceq_{lex}}$, where $\preceq_{lex}$ is the lexicographic extension
of the well-founded order:  $a \preceq b$ if and only if $a+1\le b$.

Our class of {\llrf}s  differs somewhat from other classes
of ``lexicographic-linear ranking functions''
that appeared in the literature~\cite{DBLP:conf/cav/BradleyMS05,ADFG:2010}.
Specifically, the definition in~\citeN{ADFG:2010} is more restrictive since it
requires~\eqref{eq:llrf2} to hold for all $1 \le j \le d$.
The following example illustrates the difference.

\bexm 
\label{ex:kinds-of-lrfs-1}
Consider the \slc loop
\begin{equation}
\label{eq:llrf:loop:1}
\while ( x_1 \ge 0, x_2 \ge 0, x_3 \ge -x_1 )~\wdo~
x_2'= x_2-x_1, x_3'= x_3+x_1-2\,.
\end{equation}
It has a \llrf $\llrfsym=\tuple{x_2, x_3}$ as in
Definition~\ref{def:lexlinearrf} (over both rationals and integers),
however, it does not have a \llrf according to~\citeN{ADFG:2010}.
Indeed, when $x_2$ decreases $x_3$ can be
negative (e.g., for $x_1=1$, $x_2=2$ and $x_3=-1$).
\eexm

Another difference from~\citeN{ADFG:2010} lies in the fact that they require the non-negativity
conditions \eqref{eq:llrf2} to be implied by \emph{the loop guard}.  That is, it is not possible to use
the constraints in the update part of the loop in proving this condition, when according to our
definition it is possible.

The definition of~\citeN{DBLP:conf/cav/BradleyMS05} 
requires~\eqref{eq:llrf2} to hold only for $j=i$, which adds flexibility, as we show next.

\bexm 
\label{ex:kinds-of-lrfs-2}
Consider the \mlc loop
\begin{equation}
\label{eq:llrf:loop:2}
\mlcwhile: 
\set{ x_1 \geq 0, x_1' = x_1-1 } \vee 
\set{ x_2 \geq 0, x_2'=x_2-1,  x_1' \leq x_1}\,.
\end{equation}
It has a \llrf $\llrfsym=\tuple{x_1,x_2}$ according to the
definition of~\citeN{DBLP:conf/cav/BradleyMS05}, however, it does not
have one that satisfies Definition~\ref{def:lexlinearrf}. Indeed,
in transitions where $x_2$ decreases $x_1$ may be negative, but $x_1$ must be the first component.
\eexm
Another difference is that \citeN{DBLP:conf/cav/BradleyMS05} require a fixed association of 
ranking-function components with the paths of the loop. So, for example, they cannot have
a 2-dimensional \llrf for an \slc loop, as in Example~\ref{ex:kinds-of-lrfs-1}.

\bdfn 
The decision problem \emph{Existence of a \llrf} is
defined by

\smallskip
\noindent
\begin{tabular}{ll}
 \emph{Instance}:& an \mlc loop. \\
 \emph{Question}:& does there exist a \llrf for
this loop? 
\end{tabular}

\smallskip
\noindent
The decision problem is denoted by \llinrfq and \llinrfz for rational
and integer loops respectively. 
\edfn

\section{{\linrfz} is {\lowercase{co}NP}-complete}
\label{sec:coNP}

In this section we show that the \linrfz problem is coNP-complete;
it is coNP-hard already for \slc loops that restrict the variables to a finite range.
We also show that  \lrfs can be
synthesized in deterministic exponential time.
This section is organized as follows: in
Section~\ref{sec:conp-hardness} we show that \linrfz is coNP-hard; in
Section~\ref{sec:conp-incl-slc} we show that it is in coNP for
\slc loops, and in Section~\ref{sec:conp-incl-mlc} for \mlc loops; finally, in
Section~\ref{sec:rf-synthesis}, we describe an algorithm for
synthesizing \lrfs.

\subsection{coNP-hardness}
\label{sec:conp-hardness}

We prove coNP-hardness in a strong form.  Recall that a
number problem (a problem whose instance is a matrix of integers)
$\mathbf{Prob}$ is strongly hard for a complexity class, if there are
polynomial reductions from all problems in that class to
$\mathbf{Prob}$ such that the values of all numbers created by the
reduction are polynomially bounded by the input bit-size. Assuming
NP$\ne$P, strongly NP-hard (or coNP-hard)
problems cannot even have pseudo-polynomial algorithms~\cite{garey-johnson:1979}.

\bthm
\label{th:conp-hardness}
The \linrfz problem is strongly coNP-hard, even for  \slc loops with affine-linear updates.
\ethm

\bprf 
The problem of deciding whether a polyhedron given by $B\vec x\le \vec b$ contains no integer
point is a well-known coNP-hard problem (an easy reduction from SAT~\cite{Karp72}).
We reduce this problem to \linrfz.  

Given $B\in\ints^{m\times n}$ and
$\vec{b}\in\ints^m$, we construct the following integer \slc loop
$$\mathit{while}~\begin{pmatrix}B & -I \\ 0 & -I \end{pmatrix}
\begin{pmatrix} \vec{x}\\ \vec{z}\end{pmatrix}  \le
\begin{pmatrix} \vec{b}\\ \vec{0}\end{pmatrix}
 ~\mathit{do}~
\begin{pmatrix} \vec{x}'\\ \vec{z}'\end{pmatrix}  =
\begin{pmatrix} \vec{x}\\ \vec{0}\end{pmatrix}
$$
where $\vec x = (x_1,\ldots,x_n)^\trans$, $\vec z = (z_1,\ldots,z_m)^\trans$ are integer
variables, and $I$ is an identity matrix of size $m\times m$.

Suppose $B\vec x\le\vec b$ has an integer solution
$\vec{x}$. Then, it is easy to see that 
the loop does not
terminate when starting from this $\vec{x}$ and $\vec{z}$ set to
$\vec{0}$, since the guard is satisfied and the update does not change the values.
Thus, it does not have any ranking function, let alone a \lrf.

Next, suppose $B\vec x\le\vec b$ does not have an integer solution. Then,
for any initial state for which the loop guard is enabled it must
hold that $z_1+\cdots+z_m>0$, for otherwise $z_1,\dots,z_m$ must be $0$ in which case the
constraint $B\vec{x} -I\vec{z} \le \vec{b}$ has no integer solution.
Since the updated vector $\vec{z}'$ is deterministically set to
$\vec{0}$, the guard will not be enabled in the next state, hence
the loop terminates after one iteration. Clearly $z_1 +\cdots+z_m >
z_1'+\cdots+z_m' = 0$, so we conclude that $z_1+\cdots+z_m$ is a \lrf.
\eprf

Note that in the above reduction we rely on the hardness of whether a
given polyhedron is empty. This problem is coNP-hard already for bounded
polyhedra (due to the reduction from SAT in which variables are
bounded by $0$ and $1$). This means that even for loops that only manipulate
integers in a rather small range, the problem is coNP-hard.  The parameter
``responsible" for the exponential behavior in this case is the number of variables.

Note also that the loop constructed in the reduction either has a \lrf, or fails to terminate.
Hence, one cannot hope to avoid the coNP-hardness
by using another kind of certificate instead of linear ranking
functions, as long as the certificate is sufficiently expressive to
capture the termination argument for integer loops where 
variables are limited to $[0,1]$, update is an affine linear function, and
termination follows from the fact that a sum of variables always
descends.  

\subsection{Inclusion in coNP for \slc Loops}
\label{sec:conp-incl-slc}

To prove that \linrfz is in coNP, we show that the complement
of \linrfz, the problem of \emph{nonexistence} of a \lrf, is in NP, that is, has a
polynomially-checkable witness.
In what follows we assume as input an \slc loop with a transition
polyhedron $\transitions \subseteq \rats^{2n}$. The input is given
as the set of linear inequalities $A''\vec{x}''\le\vec{c}''$
that define $\transitions$.
The proof follows the following lines:
\begin{enumerate}
\item We show that there is no \lrf for $\intpoly{\transitions}$ if and only if 
  there is a \emph{witness} that consists of two sets of
  integer points $X\subseteq\intpoly{\transitions}$ and
  $Y\subseteq\intpoly{\recess{\transitions}}$, such that a
  certain set of inequalities $\wspacecons{X}{Y}$ has no
  solution over the rationals; and
\item We show that if there is a witness then there is one with
  bit-size polynomial in the input bit-size.
\end{enumerate}
To make sense of the following definitions, think of a vector $\rfp{\rfcoeff_0}{\vect{\rfcoeff}}\in\rats^{n+1}$
as a ``candidate \lrf'' that we may want to verify~(or, in our case, to eliminate).

\bdfn 
\label{def:witness}
We say that $\vec{x}'' \in \intpoly{\transitions}$
is a witness \emph{against} $\rfp{\rfcoeff_0}{\vect{\rfcoeff}}\in\rats^{n+1}$
if it fails to satisfy at least one of %
\begin{align}
 \vect{\rfcoeff} \cdot \vec{x} + \rfcoeff_0 \ge 0 \label{eq:rf1}\\
 \vect{\rfcoeff} \cdot (\vec{x} - \vec{x}') \ge 1 \label{eq:rf2} 
\end{align}
The set of $\rfp{\rfcoeff_0}{\vect{\rfcoeff}}$ %
witnessed against by $\vec{x}''$ is denoted by $\witness(\vec{x}'')$.
\edfn

Note that conditions~(\ref{eq:rf1},\ref{eq:rf2}) are obtained
from~(\ref{eq:lrf1},\ref{eq:lrf2}) by writing $\rho$ explicitly.

\bdfn
\label{def:h-witness}
We say that $\vec{y}'' \in
\intpoly{\recess{\transitions}}$ is a \emph{homogeneous (component of a) witness} (h-witness)
against $\rfp{\rfcoeff_0}{\vect{\rfcoeff}}\in\rats^{n+1}$ if it fails to
satisfy at least one of
\begin{align}
\vect{\rfcoeff} \cdot \vec{y}  \ge 0 \label{eq:rfh1}\\
\vect{\rfcoeff} \cdot (\vec{y} - \vec{y}') \ge 0 \label{eq:rfh2}
\end{align}
The set of $\rfp{\rfcoeff_0}{\vect{\rfcoeff}}$ %
h-witnessed
against by $\vec{y}''$ is denoted by $\hwitness(\vec{y}'')$.
\edfn

The meaning of the witness of Definition~\ref{def:witness} is quite
straightforward. Let us intuitively explain the meaning of an
h-witness. 
Suppose that $\vec{x}''$ is a point in $\inthull{\transitions}$, and $\vec{y}''$ is a ray of $\inthull{\transitions}$.
Then a \lrf $\rho$ has to satisfy~\eqref{eq:rf1} for any point of the form
$\vec{x}'' + a\vec{y}''$ with integer $a>0$ since it is a point in $\intpoly{\transitions}$; letting $a$ grow to infinity, we see that~\eqref{eq:rf1} 
implies the homogeneous inequality~\eqref{eq:rfh1}. Similarly,~\eqref{eq:rf2} 
implies~\eqref{eq:rfh2}. 

\bdfn
\label{def:wspace} 
Given %
$X \subseteq \intpoly{\transitions}$
and $Y\subseteq \intpoly{\recess{\transitions}}$, define
\begin{align}
  \wspace{X}{Y} = \bigcup_{\vec{x}'' \in X} \witness(\vec{x}'') \ \cup
  \ \bigcup_{\vec{y}'' \in Y} \hwitness(\vec{y}'') \,.
\end{align}
\edfn

\blem 
\label{lem:witness-1}
Let $X \subseteq \intpoly{\transitions}$, $X\neq\emptyset$, and
$Y\subseteq \intpoly{\recess{\transitions}}$.
If $\wspace{X}{Y}=\rats^{n+1}$, then there is no \lrf for
$\intpoly{\transitions}$.
\elem

\bprf 
Let $\wspace{X}{Y}=\rats^{n+1}$. For any $\rfp{\rfcoeff_0}{\vect{\rfcoeff}} \in \rats^{n+1}$, we prove
that $\rho(\vec{x}) = \vect{\rfcoeff}\cdot\vec{x} + \rfcoeff_0$ is not a
\lrf.
If $\rfp{\rfcoeff_0}{\vect{\rfcoeff}} \in \witness(\vec{x}'')$ for some
$\vec{x}'' \in X$, then the conclusion is clear since
one of the conditions\eqref{eq:rf1} and~\eqref{eq:rf2} does not hold.
Otherwise, suppose that
$\rfp{\rfcoeff_0}{\vect{\rfcoeff}} \in \hwitness(\vec{y}'')$ for
$\vec{y}'' \in Y$.  Thus, $\vec{y}''$ fails to satisfy one of
conditions (\ref{eq:rfh1},\ref{eq:rfh2}). Next we show that, in such case,
there must exist $\vec{z}''\in \intpoly{\transitions}$ that fails
either \eqref{eq:rf1} or \eqref{eq:rf2}.
In this part of the proof, we rely on the fact that
$X\ne\emptyset$.

\medskip
\noindent
\emph{Case}~1: Suppose \eqref{eq:rfh1} is not satisfied. That is,
$\vect{\rfcoeff} \cdot \vec{y} < 0$.

Choose $\vec{x}''\in X$, and note that $\rho(\vec{x})\ge 0$, otherwise
$\rfp{\rfcoeff_0}{\vect{\rfcoeff}}\in\witness(\vec{x}'')$ which we
have assumed not true.
Note that for any integer $a\ge 0$, the integer point 
$\vec{z}''=\vec{x}'' + a\cdot\vec{y}''$
is a transition in $\intpoly{\transitions}$, and
$\vec{z}''=\tr{\vec{x}\phantom{'}+a\cdot\vec{y}}{\vec{x}'+a\cdot\vec{y}'}$.
We choose 
$a$ as an integer sufficiently large so that
$a\cdot(\vect{\rfcoeff} \cdot \vec{y}) \le -(1+\rho(\vec{x}))$.
Now,
\begin{align*}
\rho(\vec{z}) %
 &= \vect{\rfcoeff} \cdot  (\vec{x} + a\cdot\vec{y}) + \rfcoeff_0 \\ &=
\rho(\vec{x}) + a\cdot(\vect{\rfcoeff} \cdot  \vec{y})  \le  \rho(\vec{x}) - (1+\rho(\vec{x})) = -1
\end{align*}
So $\rho$ fails \eqref{eq:rf1} on
$\vec{z}''\in\intpoly{\transitions}$, and thus cannot be a \lrf. %

\medskip
\noindent
\emph{Case}~2: Suppose \eqref{eq:rfh2} is not satisfied. That is, 
$\vect{\rfcoeff} \cdot (\vec{y} - \vec{y}') < 0$.

Choose $\vec{x}''\in X$, and note that $\rho(\vec{x})-\rho(\vec{x}')
\ge 1$, otherwise
$\rfp{\rfcoeff_0}{\vect{\rfcoeff}}\in\witness(\vec{x}'')$ which we
have assumed not true.
Define
$\vec{z}''$ as above, but now choosing $a$ sufficiently large to make
$a\cdot(\vect{\rfcoeff} \cdot (\vec{y}-\vec{y}')) \le
-(1+\rho(\vec{x})-\rho(\vec{x}'))$. Now,
\begin{align*}
\rho(\vec{z})-\rho(\vec{z}') %
&=
   \vect{\rfcoeff} \cdot  ((\vec{x} + a\cdot\vec{y}) - (\vec{x}' + a\cdot\vec{y}')) \\ 
&=   \rho(\vec{x})-\rho(\vec{x}') + a\cdot(\vect{\rfcoeff} \cdot  (\vec{y}-\vec{y}')) \\
&\le \rho(\vec{x})-\rho(\vec{x}')-(1+\rho(\vec{x})-\rho(\vec{x}')) =  -1
\end{align*}
So $\rho$ fails \eqref{eq:rf2} on
$\vec{z}''\in\intpoly{\transitions}$, and thus cannot be a \lrf. %
\eprf

Note that the condition $\wspace{X}{Y}=\rats^{n+1}$ is equivalent to
saying that the conjunction of inequalities
(\ref{eq:rf1},\ref{eq:rf2}), for all $\vec{x}'' \in X$, and
inequalities (\ref{eq:rfh1},\ref{eq:rfh2}), for all $\vec{y}'' \in Y$,
has no (rational) solution.
We denote this set of inequalities by $\wspacecons{X}{Y}$.  Note that
the variables in $\wspacecons{X}{Y}$ are $\rfcoeff_0,\ldots,
\rfcoeff_n$, which range over $\rats$, and thus, the test that it has
no solution can be done in polynomial time since it is an \lp problem
over the rationals.

\bexm
\label{ex:gen:witness1}
Consider the following integer \slc loop:
\begin{equation*}
\label{loop:phasetrans}
while~(x_1 \ge 0)~do~x_1'=x_1+x_2, x_2'=x_2-1
\end{equation*}
Let 
$\vec{x}_1''=(0,2,2,1)^\trans\in\intpoly{\transitions}$ and
$\vec{y}_1''=(1,-2,-1,-2)^\trans\in\intpoly{\recess{\transitions}}$.
Then, $\wspacecons{\{\vec{x}_1''\}}{\{\vec{y}_1''\}}$ is a conjunction
of the %
inequalities
\begin{equation}
\label{eq:psi1}
\{  2\rfcoeff_2+\rfcoeff_0 \ge 0,~
   -2\rfcoeff_1+\rfcoeff_2 \ge 1,~
    \rfcoeff_1-2\rfcoeff_2 \ge 0,~
  2\rfcoeff_1 \ge 0\}
\end{equation}
The first two inequalities correspond to applying
(\ref{eq:rf1},\ref{eq:rf2}) to $\vec{x}_1''$, and the other ones
to applying (\ref{eq:rfh1},\ref{eq:rfh2}) to $\vec{y}_1''$.
It is easy to verify that~\eqref{eq:psi1} is not satisfiable, thus,
$\wspace{\{\vec{x}_1''\}}{\{\vec{y}_1''\}}=\rats^3$ and the loop does
not have a \lrf. This is a classical loop for which there
is no \lrf.
\eexm

Lemma~\ref{lem:witness-1} provides a sufficient condition for the
nonexistence of \lrf, the next lemma shows that this condition is also
necessary. In particular, it shows that if there is no \lrf for
$\intpoly{\transitions}$, then the vertices and rays of
$\inthull{\transitions}$ serve as $X$ and $Y$ of
Lemma~\ref{lem:witness-1} respectively.

\blem 
\label{lem:witness-2}
Let the integer hull of the transition polyhedron $\transitions$ be $\inthull{\transitions} = \convhull\{\vec{x}''_1,\dots,\vec{x}''_m\} +
\cone\{\vec{y}''_1,\dots,\vec{y}''_t\}$.
If there is no \lrf for $\intpoly{\transitions}$,
then
$\wspace{\{\vec{x}''_1,\dots,\vec{x}''_m\}}{\{\vec{y}''_1,\dots,\vec{y}''_t\}}=\rats^{n+1}$.
\elem

\bprf 
We prove the contra-positive. Suppose that
$$\wspace{\{\vec{x}''_1,\dots,\vec{x}''_m\}}{\{\vec{y}''_1,\dots,\vec{y}''_t\}}
\ne \rats^{n+1} \,.$$
Then, there is $\rfp{\rfcoeff_0}{\vect{\rfcoeff}}\in\rats^{n+1}$ that fulfills
(\ref{eq:rf1},\ref{eq:rf2}) for all $\vec{x}''_i$ and
(\ref{eq:rfh1},\ref{eq:rfh2}) for all $\vec{y}''_j$.
We claim that $\rho(\vec{x}) = \vect{\rfcoeff}\cdot\vec{x} + \rfcoeff_0$
is a \lrf for $\intpoly{\transitions}$.

To see this, let $\vec{x}''$ be an arbitrary
point of $\intpoly{\transitions}$.  Then $\vec{x}'' = \sum_{i=1}^m
a_i\cdot \vec{x}''_i + \sum_{j=1}^t b_j\cdot \vec{y}''_j$ for some
$a_i,b_j\ge 0$ where $\sum_{i=1}^m a_i = 1$.  Now, we show that
$\vec{x}''$ and $\rho$ satisfy (\ref{eq:rf1},\ref{eq:rf2}) which
means that $\rho$ is a \lrf for $\intpoly{\transitions}$:
\begin{align*}
\vect{\rfcoeff} \cdot \vec{x} + \rfcoeff_0  
&=
\rfcoeff_0 + 
  \sum_{i=1}^m  a_i \cdot (\vect{\rfcoeff}\cdot \vec{x}_i) + 
    \sum_{j=1}^t b_j \cdot (\vect{\rfcoeff} \cdot \vec{y}_j) \\
&=
\sum_{i=1}^m a_i \cdot (\vect{\rfcoeff} \cdot \vec{x}_i + \rfcoeff_0) + 
  \sum_{j=1}^t b_j \cdot (\vect{\rfcoeff} \cdot \vec{y}_j) \\
&\ge  0 + 0 = 0 \\
& = 
\sum_{i=1}^m a_i \cdot (\vect{\rfcoeff} \cdot (\vec{x}_i - \vec{x}'_i)) + 
  \sum_{j=1}^t b_j \cdot (\vect{\rfcoeff} \cdot (\vec{y}_j - \vec{y}'_j)) \\
&\ge
   1 + 0 = 1
\end{align*}
\eprf

Note that the solutions of
$\wspacecons{\{\vec{x}''_1,\dots,\vec{x}''_m\}}{\{\vec{y}''_1,\dots,\vec{y}''_t\}}$ in Lemma~\ref{lem:witness-2}
actually define the set of all \lrfs for $\intpoly{\transitions}$. We
will address this point later in Section~\ref{sec:rf-synthesis}, for synthesizing \lrfs.

\bexm
\label{ex:gen:witness2}
Consider again the loop of Example~\ref{ex:gen:witness1}, and recall that it does not have a \lrf. The generator
representation of $\inthull{\transitions}$ is
\[
\inthull{\transitions} = \convhull\{\vec{x}_1''\}+\cone\{\vec{y}_1'',\vec{y}_2'',\vec{y}_3''\}
\]
where $\vec{x}_1''=(0,1,1,0)^\trans$,
$\vec{y}_1''=(0,-1,-1,-1)^\trans$, $\vec{y}_2''=(0,1,1,1)^\trans$ and
$\vec{y}_3''=(1,-1,0,-1)^\trans$.
Then, $\wspacecons{\{\vec{x}_1''\}}{\{\vec{y}_1'',\vec{y}_2'',\vec{y}_3''\}}$
is a conjunction of the following inequalities
\begin{equation}
\label{eq:psi2}
\left\{
\begin{array}{@{}r@{~~~~}r@{~~~~}r@{~~~~}r@{}}
  \rfcoeff_2+\rfcoeff_0 \ge 0, &
   -\rfcoeff_2 \ge 0, &
     \rfcoeff_2 \ge 0, &
      \rfcoeff_1-\rfcoeff_2 \ge 0,\\
  -\rfcoeff_1+\rfcoeff_2 \ge 1, &
   \rfcoeff_1 \ge 0, &
  -\rfcoeff_1 \ge 0, &
  \rfcoeff_1 \ge 0\phantom{,}\\
\end{array}
\right\}
\end{equation}
The inequalities in the leftmost column correspond to applying
(\ref{eq:rf1},\ref{eq:rf2}) to $\vec{x}_1''$, and those in the other
columns to applying (\ref{eq:rfh1},\ref{eq:rfh2}) to $\vec{y}_1''$,
$\vec{y}_2''$, and $\vec{y}_3''$ respectively.
It is easy to verify that~\eqref{eq:psi2} is not satisfiable, and thus,
$\wspace{\{\vec{x}_1''\}}{\{\vec{y}_1'',\vec{y}_2'',\vec{y}_3''\}} = \rats^3$.
\eexm

Lemmas~\ref{lem:witness-1} and \ref{lem:witness-2} provide a necessary
and sufficient condition for the nonexistence of a \lrf.

\bcor
\label{cor:witness} 
There is no \lrf for $\intpoly{\transitions}$ if and only if there
are two finite sets $X \subseteq \intpoly{\transitions}$, $X\ne \emptyset$, and $Y
\subseteq\intpoly{\recess{\poly{P}}}$, such that
$\wspace{X}{Y}=\rats^{n+1}$.
\ecor

The next lemma concerns the bit-size of the witness. 

\blem
\label{lem:witness-size}
If there exists a witness for the nonexistence of a \lrf for
$\intpoly{\transitions}$,  there exists one with
$X\subseteq\intpoly{\transitions}$ and
$Y\subseteq\intpoly{\recess{\transitions}}$ such that $|X|+|Y|
\le n+2$; and its bit-size is polynomially bounded in the bit-size
of the input.
\elem

\bprf
Recall that by Lemma~\ref{lem:witness-2}, if $\intpoly{\transitions}$
has no \lrf, then
\[
\wspace{\{\vec{x}''_1,\dots,\vec{x}''_m\}}{\{\vec{y}''_1,\dots,\vec{y}''_t\}}=\rats^{n+1}
\] 
or, equivalently, %
$\wspacecons{\{\vec{x}''_1,\dots,\vec{x}''_m\}}{\{\vec{y}''_1,\dots,\vec{y}''_t\}}$
has no solution.
A corollary of Farkas' Lemma~\cite[p.~94]{Schrijver86} states that if
a finite set of inequalities over $\rats^d$, for some $d>0$, has no solution, there is a
subset of at most $d+1$ inequalities that has no solution.
Since the set of inequalities
$\wspacecons{\{\vec{x}''_1,\dots,\vec{x}''_m\}}{\{\vec{y}''_1,\dots,\vec{y}''_t\}}$
is over $\rats^{n+1}$, there is a subset of at most $n+2$ inequalities
that has no solution.
This subset involves at most $n+2$ integer points out of
$\{\vec{x}''_1,\dots,\vec{x}''_m\}$ and
$\{\vec{y}''_1,\dots,\vec{y}''_t\}$, 
because every inequality
in
$\wspacecons{\{\vec{x}''_1,\dots,\vec{x}''_m\}}{\{\vec{y}''_1,\dots,\vec{y}''_t\}}$
is defined by either one $\vec{x}''_i$ or $\vec{y}''_i$
(see~(\ref{eq:rf1}--\ref{eq:rfh2})).
Let these points be $X$ and $Y$, then $|X|+|Y|\le n+2$ and
$\wspacecons{X}{Y}$ has no solution, i.e., $\wspace{X}{Y}=\rats^{n+1}$. 
Moreover it must be that case that $X\neq\emptyset$, since all
constraints of the type (\ref{eq:rfh1},\ref{eq:rfh2}) are satisfied by
$\rfp{\rfcoeff_0}{\vect{\rfcoeff}} = \vec{0}^\trans$.

Now we show that $X$ and $Y$ may be chosen to have bit-size polynomial
in the size of the input.  Recall that the input is the set of
inequalities $A''\vec{x}'' \le \vec{c}''$ that define $\transitions$,
and its bit-size is $\bitsize{\transitions}$.
Recall that the points of $X$ and $Y$ in Lemma~\ref{lem:witness-2} come from the generator
representation, and that there is a generator representation in which
each vertex/ray can fit in $\vtxsize{\inthull{\transitions}}$
bits. 
Thus, the bit-size of $X$ and $Y$ may be bounded by
$(n+2)\cdot\vtxsize{\inthull{\transitions}}$. 
By Theorem~\ref{thm:PIsize}, since the dimension of $\transitions$ is $2n$,
\[
(n+2)\cdot\vtxsize{\inthull{\transitions}} \le
(n+2)\cdot (6\cdot (2n)^3\cdot\fctsize{\transitions}) \le 
(48n^4 + 96n^3)\cdot\bitsize{\transitions}
\]
which is polynomial in the bit-size of the input.
\eprf

\bexm
\label{ex:gen:sizeofwitness}
Consider
$\wspacecons{\{\vec{x}_1''\}}{\{\vec{y}_1'',\vec{y}_2'',\vec{y}_3''\}}$
of Example~\ref{ex:gen:witness2}. It is easy to see that the
inequalities $-\rfcoeff_2 \ge 0$, $\rfcoeff_1 \ge 0$ and
$-\rfcoeff_1+\rfcoeff_2 \ge 1$ are enough for unsatisfiability
($n+1$ inequalities, since $n=2$).
These inequalities correspond to $\vec{x}_1''$ and $\vec{y}_1''$, and
thus, these two points witness the nonexistence of a \lrf
(note that this witness consists, in this example, of less than $n+2$ points).
\eexm

\bthm
\label{th:conp-inc-spl}
$\linrfz \in \mathrm{coNP}$ for \slc loops.
\ethm

\bprf 
We show that the complement of \linrfz has a polynomially checkable
witness. The witness is a listing of sets $X$ and $Y$ of at most $n+2$
elements and has a polynomial bit-size (specifically, a bit-size bounded as in
Lemma~\ref{lem:witness-size}).
Verifying a witness consists of the following steps:

\paragraph{Step 1} Verify that each $\vec{x}''\in X$ is in
  $\intpoly{\transitions}$, which can be done by verifying  $A''
  \vec{x}'' \le \vec{c}''$; and that each $\vec{y}''\in Y$ is in
  $\intpoly{\recess{\transitions}}$, which can be done by verifying
   $A'' \vec{y}'' \le \vec{0}$. This is done in polynomial time.
  Note that according to Lemma~\ref{lem:witness-1} it is not necessary to
  check that $X$ and $Y$ come from a particular generator representation.

\paragraph{Step 2} Verify that $\wspace{X}{Y}=\rats^{n+1}$. This can be done by
  checking that $\wspacecons{X}{Y}$ has no solutions, which can be
  done in polynomial time since it is an \lp problem
  over $\rats^{n+1}$.
\eprf

\subsection{Inclusion in coNP for \mlc Loops}
\label{sec:conp-incl-mlc}

In this section we consider the inclusion in coNP for \mlc loops.  For
this, we assume an input \mlc loop with transition polyhedra
$\transitions_1,\ldots,\transitions_k$ where each $\transitions_i$ is
specified by $A_i''\vec{x}''\le\vec{c}_i''$.

The proof follows the  structure of the \slc case. The main
difference is that points of the witness may come from
different transition polyhedra. Namely, $X=X_1\cup\cdots\cup X_k$ and
$Y=Y_1\cup\cdots\cup Y_k$ where each
$X_i\subseteq\intpoly{\transitions_i}$ and
$Y_i\subseteq\intpoly{\recess{\transitions_i}}$.
Lemmas~\ref{lem:witness-1}, \ref{lem:witness-2}, and
\ref{lem:witness-size}, Corollary~\ref{cor:witness}, and
Theorem~\ref{th:conp-inc-spl}, are rewritten in terms of such witnesses
as follows (the proofs are the same unless stated otherwise).

\blem 
\label{lem:witness-1-mlc}
Let $X=X_1\cup\cdots\cup X_k$ and $Y=Y_1\cup\cdots\cup Y_k$, where
$X_i \subseteq \intpoly{\transitions_i}$, $Y_i\subseteq
\intpoly{\recess{\transitions_i}}$ and $Y_i\neq\emptyset \Rightarrow X_i\neq\emptyset$.
If $\wspace{X}{Y}=\rats^{n+1}$, then there is no \lrf for
$\intpoly{\transitions_1},\ldots,\intpoly{\transitions_k}$.
\elem

Note that $\wspace{X}{Y} = \bigcup_{i=1}^k \wspace{X_i}{Y_i}$ and that in 
the proof  of Lemma~\ref{lem:witness-1} (when re-used to obtain the above lemma) 
it is necessary to use the condition $Y_i\neq\emptyset \Rightarrow X_i\neq\emptyset$.

\blem 
\label{lem:witness-2-mlc}
For $1\le i\le k$, let $\inthull{\transitions_i} = \convhull\{X_i\} +
\cone\{Y_i\}$ be the integer hull of $\transitions_i$, and 
define $X=X_1\cup\cdots\cup X_k$ and $Y=Y_1\cup\cdots\cup Y_k$.
If there is no \lrf for
$\intpoly{\transitions_1},\ldots,\intpoly{\transitions_k}$, then
$\wspace{X}{Y}=\rats^{n+1}$.
\elem

\bprf 
The proof follows that of Lemma~\ref{lem:witness-2}. We 
pick $\rfp{\rfcoeff_0}{\vect{\rfcoeff}}\in\rats^{n+1} \setminus \wspace{X}{Y}$ and
show that $\rho(\vec{x}) = \vect{\rfcoeff}\cdot\vec{x} + \rfcoeff_0$
is a \lrf for all $\intpoly{\transitions_i}$. This is accomplished by performing the
same calculation, however referring to $X_i$ and $Y_i$ when proving that $\rho$ is
a \lrf for
$\intpoly{\transitions_i}$.
\eprf

\bcor
\label{cor:witness-mlc} 
There is no \lrf for
$\intpoly{\transitions_1},\ldots,\intpoly{\transitions_k}$, if and only if
 there are two finite sets $X=X_1\cup\cdots\cup X_k$ and
$Y=Y_1\cup\cdots\cup Y_k$, where $X_i \subseteq
\intpoly{\transitions_i}$ and $Y_i\subseteq
\intpoly{\recess{\transitions_i}}$, 
 and $Y_i\neq\emptyset \Rightarrow X_i\neq\emptyset$, such that
$\wspace{X}{Y}=\rats^{n+1}$.
\ecor

\blem 
\label{lem:witness-size-mlc}
If there exists a witness for the nonexistence of a \lrf for
$\intpoly{\transitions_1},\ldots,\intpoly{\transitions_k}$, then there
exists one with $X=X_1\cup\cdots\cup X_k$ and $Y=Y_1\cup\cdots\cup
Y_k$, where $X_i \subseteq \intpoly{\transitions_i}$ and $Y_i\subseteq
\intpoly{\recess{\transitions_i}}$, such that $\sum_{i=1}^k(|X_i|+|Y_i|) \le 2n+3$;
and its bit-size is polynomially bounded in the bit-size
of the input.
\elem

\bprf 
Let $\hat X_i, \hat Y_i$ be the generators of $\inthull{\transitions_i}$.
First, as in Lemma~\ref{lem:witness-size}, we argue that there is a set of at most $n+2$ inequalities
out of $\wspacecons{\bigcup\hat X_i}{\bigcup\hat Y_i}$ that have no solution. These inequalities correspond to
$n+2$ points out of the sets $\hat X_i$, $\hat Y_i$. Let $X_i$ (respectively $Y_i$) be the set of points that come from $\hat X_i$
(respectively $\hat Y_i$). Since $\rfp{\rfcoeff_0}{\vect{\rfcoeff}} = \vec{0}^\trans$ is not a solution, at least one of the points
must come from a set $\hat X_i$. But $n+1$ other points might come from sets $\hat Y_i$. Since a witness must satisfy
$Y_i\ne \emptyset \Rightarrow X_i\ne \emptyset$, we may have to add $n+1$ points to form a valid witness,
for a total of $2n+3$ (clearly, $n+1$ can be replaced by $k$ when $k<n+1$).
Bounding the bit-size of the witness is done as in
Lemma~\ref{lem:witness-size}, but using the $2n+3$ instead of $n+2$,
and $\max_i\bitsize{\transitions_i}$ instead of
$\bitsize{\transitions}$.
\eprf

\bthm
\label{th:conp-inc-mlc}
$\linrfz \in \mathrm{coNP}$.
\ethm

\bprf 
Almost identical to
the proof of Theorem~\ref{th:conp-inc-spl}. Note 
that the witness is given as $X=X_1\cup\cdots\cup X_k$ and
$Y=Y_1\cup\cdots\cup Y_k$, and
the verifier should use the appropriate set of constraints to
check that each $\vec{x}''\in X_i$ is in $\intpoly{\transitions_i}$,
and that each $\vec{y}''\in Y_i$ is in $\intpoly{\recess{\transitions_i}}$.
\eprf

\bexm
\label{ex:ack}
Consider again the integer \mlc loop~\eqref{intro:ex:llrf} from 
Section~\ref{sec:introduction}.
It is a classical \mlc loop for which there is no \lrf.
The integer hulls of the corresponding transition polyhedra are
\[
\begin{array}{rcl}
\inthull{\transitions_1}&=&\convhull\{\vec{x}_1''\}+\cone\{\vec{y}_1'',\vec{y}_2'',\vec{y}_3'',\vec{y}_4''\}\\
\inthull{\transitions_2}&=&\convhull\{\vec{x}_2''\}+\cone\{\vec{y}_5'',\vec{y}_6''\}
\end{array}
\]
where
\[
\begin{array}{llll}
\vec{x}_1''=(0,0,-1,0)^\trans &\quad \vec{y}_1''=(0,0,0,-1)^\trans &\quad \vec{y}_3''=(0,1,0,0)^\trans &\quad \vec{y}_5''=(0,1,0,1)^\trans \\ 
\vec{x}_2''=(0,0,0,-1)^\trans &\quad \vec{y}_2''=(0,0,0,1)^\trans &\quad \vec{y}_4''=(1,0,1,0)^\trans &\quad \vec{y}_6''=(1,0,1,0)^\trans
\end{array}
\]
Let us first consider each path separately.
We get 
\begin{eqnarray}
\wspacecons{\{\vec{x}_1''\}}{\{\vec{y}_1'',\vec{y}_2'',\vec{y}_3''\}} & = & 
\set{ \rfcoeff_0 \ge 0,~ \rfcoeff_1 \ge 1,~ \rfcoeff_2 \ge 0,~ -\rfcoeff_2 \ge 0 }
\label{eq:mlc:psi1}\\
\wspacecons{\{\vec{x}_2''\}}{\{\vec{y}_4'',\vec{y}_5'',\vec{y}_6''\}} & = &
\set{ \rfcoeff_0 \ge 0,~ \rfcoeff_1 \ge 0,~ \rfcoeff_2 \ge 1 } 
\label{eq:mlc:psi2}
\end{eqnarray} 
Both~\eqref{eq:mlc:psi1} and \eqref{eq:mlc:psi2} are
satisfiable. In fact, their solutions define the corresponding \lrfs for each path when
considered separately.
For the \mlc loop, %
we have that
$\wspacecons{\{\vec{x}_1'',\vec{x}_2''
  \}}{\{\vec{y}_1'',\ldots,\vec{y}_6''\}}$ is the conjunction
of the inequalities in~\eqref{eq:mlc:psi1} and \eqref{eq:mlc:psi2},
which is not satisfiable. Thus, while each path has a \lrf, the \mlc
loop does not.
Note that the inequalities $\rfcoeff_2 \ge 1$ and $-\rfcoeff_2 \ge
0$ are enough to get unsatisfiability
of~(\ref{eq:mlc:psi1},\ref{eq:mlc:psi2}), thus, a possible witness is
$X_1=\{\vec{x}_1''\}$, $Y_1=\{\vec{y}_2''\}$,  $X_2=\{\vec{x}_2''\}$, $Y_2=\emptyset$.  Note that it consists of less than
 $2n+3$ points (as $n=2$).
\eexm

\subsection{Synthesizing a Linear Ranking Function}
\label{sec:rf-synthesis}

Although the existence of a \lrf suffices for proving termination,
generating a complete representation of the \lrf is important in some contexts, for instance complexity
analysis where a ranking function provides an upper bound on the number of
iterations that a loop can perform. 
In this section we give a complete algorithm that generates \lrfs for  \mlc
loops given by transition polyhedra
$\transitions_1,\ldots,\transitions_k$.
The following result is directly implied by lemmas~\ref{lem:witness-1-mlc} and
\ref{lem:witness-2-mlc}.

\bthm
\label{thm:rf-synth}
For $1\le i\le k$, let 
$\inthull{\transitions_i} = \convhull\{X_i\} +
\cone\{Y_i\}$
be the integer hull of $\transitions_i$, and 
define $X=X_1\cup\cdots\cup X_k$ and $Y=Y_1\cup\cdots\cup Y_k$.
Then, $\rho(\vec{x})=\vect{\rfcoeff}\cdot\vec{x}+\rfcoeff_0$ is a \lrf
for $\intpoly{\transitions_1},\ldots,\intpoly{\transitions_k}$, if and only if
 $\rfp{\rfcoeff_0}{\vect{\rfcoeff}}$ %
is a
solution of $\wspacecons{X}{Y}$.
\ethm

The following algorithm follows:
(1) Compute the generator representation for each
$\inthull{\transitions_i}$;
(2) Construct $\wspacecons{X}{Y}$; and
(3) Use \lp to find a solution
$\rfp{\rfcoeff_0}{\vect{\rfcoeff}}$ %
for $\wspacecons{X}{Y}$.

\bexm
\label{ex:gen:rfsyn}
Consider again Loop~\eqref{eq:intro:loop1} from Section~\ref{sec:introduction}. 
The integer hull of the transition polyhedron is
\[
\inthull{\transitions} = \convhull\{\vec{x}_1'',\vec{x}_2''\}+\cone\{\vec{y}_1'',\vec{y}_2''\}
\]
where $\vec{x}_1''=(1,1,1,0)^\trans$, $\vec{x}_2''=(1,0,1,-1)^\trans$,
$\vec{y}_1''=(1,1,1,-1)^\trans$, and $\vec{y}_2''=(1,-1,1,-3)^\trans$. 
The formula $\wspacecons{\{\vec{x}_1'',\vec{x}_2''\}}{\{\vec{y}_1'',\vec{y}_2''\}}$
is the conjunction of the following inequalities (we eliminated 
clearly redundant inequalities)
\begin{equation}
\label{eq:synth:psi1}
\set{
\rfcoeff_1+\rfcoeff_2+\rfcoeff_0 \ge 0,~
\rfcoeff_1 +\rfcoeff_0 \ge 0,~
\rfcoeff_1+\rfcoeff_2 \ge 0,~
\rfcoeff_1-\rfcoeff_2 \ge 0,~
\rfcoeff_2 \ge 1
}
\end{equation}
which is satisfiable for $\rfcoeff_1=\rfcoeff_2=1$ and $\rfcoeff_0=-1$,
and therefore, $f(x_1,x_2)=x_1+x_2-1$ is a \lrf.
Note that the loop does not terminate when the variables range
over $\rats$, e.g., for $x_1=x_2=\frac{1}{2}$ (see Figure~\ref{fig:intpoly}(A)).
\eexm

\bexm
\label{ex:gen:rfsyn:2}
Let us consider now Loop~\eqref{eq:intro:loop2} from Section~\ref{sec:introduction}. 
The integer hull of the transition polyhedron is
\[
\inthull{\transitions} =
\convhull\{\vec{x}_1'',\vec{x}_2'',\vec{x}_3'',\vec{x}_4'',\vec{x}_5'',\vec{x}_6''\}+\cone\{\vec{y}_1'',\vec{y}_2''\}
\]
where 
\[
\begin{array}{llll}

\vec{x}_1''=(4,16,1,16)^\trans &\quad \vec{x}_3''=(2,8,1,8)^\trans  &\quad \vec{x}_5''=(4,1,1,1)^\trans  &\quad \vec{y}_1''=(5,0,2,0)^\trans  \\
\vec{x}_2''{}=(1,4,0,4)^\trans &\quad \vec{x}_4''=(1,1,0,1)^\trans  &\quad \vec{x}_6''=(2,1,1,1)^\trans  &\quad  \vec{y}_2''=(5,20,2,20)^\trans \\
\end{array}
\]
The formula
$\wspacecons{\{\vec{x}_1'',\ldots,\vec{x}_6''\}}{\{\vec{y}_1'',\vec{y}_2''\}}$
is the conjunction of the following inequalities (we eliminated
clearly redundant ones)
\begin{equation}
\label{eq:synth:psi2}
\left\{
\begin{array}{llll}
 \rfcoeff_1 \ge 1, &
4\rfcoeff_1 + \rfcoeff_2 + \rfcoeff_0 \ge 0, &
4\rfcoeff_1 + 16\rfcoeff_2 + \rfcoeff_0 \ge 0, &
2\rfcoeff_1 + \rfcoeff_2 + \rfcoeff_0 \ge 0,\\
5\rfcoeff_1+20\rfcoeff_2 \ge 0, &
2\rfcoeff_1 + 8\rfcoeff_2 + \rfcoeff_0 \ge 0, &
\rfcoeff_1 + 4\rfcoeff_2 + \rfcoeff_0 \ge 0, &
\rfcoeff_1+\rfcoeff_2+\rfcoeff_0 \ge 0 \\
\end{array}
\right\}
\end{equation}
which is satisfiable for $\rfcoeff_1=1$,
$\rfcoeff_2=0$ and $\rfcoeff_0=-1$, and therefore, $f(x_1,x_2)=x_1-1$ is a \lrf.
Note that this loop, too, does not terminate when the variables range over $\rats$, e.g., for
$x_1=\frac{1}{4}$ and $x_2=1$ (see Figure~\ref{fig:intpoly}(C)).

If we consider both loops~\eqref{eq:intro:loop1} and~\eqref{eq:intro:loop2} as two paths in an \mlc loop,
then to synthesize \lrfs we use the conjunction of the
inequalities in~\eqref{eq:synth:psi1} and~\eqref{eq:synth:psi2}.
In this case, $\rfcoeff_1=\rfcoeff_2=1$ and $\rfcoeff_0=-1$, is a
solution, but $\rfcoeff_1=1$, $\rfcoeff_2=0$ and $\rfcoeff_0=-1$ is not.
Therefore, $f(x_1,x_2)=x_1+x_2-1$ is a \lrf for both paths, and thus for the \mlc loop, but not
$f(x_1,x_2)=x_1-1$.
\eexm

Given our hardness results, one cannot expect a polynomial-time algorithm. Indeed,
constructing the generator representation of the integer hull of a polyhedron
from the corresponding set of
inequalities $A''_i\vec{x}\le\vec{c}_i''$ 
may require exponential time---the number
of generators itself may be exponential. %
Their bit-size, on the other hand, is polynomial by Theorem~\ref{thm:PIsize}. This is interesting,
since it yields:

\bcor 
\label{cor:bitsize}
Consider an \mlc loop specified by the transition polyhedra
$\transitions_1, \ldots, \transitions_k$, where each $\transitions_i$ is specified by $A_i''\vec{x}\le\vec{c}_i''$.
If there is a \lrf for $\intpoly{\transitions_1},\ldots,\intpoly{\transitions_k}$, there is one whose bit-size is polynomial in the bit-size of $\{A_i''\vec{x}\le\vec{c}_i'' \}$, namely
in $\max_i\bitsize{\transitions_i}$.
\ecor

\bprf
As in the last section, we bound the bit-size of each of the generators of $\inthull{\transitions_i}$ by 
$\vtxsize{\inthull{\transitions_i}} \le 6(2n)^3\cdot\fctsize{\transitions_i} \le 48n^3\cdot\bitsize{\transitions_i}$  for an appropriate $i$.
This means that the bit-size of each equation in $\wspacecons{X}{Y}$, having one of the
forms \eqref{eq:rf1},  \eqref{eq:rf2}, \eqref{eq:rfh1}, or \eqref{eq:rfh2}
is at most%
\footnote{According to Section~\ref{sec:prelim:polyhedra}, the
  bit-size of inequality~\eqref{eq:rf1} is $\bitsize{\vect{\rfcoeff}
    \cdot \vec{x} + \rfcoeff_0 \ge 0}=
  1+\bitsize{(1,\vec{x})}+\bitsize{0} = 5+\bitsize{\vec{x}} \le 5+
  \vtxsize{\inthull{\transitions_i}} \le
  5+48n^3\cdot(\max_i\bitsize{\transitions_i})$. The bit-size of
  \eqref{eq:rf2}-\eqref{eq:rfh2} is similar. }%
~$5+48n^3\cdot(\max_i\bitsize{\transitions_i})$.
Let $\poly{P}$ be the polyhedron defined by $\wspacecons{X}{Y}$, then
$\fctsize{\poly{P}} \le 5+48n^3\cdot(\max_i\bitsize{\transitions_i})$.
If $\wspacecons{X}{Y}$ has a solution, then any vertex of $\poly{P}$ is such a solution, and
yields a \lrf. Using Theorem~\ref{thm:size}, together with the
above bound for $\fctsize{\poly{P}}$ and the fact that 
the dimension of $\poly{P}$ is $n+1$, 
we conclude that there is a
generator representation for $\poly{P}$ in which the bit-size $\vtxsize{\poly{P}}$ of the
vertices is bounded as follows:
\[
\vtxsize{\poly{P}} \le 4\cdot(n+1)^2\cdot\fctsize{\poly{P}} \le 4\cdot(n+1)^2 \cdot (5+48n^3\cdot(\max_i \bitsize{\transitions_i}))
\]
This also bounds the bit-size of the corresponding \lrf.
\eprf

We conclude this section by noting that %
Theorem~\ref{thm:rf-synth} works also for \linrfq, if we
consider $\transitions_i$ instead of $\inthull{\transitions_i}$.
This can be easily proven by reworking the proofs of
Lemmas~\ref{lem:witness-1-mlc} and~\ref{lem:witness-2-mlc} for the
case of $\transitions_i$ instead of $\inthull{\transitions_i}$. We did
not develop this line since the main use of these definitions is
proving the coNP-completeness for \linrfz.
This, however, has an interesting consequence: 
 \linrfq is still PTIME even if the input loop is given in the
generator representations form instead of the constraints form. 
Practically, implementations of polyhedra that use the \emph{double
  description method}, such as PPL~\cite{BagnaraHZ08}, in which both
the generators and constraint representations are kept at the same
time, can use the algorithm of Theorem~\ref{thm:rf-synth} judiciously
when it seems better than algorithms that use the constraints
representation~\cite{DBLP:conf/vmcai/PodelskiR04,DBLP:journals/tplp/MesnardS08}.

\section{Special Cases in PTIME}
\label{sec:ptime}

In this section we discuss cases in which the \linrfz problem is
PTIME-decidable.
We start by a basic observation: when the transition
polyhedron of an \slc loop is \emph{integral},  the
\linrfz and \linrfq problems are equivalent 
(a very similar statement stated by~\citeN[Lemma 3]{CookKRW10}).

\blem
\label{lem:basic}
Let $\transitions$ be a transition polyhedron of a given \slc loop,
and let $\rho$ be an affine linear function.
If $\transitions$ is integral, %
then $\rho$ is a \lrf for
$\transitions$ if and only if $\rho$ is a \lrf for
$\intpoly{\transitions}$.
\elem

\bprf
Let $\transitions$ be an integer polyhedron.
($\Rightarrow$) Suppose that $\rho$ is a \lrf for $\transitions$, then clearly it is
also a \lrf for $\intpoly{\transitions}$ since
$\intpoly{\transitions}\subseteq\transitions$.
($\Leftarrow$) Suppose that $\rho$ is a \lrf for
$\intpoly{\transitions}$, it thus satisfies
(\ref{eq:lrf1},\ref{eq:lrf2}) of Definition~\ref{def:linearrf} for any
integer point in $\transitions$.
However, by definition of an integer polyhedron, every
rational point in $\transitions$ is a convex combination of integer
points from $\intpoly{\transitions}$, this proves that $\rho$
satisfies conditions (\ref{eq:lrf1},\ref{eq:lrf2}) for any rational
point in $\transitions$, as follows.
Choose an arbitrary rational point
$\vec{x}''\in \transitions$. It can be
written as $\vec{x}''= \sum a_i\cdot\vec{x}_i''$ where $a_i>0$, $\sum a_i=1$ and
$\vec{x}_i''\in\intpoly{\transitions}$. Thus,
$\vec{x}''=\tr{\sum a_i\cdot\vec{x}_i}{\strut\sum
  a_i\cdot\vec{x}_i'}$, and 
\begin{align*}
  \rho(\vec{x}) &= (\vect{\rfcoeff}\cdot\sum a_i\cdot\vec{x}_i)+\rfcoeff_0 
                = \sum a_i\cdot(\vect{\rfcoeff}\cdot\vec{x}_i+\rfcoeff_0) \ge 0\\
  \diff{\rho}(\vec{x}'') &=
    (\vect{\rfcoeff}\cdot\sum a_i\cdot\vec{x}_i) - 
      (\vect{\rfcoeff}\cdot\sum a_i\cdot\vec{x}_i') 
 = 
    \sum a_i\cdot\vect{\rfcoeff}\cdot(\vec{x}_i -\vec{x}_i') \ge 1
\end{align*}
\eprf

The above lemma provides an alternative, and \emph{complete},
procedure for \linrfz, namely, compute a constraint representation of its integer hull $\inthull{Q}$ 
and solve \linrfq. 
Note that computing the integer hull might require exponential time,
and might also result in a polyhedron with an exponentially larger description.
This means that the above procedure is exponential in
general;
but this concern is circumvented if the
transition polyhedron is integral to begin with; and in special cases where it is 
known that computing the integer hull is easy. Formally, we call a class of polyhedra
\emph{easy}  if computing its integer hull can be done in polynomial time.

\begin{figure}

\begin{center}
\begin{minipage}{3.3cm}
\begin{tikzpicture}

  \coordinate (a_1) at (0.3,2.3);
  \coordinate (a_2) at (2.4,4.4);
  \coordinate (b_1) at (0.3,2.7);
  \coordinate (b_2) at (2.4,0.6);
  \coordinate (c_1) at (1,4);
  \coordinate (c_2) at (1,1.75);

  \coordinate (c) at (intersection of a_1--a_2 and b_1--b_2);
  \coordinate (d) at (intersection of c_1--c_2 and a_1--a_2);
  \coordinate (e) at (intersection of c_1--c_2 and b_1--b_2);

  \fill[pattern=dots] (e) -- (b_2) -- (2.4,4.4) -- (a_2) -- (d) -- cycle;
  \fill[fill=black!20] (c) -- (d) -- (e) -- cycle;
  \draw [->,thick] (0,2) node  {} -- (2.5,2) node (xaxis) [right] {$x_1$};
  \draw [<->,thick] (0,0) node  {} -- (0,5) node (yaxis) [above] {$x_2$};

  \draw [thick]  (a_1) -- (a_2) {};
  \node[rotate=45] () at (1.8,4.1) {\small \textcolor{DarkBlue}{$x_2{-}x_1{\le}0$}}; 

  \draw [thick]  (b_1) -- (b_2) {};
  \node[rotate=-45] () at (1.8,0.9) {\small \textcolor{DarkBlue}{${-}x_1{-}x_2{\le}{-}1$}};

  \draw [dashed,thick]  (c_1) -- (c_2) {};
  \node[rotate=90] () at (0.8,3.8) {\small \textcolor{blue}{$x_1{\ge}1$}};
 
  \draw[dotted] (yaxis |- c) node[left] {\scriptsize $\mathbf{\frac{1}{2}}$}
        -| (xaxis -| c) node[below] {\scriptsize $\mathbf{\frac{1}{2}}$};
  \draw[dotted] (yaxis |- d) node[left] {\scriptsize $\mathbf{1}$} -- (d);
  \draw[dotted] (yaxis |- e) node[left] {\scriptsize $\mathbf{0}$}
        -| (xaxis -| e) node[below] {\scriptsize $\mathbf{1}$};

  \fill[] (c) circle (2pt);
  \fill[] (d) circle (2pt);
  \fill[] (e) circle (2pt);

  \node[] at (2.8,0.18) {\textcolor{purple}{\textbf{A}}};
\end{tikzpicture}
\end{minipage}
\begin{minipage}{3.4cm}
\vspace*{0.08cm}
\begin{tikzpicture}
  \coordinate (a_1) at (0.1,2.1);
  \coordinate (a_2) at (2.4,4.4);
  \coordinate (b_1) at (0.1,2.9);
  \coordinate (b_2) at (1.5,0);
  \coordinate (c_1) at (1,4);
  \coordinate (c_2) at (1,0.85);
  \coordinate (c) at (intersection of a_1--a_2 and b_1--b_2);
  \coordinate (d) at (intersection of c_1--c_2 and a_1--a_2);
  \coordinate (e) at (intersection of c_1--c_2 and b_1--b_2);

  \fill[pattern=dots] (e) -- (b_2) -- (2.45,0) -- (2.45,4.4) -- (a_2) -- (d) -- cycle;
  \fill[fill=black!20] (c) -- (d) -- (e) -- cycle;

  \draw [->,thick] (0,2) node  {} -- (2.5,2) node (xaxis) [right] {$x_1$};
  \draw [<->,thick] (0,0) node  {} -- (0,5) node (yaxis) [above] {$x_2$};
  \node[] () at (-0.23,2) {\scriptsize $\mathbf{0}$};

  \draw [thick] (a_1) -- (a_2) {};
  \node[rotate=45] () at (1.8,4.1) {\small \textcolor{DarkBlue}{$x_2{-}x_1{\le}0$}}; 
 
  \draw [thick] (b_1) -- (b_2) {};
  \node[rotate=-64, inner sep=0] () at (0.88,0.9) {\small \textcolor{DarkBlue}{${-}{2}x_1{-}x_2{\le}{-}1$}};

  \draw [dashed,thick] (c_1) -- (c_2) {};
  \node[rotate=90] () at (0.8,3.8) {\small \textcolor{blue}{$x_1{\ge}1$}};
 
  \draw[dotted] (yaxis |- c) node[left] {\scriptsize $\mathbf{\frac{1}{3}}$}
        -| (xaxis -| c) node[below] {\scriptsize $\mathbf{\frac{1}{3}}$};
  \draw[dotted] (yaxis |- d) node[left] {\scriptsize $\mathbf{1}$} -- (d);

  \draw[dotted] (yaxis |- e) node[left] {\scriptsize $\mathbf{-1}$} -- (e);

  \fill[] (c) circle (2pt);
  \fill[] (d) circle (2pt);
  \fill[] (e) circle (2pt);

  \node[] at (2.8,0.18) {\textcolor{purple}{\textbf{B}}};
\end{tikzpicture}
\end{minipage}
\begin{minipage}{6.9cm}
\vspace*{-0.15cm}
\begin{tikzpicture}
  \coordinate (a_1) at (0.35,4.5);
  \coordinate (a_2) at (0.35,1.2);
  \coordinate (b_1) at (0.2,1.4111111);
  \coordinate (b_2) at (6.1,4);
  \coordinate (c_1) at (0.2,2.25);
  \coordinate (c_2) at (5.8,4.6);
  \coordinate (d_1) at (1.1,2);
  \coordinate (d_2) at (3.4,4.3333333);
  \coordinate (e_1) at (1.1,2);
  \coordinate (e_2) at (5.544444,3.653);
  \coordinate (a) at (intersection of a_1--a_2 and b_1--b_2);
  \coordinate (b) at (intersection of a_1--a_2 and c_1--c_2);
  \coordinate (c) at (intersection of d_1--d_2 and e_1--e_2);
  \coordinate (d) at (intersection of c_1--c_2 and d_1--d_2);
  \coordinate (e) at (intersection of b_1--b_2 and e_1--e_2);
  \coordinate (f) at (3.1,3.06666);
  \coordinate (g) at (4.8,4);

  \fill[pattern=dots] (c) -- (e) -- (b_2) -- (6.1,4.75) -- (c_2) -- (d) -- cycle;
  \fill[fill=black!20] (a) -- (e) -- (c) -- (d) -- (b) -- cycle;

  \draw [->,thick] (0.1,2) node  {} -- (5.9,2) node (xaxis) [right] {$x_1$};
  \draw [<->,thick] (0.1,0) node  {} -- (0.1,5) node (yaxis) [above] {$x_1'$};

  \draw [thick] (a_1) -- (a_2) {};
  \node[rotate=90] () at (0.55,3.7) {\small \textcolor{DarkBlue}{$4x_1{\ge}1$}};

  \draw [thick] (b_1) -- (b_2) {};
  \node[rotate=21,anchor=west] () at (0.35,1.25) {\small \textcolor{DarkBlue}{$2x_1{-}5x_1'{\le}3$}};

  \draw [thick] (c_1) -- (c_2) {};
  \node[rotate=23,anchor=west] () at (0.3,2.55) {\small \textcolor{DarkBlue}{${-}2x_1{+}5x_1'{\le}1$}};

  \draw [dashed,thick] (d_1) -- (d_2) {};
  \node[rotate=48,anchor=west] () at (2.3,3.45) {\small \textcolor{blue}{${-}x_1{+}x_1'{\le}{-}1$}};

  \draw [dashed,thick] (e_1) -- (e_2) {};
  \node[rotate=19,anchor=west] () at (4.7,3.0) {\small \textcolor{blue}{$\frac{1}{3}x_1{-}x_1'{\le}\frac{1}{3}$}};

  \draw[dotted] (yaxis |- c) node[left] {\scriptsize $\mathbf{0}$}
        -| (xaxis -| c) node[below] {\scriptsize};
  \draw[dotted] (yaxis |- d) node[left] {\scriptsize $\mathbf{1}$}
        -| (xaxis -| d) node[below] {\scriptsize $\mathbf{2}$};
  \draw[dotted] (yaxis |- f) node[left] {\scriptsize}
        -| (xaxis -| f) node[below] {\scriptsize $\mathbf{3}$};
  \draw[dotted] (yaxis |- e) node[left] {\scriptsize}
        -| (xaxis -| e) node[below] {\scriptsize $\mathbf{4}$};
  \draw[dotted] (yaxis |- g) node[left] {\scriptsize $\mathbf{2}$}
        -| (xaxis -| g) node[below] {\scriptsize $\mathbf{5}$};

  \draw[dotted] (yaxis |- a) node[left] {\scriptsize $\mathbf{-\frac{1}{2}}$} -- (a);
  \draw[dotted] (yaxis |- b) node[left] {\scriptsize $\mathbf{\frac{3}{10}}$} -- (b);

  \fill[] (a) circle (2pt);
  \fill[] (b) circle (2pt);
  \fill[] (c) circle (2pt);
  \fill[] (d) circle (2pt);
  \fill[] (e) circle (2pt);
  \fill[] (f) circle (2pt);
  \fill[] (g) circle (2pt);

  \node[] at (6.2,0.185) {\textcolor{purple}{\textbf{C}}};

\end{tikzpicture}
\end{minipage}
\end{center}
 
\caption{The polyhedra associated with three of our examples,
  projected to two dimensions: \textbf{(A)} corresponds to 
  Loop~\eqref{eq:intro:loop1} at Page~\pageref{eq:intro:loop1};
  \textbf{(B)} corresponds to the loop in
   Example~\ref{ex:sc:tvpi-2} at Page~\pageref{ex:sc:tvpi-2}; and
  \textbf{(C)} corresponds Loop~\eqref{eq:intro:loop2} at
  Page~\pageref{eq:intro:loop2}.
  Dashed lines are added when computing the integer hull; dotted areas
  represent the integer hull; gray areas are rational points
  eliminated when computing the integer hull.}
\label{fig:intpoly}
\end{figure}

\bexm
\label{ex:sc:general}
Consider again the \slc loop~\eqref{eq:intro:loop2} of
Section~\ref{sec:introduction}.
The transition polyhedron is not integral, computing its integer hull
adds the inequalities $-x_1+x_1' \le -1$ and $\frac{1}{3}x_1-x_1' \le \frac{1}{3}$. This
is depicted in Figure~\ref{fig:intpoly}(C).
Applying \linrfq on this loop does not find a \lrf since it does
not terminate when the variables range over \rats, however, applying it on the integer hull
finds the \lrf $f(x_1,x_2)=x_1-1$.
\eexm

\bcor
\label{cor:basic-1}
The \linrfz problem is PTIME-decidable for \slc loops in which the
transition polyhedron $\transitions$ is guaranteed
 to be integral.
 This also applies to any easy class of polyhedra, namely a class where the
integer hull is PTIME-computable.
\ecor

\bprf
Immediate from Lemma~\ref{lem:basic} and the fact that \linrfq is
PTIME-decidable.
\eprf

\bcor
\label{cor:basic-2}
The \linrfz problem is PTIME-decidable for \slc loops in which the
condition polyhedron $\states$
is guaranteed to be integral, or belongs to an easy class, and the
update is affine linear with integer coefficients. 
\ecor

\bprf 
We show that if $\states$ is integral, the transition polyhedron $\transitions$
is also integral, and thus
Corollary~\ref{cor:basic-1} applies.
Let the condition polyhedron $\states$ be integral, and
the update be $\vec{x}' = A'\vec{x} + \vec{c}'$ where the entries
of $A'$ and $\vec{c}'$ are integer.
Let $\vec{x}'' \in \transitions$, that is,
$\vec{x} \in \states$ and $\vec{x}' = A'\vec{x} + \vec{c}'$.
Since $\states$ is integral,  $\vec{x}$ is a convex
combination of some integer points. I.e., $\vec{x} = \sum a_i\cdot
\vec{x}_i$ where $a_i>0$, $\sum a_i=1$ and $\vec{x}_i\in\intpoly{\states}$.
Hence, $\vec{x}' = A'(\sum a_i\cdot\vec{x}_i) + \vec{c}' =\sum
a_i\cdot(A'\vec{x}_i + \vec{c}')$ and
\begin{align*}
 \vec{x}''= 
 \trdisp{\vec{x}\phantom{'}}{\vec{x}'} =  
 \trdisp{\sum a_i\cdot \vec{x}_i}{\sum a_i\cdot(A'\vec{x}_i + \vec{c}')}
=\sum a_i\cdot\trdisp{\phantom{A'}\vec{x}_i}{A'\vec{x}_i + \vec{c}'}
\end{align*}
Now note that $\tr{\phantom{A'}\vec{x}_i}{A'\vec{x}_i + \vec{c}'}$ are integer points from
$\intpoly{\transitions}$, which implies that $\vec{x}''$ is a convex
combination of integer points in $\transitions$.  Hence,
$\transitions$ is integral.
\eprf

Corollaries~\ref{cor:basic-1} and \ref{cor:basic-2}  suggest looking for
classes of
\slc loops where we can easily ascertain that $\transitions$ is integral,
or that its integer hull can be computed in polynomial time.
In what follows we address such cases:
Section~\ref{sec:special:intpoly} discusses special cases in which
the transition or condition polyhedron is integral by construction;
Section~\ref{sec:special:bnv} considers cases in which the 
the transition or condition polyhedron can be separated into independent groups of constraints,
each involving few variables;
Section~\ref{sec:special:octagons} discusses the case of
octagonal relations;
Section~\ref{sec:special:strongpoly} shows that for some cases \linrfz
is even strongly polynomial; and
Section~\ref{sec:special:mlc} extends the results to \mlc loops.

\subsection{Loops Specified by Integer Polyhedra}
\label{sec:special:intpoly}

There are some well-known examples of polyhedra that are known to be
integral due to some structural property. This gives us classes of
\slc loops where \linrfz is in PTIME.  The examples below follows %
\citeN{Schrijver86}, where the proofs of the lemmas can be found.

\blem[\normalfont{\cite[Eq.~(9), p.~230]{Schrijver86}}]
\label{lem:cone}
For any rational matrix $B$, the \emph{cone} $\{ \vec{x}
\mid B\vec{x} \le \vec{0} \}$ is an integer polyhedron. 
\elem

\bcor
The \linrfz problem is PTIME-decidable for \slc loops of the form
\begin{align*}
\mathit{while}~(B\vec{x} \le  \vec{0})~\mathit{do}~ \vec{x}' = A'\vec{x} + \vec{c}'
\end{align*}
where the entries in $A'$ and $\vec{c}'$ are integer.
\ecor

Recall that a matrix $A$ is totally unimodular if each subdeterminant
of $A$ is in $\{0,\pm 1\}$. In particular, the entries of such matrix
are from $\{0,\pm 1\}$.

\blem[\normalfont{\cite[Th.~19.1, p.~266]{Schrijver86}}]
\label{lem:tum}
For any totally unimodular matrix $A$ and integer vector $\vec b$, the
polyhedron $\poly{P}=\{ \vec{x} \mid A\vec{x} \le \vec{b} \}$ is integral. 
\elem

For brevity, if a polyhedron
$\poly{P}$ is specified by $A\vec{x} \le \vec{b}$ in which $A$ is a
totally unimodular matrix and $\vec{b}$ an integer vector,  we say that
$\poly{P}$ is totally unimodular.

\bcor
The \linrfz problem is PTIME-decidable for \slc loops in which (1) the
transition polyhedron $\transitions$ is totally unimodular; or (2)
the condition polyhedron $\states$ is totally unimodular and the
update is affine linear with integer coefficients.
\ecor
As a notable example,
difference bound constraints~\cite{DBLP:journals/toplas/Ben-Amram08,DBLP:conf/vmcai/BozzelliP12,DBLP:conf/tacas/BozgaIK12} 
are defined
by totally unimodular matrices. Such constraints have the
form  $x-y\le d$ with $d\in\rats$; constraints of the form $\pm x\le d$ can also be admitted.
In the integer case we can always tighten $d$ to $\lfloor d \rfloor$ and thus
get an integer polyhedron.
It might be worth mentioning that checking if a matrix is totally
unimodular can be done in polynomial
time~\cite[Th.~20.3, p.~290]{Schrijver86}.

On the other hand, highlighting the gap between linear-ranking proofs and termination proofs
in general, we may note that \mlc loops with difference bounds, even restricted to the forms
$x_i \ge 0$ and $x_i' \le x_j+c$, already have an undecidable
termination problem~\cite{DBLP:journals/toplas/Ben-Amram08}.

\subsection{Bounded Number of Variables}
\label{sec:special:bnv}

In this section we consider cases in which the input loop can be
decomposed into different components that do not share variables, and
each involves at most $N$ variables for an arbitrary \emph{fixed}
$N$. We start with $N=2$, and towards the end of this section we
consider larger values of $N$.

\emph{Two variable per inequality} constraints (\tvpi for short) are
inequalities of the form $ax+by \le d$ with $a,b,d\in\rats$. Clearly,
polyhedra defined by such inequalities are not guaranteed to be
integral. See, for example, Figure~\ref{fig:intpoly}(B).
\citeN{DBLP:journals/siamcomp/Harvey99} showed that for
\emph{two-dimensional} polyhedra, which are specified by \tvpi
constraints by definition, the integer hull can be computed in
$O(m\log A_{max})$ where $m$ is the number of inequalities and
$A_{max}$ is the magnitude of the largest coefficient.

\bdfn 
\label{def:ptvpi}
Let $T$ be a set of constraints. We say that the polyhedron specified by $T$ is a
\emph{product of independent two-dimensional \tvpi polyhedra}
(\ptvpi for short), if $T$ can be partitioned into $T_1,\ldots,T_n$
such that (1) each $T_i$ is two-dimensional, i.e., involves at most
two variables; and (2) each distinct $T_i$ and $T_j$ do not share
variables.
\edfn

\blem
\label{lem:ptvpi}
The integer hull of \ptvpi polyhedra can be computed in polynomial time.
\elem

\bprf 
Recall that a polyhedron
$\poly{P}$ is integral if and only if each of its \emph{faces} has an
integer point. A~face~of $\poly{P}$ is obtained by turning some
inequalities to equalities such that the resulting polyhedron in not
empty (over the rationals).
First we claim that if $T_1$ and $T_2$ are two sets of inequalities that do not
share variables, and the corresponding polyhedra $\poly{T}_1,\poly{T}_2$ are integral,
then $T_1\cup T_2$ specifies an integral polyhedron $\poly{T}$ over the combined set of variables.
Note that $\poly{T} = \poly{T}_1\times \poly{T}_2$.
To prove our claim, note that
 a face of $\poly{T}$ is specified by some constraints defining a face of $\poly{T}_1$ and some constraints
defining a face of $\poly{T}_2$. Since each has an integer point,
we get an integer point (in the combined set of variables) satisfying all constraints, i.e., belonging
to a face of $\poly{T}$.

To compute the integer hull of a \ptvpi polyhedron $\poly{T}$, we
partition its constraints $T$ into independent sets $T_1,\ldots,T_n$, and
compute the integer hull of each $\poly{T}_i$ in polynomial time 
using Harvey's method. The above argument shows that $\inthull{\poly{T}_1}\times\dots\times\inthull{\poly{T}_n}$
 is integral.
Moreover, every integer point of $\poly{T}$, when projected into the set of variables associated with $T_i$,
is still integer, hence in $\inthull{\poly{T}_i}$,
which shows that $\inthull{\poly{T}_1}\times\dots\times\inthull{\poly{T}_n}$ is the integer hull of $\poly{T}$.
\eprf

The above approach can easily be generalized. Given any 
polyhedron, we first decompose it into independent sets of inequalities, in
polynomial time (these are the connected components of an obvious graph),
 and then check if each set is covered by any of the
special cases for which the integer hull can be efficiently computed.

\bcor 
\label{cor:ptvpi}
The \linrfz problem is PTIME-decidable for \slc loops in which: (1)
the transition polyhedron $\transitions$ is \ptvpi; or (2) the
condition polyhedron $\states$ is \ptvpi, and the update is affine linear
with integer coefficients.
\ecor

\bexm
\label{ex:sc:tvpi-1}
Consider the following \slc loop, as an example for case~(1) of
Corollary~\ref{cor:ptvpi}
\begin{equation}
\label{loop:tvpi-1}
\begin{array}{l}
\while~(4x_1 \ge 1,~\ x_2\ge 1)~\wdo~\\
~~~~~~2x_1-5x_1' \le 3,~\ -2x_1+5x_1' \le 1,~\ x_2'=x_2+1
\end{array}
\end{equation}
Applying \linrfq does not find a \lrf since the loop does not
terminate when the variables range over \rats, e.g., for $x_1=\frac{1}{4}$ and $x_2=1$.
The transition polyhedron is not integral, however, it is \ptvpi since
the constraints can be divided into $T_1=\{{4x_1 \ge 1}, \ {2x_1-5x_1' \le 3}, \ {-2x_1+5x_1'
\le 1}\}$ and $T_2=\{{x_2\ge 1}, {x_2'=x_2+1}\}$.
It is easy to check that $\poly{T}_2$ is already integral. Computing the integer
hull of $\poly{T}_1$ adds the inequalities $-x_1+x_1' \le -1$ and $\frac{1}{3}x_1-x_1' \le \frac{1}{3}$. See
Figure~\ref{fig:intpoly}(C).
Now \linrfq finds the \lrf $f(x_1,x_2)=x_1-1$.
\eexm

\bexm
\label{ex:sc:tvpi-2}
Consider the following \slc loop, as an example for case~(2) of
Corollary~\ref{cor:ptvpi}
\begin{equation}
\begin{array}{@{}l@{}}
\while~( -x_1+x_2 \le 0,~\ -2x_1-x_2 \le -1,~\ x_3\le 1 ) ~do\\
~~~~~~x_1' = x_1,~\ x_2' = x_2-2x_1+x_3,~\ x_3'=x_3
\end{array}
\end{equation}
Applying \linrfq does not find a \lrf since it does not terminate
over \rats, e.g., for $x_1=x_2=\frac{1}{2}$ and $x_3=1$.
The condition polyhedron is not integral, but it is \ptvpi since the constraints can
be divided into $T_1=\{-x_1+x_2 \le 0 ,~\  -2x_1-x_2 \le -1\}$ and
$T_2=\{x_3\le 1\}$. 
It is easily seen that $\poly{T}_2$ is already integral; computing the
integer hull of $\poly{T}_1$ adds $x_1\ge 1$. See Figure~\ref{fig:intpoly}(B).
Now \linrfq finds the \lrf $f(x_1,x_2,x_3)=2x_1+x_2-1$.
Note that the update in this loop involves constraints which are not
\tvpi.
\eexm

The special case described above is based on the fact that \linrfz for
two-dimensional polyhedra is PTIME. In the rest of this section we
show that it is PTIME for $N$-dimensional polyhedra, for a
\emph{fixed} constant $N$, as well.
Given a polyhedron $\poly{P}$, as a set of linear inequalities
$A\vec{x}\leq\vec{b}$ with $n$ variables and $m$ inequalities,
\citeN[Sec.~4.2]{Hartmann88} describes an algorithm for computing the
vertices $\vec{v}_1,\ldots,\vec{v}_\ell$ of $\inthull{\poly{P}}$.
This algorithm is exponential in the the number of variables $n$ (for fixed $n$,
it is polynomial in the bit-size of $\poly{P}$).
This means
that if we require $n\leq N$, for an arbitrary \emph{fixed} $N$, 
we get a polynomial-time algorithm. Note that in such case the number
of vertices, $\ell$, and the bit-size of each one, are both polynomial in the
bit-size of $\poly{P}$.

Assuming that $\poly{P}$ represents a transition or condition
polyhedron, in order to apply \linrfq it is not enough to have the
vertices of $\inthull{\poly{P}}$, what we need is a complete 
representation of $\inthull{\poly{P}}$
by constraints or by generators .
The latter is excluded since the recession cone of $\inthull{\poly{P}}$ (which is
the same as the one of $\poly{P}$) can have an exponential number of
generators. We next explain how to make use of the constraints representation.

First note that
$\inthull{\poly{P}}=\convhull\{\vec{v}_1,\ldots,\vec{v}_\ell\}+\recess{\poly{P}}$,
where $\recess{\poly{P}}$ is the recession cone of $\poly{P}$, and
recall that $\recess{\poly{P}} = \{ \vec{y}\in\rats^n \mid A\vec{y}
\le \vec{0}\}$. Define the polyhedron $\poly{P}'$ as:
\[
\poly{P}' = \left\{ (\vec{x},\vec{a},\vec{y}) ~\left|~ a_1\geq 0\land
    \cdots \land a_\ell\geq 0 \land \sum_{i=1}^\ell a_i=1 \land
    \vec{x}=\left(\sum_{i=1}^\ell a_i\vec{v}_i \right) + \vec{y} \land
    A\vec{y} \le \vec{0} \right. \right\}
\]
It is easy to see that $\vec{x} \in \inthull{\poly{P}}$ if and only if
$(\vec{x},\vec{a},\vec{y}) \in \poly{P}'$ for some $\vec{a}$ and
$\vec{y}$.
The constraint representation for $\inthull{\poly{P}}$ can be
computed by projecting $\poly{P}'$ on its first $n$ components
$\vec{x}$, however, this may take an exponential time.  
The projection can be avoided by directly using ${\poly{P}}'$,
and constraining the \lrf to not use variables from
$(\vec{a},\vec{y})$\footnote{%
Such a constraint can be easily imposed when using the Podelski-Rybalchenko procedure, as described
in Sections \ref{sec:special:strongpoly} and \ref{sec:special:mlc}.}.
This yields a polynomial-time algorithm for \linrfz, for the case of
$N$-dimensional polyhedra, since the bit-size of $\poly{P}'$ is
polynomial in the bit-size of $\poly{P}$.
Clearly, the special case of \ptvpi constraints can be generalized
such that each component is an $N$-dimensional polyhedron.

\subsection{Octagonal Relations}
\label{sec:special:octagons}

\tvpi constraints in which the coefficients are from $\{0,\pm 1\}$ have
received considerable attention in the area of program analysis. Such
constraints are called \emph{octagonal relations}~\cite{Mine:2006}.
A particular interest was in developing efficient algorithms for
checking satisfiability of such relations, 
as well as inferring all
implied \emph{octagonal} inequalities, for
variables ranging either over $\rats$ or over $\ints$.

Over $\rats$, this is done by computing the \emph{transitive closure}
of the relation, which basically adds inequalities that result from
the addition of two existing inequalities, and possibly scaling to
obtain coefficients of $\pm1$. For example: starting from the set of inequalities $\{ {-x_1+x_2
\le 0},\ -x_1-x_2 \le -1 \}$, we add $-2x_1 \le -1$, or, after
scaling, $-x_1 \le -\frac{1}{2}$.
Over $\ints$, this is done by computing the \emph{tight closure},
which in addition to transitivity, is closed also under tightening. This operation
replaces  $a x + b y\le d$ by $a x + b y\le \lfloor
d\rfloor$. For example, tightening $-x_1 \le -\frac{1}{2}$ yields
$-x_1 \le -1$.
The \emph{tight closure} can be computed in polynomial time~\cite{HarveyStuckey97,Bagnara08,Revesz-SARA09}.
Since the tightening eliminates some non-integer points, it is tempting 
to expect that it actually computes the integer hull. It is easy to
show that this is true for two-dimensional relations, but it is
false already in three dimensions,  as we show in the following example.

\bexm
\label{ex:sc:oct-1}
Consider the following \slc loop
\begin{equation}
\label{loop:tclosure-1}
\begin{array}{@{}l@{}}
\while~( x_1+x_2 \le 2,~\ x_1+x_3 \le 3,~\ x_2+x_3\le 4 ) ~\wdo~ \\
~~~~~~x_1' = 1-x_1,~\  x_2' = 1+x_1,~\ x_3'=1+x_2
\end{array}
\end{equation}
Note that the transition polyhedron is octagonal, but not integral.
Applying \linrfq does not find a \lrf, since the loop does not terminate
over $\rats$, e.g., for $x_1=\frac{1}{2}$, $x_2=\frac{3}{2}$, and
$x_3=\frac{5}{2}$.
Computing the tight closure does not change the transition (or condition) polyhedron,
and thus, it is of no help in finding the \lrf.
In order to obtain the integer hull of the transition (or condition) polyhedron we
should add $x_1+x_2+x_3 \le 4$, which is not an octagonal
inequality. 
Having done so, \linrfq finds the \lrf $f(x_1,x_2,x_3)=-3x_1-4x_2-2x_3+12$.
\eexm

Although it is not guaranteed that the tight closure of an octagonal
relation corresponds to its integer hull, in practice, it does in many
cases.
Thus, since it can be computed in polynomial time, we suggest
computing it before applying \linrfq on loops that involve such
relations. The above example shows that this does not give us a complete
polynomial-time algorithm for \linrfz over octagonal relations.

\bexm
\label{ex:sc:oct-2}
Consider Loop~\eqref{eq:intro:loop1} of Section~\ref{sec:introduction} in which
the condition is an octagonal relation. \linrfq fails to find a
\lrf since 
the loop may fail to terminate for rational-valued variables.
Computing the tight closure of the condition polyhedron adds the inequality $x_1 \ge 1$, making the polyhedron integral. See 
Figure~\ref{fig:intpoly}(A). Now \linrfq finds the \lrf
$f(x_1,x_2) = x_1+x_2-1$.
Let us consider an example with higher dimensions
\begin{equation*}
\label{loop:tclosure-2}
\begin{array}{@{}l@{}}
\while~(-x_1{+}x_2 \le 0,~\ -x_1-x_2 \le -1,~\ x_2-x_3 \le 0,~\ -x_2-x_3 \le -1)~\wdo\\
~~~~~~x_1'=x_1,~\ x_2'=x_2-x_1-x_3+1,~\ x_3'=x_3
\end{array}
\end{equation*}
The condition polyhedron is octagonal, but not integral;
moreover, it is not \ptvpi.
\linrfq does not find a \lrf (indeed the loop fails to terminate 
for $x_1=x_2=x_3=\frac{1}{2}$). 
Computing the tight closure of the condition adds $-x_1 \le -1$ and
$-x_3 \le -1$, which results in the integer hull. Now \linrfq
finds the \lrf $f(x_1,x_2,x_3)=x_1+x_2-1$.
\eexm

A polynomial-time algorithm for computing the integer hull of
octagonal relations is, unfortunately, ruled out by examples of
such relations whose integer hulls have exponentially many
facets. 

\bthm
There is no polynomial-time algorithm for computing the integer hull
of general octagonal relations.
\ethm

\bprf
We build an octagonal relation $\poly{O}$, such that the minimum
number of inequalities required to describe its integer hull
$\inthull{\poly{O}}$ is not polynomial in the number of inequalities
in $\poly{O}$.
For a complete graph $K_n=\tuple{V,E}$, we let
$\poly{P}$ be defined by the set of inequalities $\{ x_e \ge 0 \mid
e\in E\}\cup\{ \sum_{v\in e} x_e \le 1 \mid v\in V\}$. Here every edge
$e\in E$ has a corresponding variable $x_e$, and the notation $v\in e$
means that $v$ is a vertex of edge $e$. Note that $\poly{P}$ is not
octagonal.
It is well-known that $\inthull{\poly{P}}$, the matching polytope of
$K_n$, has at least $\binom{n}{2}+2^{n-1}$ facets~\cite[Sec.~18.2,
p.~251]{Schrijver86}, and thus any set of inequalities that defines
$\inthull{\poly{P}}$ must have at least the same number of
inequalities.
Now let $\poly{O}$ be defined by
$\{ x_e \ge 0 \mid e\in E\}\cup\{ x_{e_1}+x_{e_2} \le 1 \mid v\in e_1,
v\in e_2 \}$, which includes $n+n\cdot\binom{n-1}{2}$ \emph{octagonal
  inequalities}.
It is easy to see that the integer solutions of $\poly{P}$ and
$\poly{O}$ are the same, 
and thus $\inthull{\poly{P}} =
\inthull{\poly{O}}$. This means that any set of inequalities that
define $\inthull{\poly{O}}$ must have at least $\binom{n}{2}+2^{n-1}$
inequalities. 
Therefore, any algorithm that computes such a representation must add at
least $\binom{n}{2}+2^{n-1}-n-n\cdot\binom{n-1}{2}$ inequalities to
$\poly{O}$, which is super-polynomial in the size of $\poly{O}$.
Unsurprisingly, the tight closure of $\poly{O}$ does not yield its integer
hull  (it only adds $x_e \le 1$ for each $x_e$).
\eprf

Note that the above theorem does not rule out a
polynomial-time algorithm for \linrfz,  for \slc loops in which the transition polyhedron
$\transitions$ is octagonal, or where the condition polyhedron is octagonal and
the update is affine linear with integer coefficients. It just rules out an algorithm that is based
on computing the integer hull of the polyhedra.
However, the coNP-hardness proof of
Section~\ref{sec:conp-hardness} could be also carried out by a reduction
from $3$SAT that produces an \slc loop where the condition is
octagonal and the update is affine linear with integer coefficients---so at least for this class there is,
presumably,
no polynomial solution. We present this reduction next.

\bthm
\label{th:conp-hardness-oct}
The \linrfz problem is strongly coNP-hard, even for deterministic \slc loops where the guard is octagonal.
\ethm

\bprf 
We exhibit a polynomial-time reduction 
from 3SAT to the complement of \linrfz 
(keeping all the numbers
in the resulting instance polynomially bounded, to obtain strong coNP-hardness).

Consider a 3SAT instance 
given as a collection of $m$ clauses, $C_1,\dots,C_{m}$, each clause $C_i$ consisting of 
three literals $L_i^j \in \{ x_1,\dots,x_n,\, \bar x_1,\dots, \bar x_n\}$.
We construct a loop over $4m$ variables. Variable $x_{ij}$ corresponds to $L_i^j$. Variable 
$x_{i0}$ is a control variable to ensure the satisfaction of clause $i$, as will be seen below.
Let $C$ be the set of all conflicting pairs, that is, pairs $((i,j),(r,s))$ such that $L_i^j$ is the complement
of $L_r^s$, and also pairs $((i,j),(i,j'))$ with $1\le j < j' \le 3$.
The loop we construct is:
\begin{align*}
\mathit{while} &  \left(\bigwedge_{((i,j),(r,s)) \in C}  x_{ij}+x_{rs}\le 1 \right) 
 \land \left(\bigwedge_{1\le i \le m ,\ 0\le j\le 3} 0\le  x_{ij}  \le 1 \right) \\
 \mathit{do} 
&   \left(\bigwedge_{\ 1\le i \le m ,\ 1\le j\le 3 } x'_{ij} = x_{ij} \right) 
 \land \left(\bigwedge_{ 1\le i \le m } x'_{i0} = x_{i0} + x_{i1} + x_{i2} + x_{i3} -1 \right) 
\end{align*}

Suppose the formula is satisfiable. 
For every clause, choose  a satisfied literal, and set the corresponding variable
$x_{ij}$ to 1; let all other variables be zero.
Observe that all the inequality constraints are fulfilled, and that the value of each $x_{i0}'$
does not change. Hence, the loop does not terminate, and does not have any ranking function, let alone a \lrf.

Next, suppose the formula is unsatisfiable.
An initial state for which the loop guard is enabled 
may be interpreted as a selection of non-conflicting literals. Since no such selection can satisfy all clauses,
looking at the update of the $x_{i0}$ variables, we see that some may stay unchanged, while some (and at least one)
will decrease. It follows that $\sum_i x_{i0}$ is a \lrf.
\eprf

\subsection{Strongly Polynomial Cases}
\label{sec:special:strongpoly}

Polynomial-time algorithms for
\linrfq~\cite{DBLP:conf/vmcai/PodelskiR04,DBLP:journals/tplp/MesnardS08,ADFG:2010}
inherit their complexity from that of \lp.
While it is known that \lp can be solved by a polynomial-time algorithm,
it is an open problem whether it has a \emph{strongly polynomial} algorithm.
Such an algorithm should perform a number of elementary arithmetic
operations polynomial in the \emph{dimensions} of the input matrix
instead of its bit-size (which accounts for the size of the matrix entries),
and such operations should be performed on numbers of
size which is polynomial to the input bit-size.
However, there are some cases for which \lp is known to have a strongly polynomial
algorithm. We first use these cases to define classes of \slc loops
for which \linrfq has a strongly polynomial algorithm, which we then
use to show that \linrfz has a strongly polynomial algorithm for some corresponding classes of \slc loops.
Our results are based on the following result by~\citeN{Tardos86} (quoting~\citeN[p.~196]{Schrijver86}).

\bthm[Tardos]
\label{th:tardos}
There is an algorithm which solves a rational \lp problem
$\max \{\vec{c}\cdot\vec{x} \mid A \vec{x} \le \vec{b}\}$ with at most
$P(size(A))$ elementary arithmetic operations on numbers of size
polynomially bounded by $size(A,\vec{b},\vec{c})$, for some polynomial
$P$.
\ethm

Note that the number of arithmetic operations
required by the \lp algorithm only depends on the bit-size of $A$. 
Clearly, if we restrict the \lp problem to cases in which the
bit-size of the entries of $A$ is bounded by a constant, then
$size(A)$ depends only on its dimensions, and we get a strongly
polynomial time algorithm. In
particular we can state the following.

\bcor
\label{cor:tardos}
There exists a strongly polynomial algorithm to solve an \lp problem
$\max \{\vec{c}\cdot\vec{x} \mid A \vec{x} \le \vec{b}\}$ where the entries of
$A$ are $\{0,\pm 1, \pm 2\}$.
\ecor

We can use this to show that \linrfq can sometimes be implemented with
strongly polynomial complexity. To do this, we use the
Podelski-Rybalchenko formulation of the
procedure~\cite{DBLP:conf/vmcai/PodelskiR04}, slightly modified to
require that the \lrf decreases at least by $1$ instead of by some $\delta>0$
(this modification only affects \eqref{eq:pr:5} below; the right-hand side of the
inequality is $-\delta$, so in their formulation the inequality was 
$\vect{\prcoeffb} \cdot\vec{c}'' < 0$).

\bthm[Podelski-Rybalchenko]
\label{th:pr04}
Given an \slc loop with a transition polyhedron
$\transitions\subseteq\rats^{2n}$, specified by
$A''\vec{x}''\le\vec{c}''$, let $A''=(A\ A')$ where each $A$ and $A'$
has $n$ columns and $m$ rows each, and let $\vect{\prcoeffa},
\vect{\prcoeffb}$ be row vectors of different $m$ rational variables
each.
A \lrf for $\transitions$ exists if and only if there is a (rational)
solution to the following set of constraints

\begin{subequations} \label{eq:pr}
\begin{align} 
\label{eq:pr:1} 
\vect{\prcoeffa},\vect{\prcoeffb}&\ge\vec{0}^{\trans} \,,\\ 
\label{eq:pr:2} 
\vect{\prcoeffa} \cdot A' &= \vec{0}^{\trans} \,,\\
\label{eq:pr:3} 
(\vect{\prcoeffa} - \vect{\prcoeffb})\cdot A &= \vec{0}^{\trans} \,,\\
\label{eq:pr:4} 
\vect{\prcoeffb}\cdot (A+A') &= \vec{0}^{\trans} \,,\\
\label{eq:pr:5} 
\vect{\prcoeffb} \cdot\vec{c}'' &\le -1 \,. 
\end{align}
\end{subequations} 
\ethm

\bthm 
\label{th:stongpoly}
The \linrfq problem is decidable in strongly polynomial time for \slc
loops specified by $A''\vec{x}''\le\vec{c''}$ where the coefficients of
$A''$ are from $\{0,\pm 1\}$.
\ethm

\bprf 
First observe that, in Theorem~\ref{th:pr04}, when the matrix $A''$
has only entries from $\{0,\pm 1\}$, then all coefficients in the
constraints (\ref{eq:pr:1}--\ref{eq:pr:4}) are from $\{0,\pm 1, \pm
2\}$. Moreover, the number of inequalities and variables in
(\ref{eq:pr:1}--\ref{eq:pr:4}) is polynomial in the dimensions of
$A''$.
Now let us modify the Podelski-Rybalchenko procedure such that instead
of testing for feasibility of the constraints
(\ref{eq:pr:1}--\ref{eq:pr:5}), we consider the minimization of
$\vect{\prcoeffb}\cdot\vec{c}''$ under the other constraints
(\ref{eq:pr:1}--\ref{eq:pr:4}). 
Clearly, this answers the same question since:
(\ref{eq:pr:1}--\ref{eq:pr:5}) is feasible, if and only if the
minimization problem is unbounded, or the minimum is negative.
This brings the problem to the form required by
Corollary~\ref{cor:tardos} and yields our result.
\eprf

\bcor
\label{cor:linrfz:stongpoly}
The \linrfz problem is decidable in strongly polynomial time for \slc
loops, specified by $A''\vec{x}''\le\vec{c}''$ where the coefficients
of $A''$ are from $\{0,\pm 1\}$, that are covered by any of the
special cases of Section~\ref{sec:special:intpoly} and the special
case of $\ptvpi$ constraints of Section~\ref{sec:special:bnv}.
\ecor

\bprf 
In the cases of Section~\ref{sec:special:intpoly}, the transition
polyhedron is guaranteed to be integral.
In the \ptvpi case of Section~\ref{sec:special:bnv}: (1) the
integer hull can be computed using Harvey's procedure, which is
strongly polynomial in this case since the entries of $A$ are from
$\{0,\pm 1\}$. This can be done also using the tight closure of
2-dimensional octagons; and
(2) the \tvpi constraints that we add when computing the integer hull
have coefficients from $\{0,\pm 1\}$, and the number of such
constraints is polynomially bounded by the number of the original
inequalities.
Thus, by Theorem~\ref{th:stongpoly}, we can apply a strongly
polynomial-time algorithm for \linrfq.
\eprf

\subsection{Multipath Loops}
\label{sec:special:mlc}

It follows immediately from the definitions that an affine linear function $\rho$ is a \lrf for an \mlc loop with
transition polyhedra $\transitions_1,\ldots,\transitions_k$ if and only if
 it is a \lrf for each $\transitions_i$. Thus, if we have the
set of \lrfs for each $\transitions_i$, we can simply take the
intersection and obtain the set of \lrfs for
$\transitions_1,\ldots,\transitions_k$.
In the Podelski-Rybalchenko procedure, the set of solutions for the
inequalities (\ref{eq:pr:1}--\ref{eq:pr:5}) defines the set of \lrfs
for the corresponding \slc loop as follows.

\blem %
\label{lemma:pr04}
Given an \slc loop with a transition polyhedron $\transitions$, 
specified by $A''\vec{x}''\le\vec{c}''$, let
$\prcs{\vect{\prcoeffa}}{\vect{\prcoeffb}}{A''}{c''}$ be the
conjunction of (\ref{eq:pr:1}--\ref{eq:pr:5}).
Then, $\rho(\vec{x})=\vect{\rfcoeff}\cdot\vec{x}+\rfcoeff_0$ is a \lrf for
$\transitions$ if and only if
$\prcs{\vect{\prcoeffa}}{\vect{\prcoeffb}}{A''}{c''}$ has a solution such that
$\vect{\rfcoeff}=\vect{\prcoeffb}\cdot A'$ and
$\rfcoeff_0\ge\vect{\prcoeffa}\cdot\vec{c}''$.
\elem

Next we show how to compute, using
the above lemma, the intersection of sets of \lrfs for several
transition polyhedra, and thus obtain the set of \lrfs for a given
\mlc loop (a very similar statement stated by~\citeN[Lemma 3]{CookKRW10}).	

\bthm
\label{thm:linrfq:mlc}
Given an \mlc loop with transition polyhedra
$\transitions_1,\ldots,\transitions_k$, each specified by
$A''_i\vec{x}''\le\vec{c}''_i$, let
$\prcs{\vect{\prcoeffa}_i}{\vect{\prcoeffb}_i}{A''_i}{c''_i}$ be the
constraints (\ref{eq:pr:1}--\ref{eq:pr:5}) for the $i$-th path, and
$\rfp{\rfcoeff_0}{\vect{\rfcoeff}}$ be $n+1$ rational variables.
Then there is a \lrf for $\transitions_1,\ldots,\transitions_k$ if and only if
the following is feasible (over the rationals)
\begin{equation}
\label{eq:prmlc}
\bigwedge_{i=1}^k 
  \prcs{\vect{\prcoeffa}_i}{\vect{\prcoeffb}_i}{A''_i}{c''_i}\wedge
  \vect{\rfcoeff}=\vect{\prcoeffb}_i\cdot A'_i\wedge
  \rfcoeff_0\ge\vect{\prcoeffa}_i\cdot\vec{c}''_i
\end{equation}
Moreover, the values of $\rfp{\rfcoeff_0}{\vect{\rfcoeff}}$ in the
solutions of \eqref{eq:prmlc} define the set of all \lrfs for
$\transitions_1,\ldots,\transitions_k$.
\ethm

\bprf 
Immediate by Lemma~\ref{lemma:pr04}, noting that for each $1 \le i \le k$
the constraints
$\prcs{\vect{\prcoeffa}_i}{\vect{\prcoeffb}_i}{A''_i}{c''_i}$ uses
different $\vect{\prcoeffa}_i$ and $\vect{\prcoeffb}_i$, while
$\rfp{\rfcoeff_0}{\vect{\rfcoeff}}$ are the same for all $i$.
\eprf

\bcor
The \linrfq problem for \mlc loops is PTIME-decidable.
\ecor

\bprf
The size of the set of inequalities~\eqref{eq:prmlc} is polynomial in
the size of the input \mlc loop, and checking if it has a rational
solution can be done in polynomial time.
\eprf

\bcor
The \linrfz problem for \mlc loops is PTIME-decidable when each path
corresponds to one of the special cases, for \slc loops, discussed in
sections~\ref{sec:special:intpoly} and \ref{sec:special:bnv}.
\ecor

\bprf
Immediate, since if the transition polyhedra are integral,
\linrfz and \linrfq are equivalent.
\eprf

\bexm
\label{ex:sc:mlc}
Consider an \mlc loop with the following two paths: Loop~\eqref{eq:intro:loop1} of
Section~\ref{sec:introduction}; and the loop of
Example~\ref{ex:sc:tvpi-1}.
Applying \linrfq (as in Theorem~\ref{thm:linrfq:mlc}) does not find a
\lrf since both paths do not terminate when the variables range over
$\rats$.
If we first compute the integer hull of both paths, \linrfq finds
the \lrf $f(x_1,x_2)=3x_1+x_2-2$.
Note that the integer hull of the first path is computable in
polynomial time since the condition is \ptvpi and the update is affine
linear with integer coefficients. That of the second path has been
computed in Example~\ref{ex:sc:tvpi-1}.
\eexm

\newcommand{\casedesc}[2]{
\smallskip
\begin{minipage}[t]{0.3cm}\textbf{(#1)}\end{minipage}\hspace*{0.2cm}
\begin{minipage}[t]{12.5cm}#2\end{minipage}
\\
}
\newcommand{\caseex}[1]{
\medskip
\hspace*{1cm}
\begin{minipage}{12.3cm}
#1
\end{minipage}
\\
}
\newcommand{\casecomment}[1]{
\medskip
\hspace*{0.5cm}
\begin{minipage}{12.5cm}
#1
\end{minipage}\\
}

\begin{figure}

\begin{center}
\begin{tabular}{|@{~}>{}m{13cm}|}
\hline
\casedesc{1}{$\transitions$ is \emph{totally unimodular} (e.g., DBM). 
In this case $\transitions$ is already integral.}
\caseex{
\( 
\begin{array}{l}
\while~( x_1 \le x_2, x_2 \le x_3) ~\wdo~ x_1' \ge x_1+1, x_3' \leq x_3 
\end{array}
\)
}
\casecomment{We compute the \lrf $f(x_1,x_2,x_3)=x_3-x_1$. }
\hline\hline
\casedesc{2}{$\transitions$ is \emph{$N$-dimensional}.
In this case we compute the integer 
hull of $\transitions$.}
\caseex{
\( 
\begin{array}{l}
\while~(4x_1 \ge 1) ~\wdo~ 5x_1' \le 2x_1+1, 5x_1' \ge 2x_1-3
\end{array}
\)
}
\casecomment{Computing the integer hull of $\transitions$ adds $-x_1+x_1' \le -1$ and $\frac{1}{3}x_1-x_1' \le \frac{1}{3}$. Then we compute the \lrf $f(x_1)=x_1-1$. }
\hline\hline
\casedesc{3}{The update is affine linear with integer 
coefficients, and $\states$ is a \emph{cone}. In this 
case $\transitions$ is already integral.}
\caseex{
\( 
\begin{array}{l}
\while~( x_1+x_2 \ge 0, 2x_2+x_3 \ge 0) ~\wdo~ \\
~~~~~x_1' = x_1-2x_2-x_3-1, x_2'=x_2, x_3'=x_3
\end{array}
\)
}
\casecomment{We compute the \lrf $f(x_1,x_2,x_3)=x_1+x_2$.}
\hline\hline
\casedesc{4}{The update is affine linear with integer 
coefficients, and $\states$ is \emph{totally unimodular}. In this  case $\transitions$ is already integral.}
\caseex{
\( 
\begin{array}{l}
\while~( x_1 \le x_2, x_3-x_2 \geq 1) ~\wdo~  x_1' = x_1+x_3-x_2, x_2'=x_2, x_3'=2x_3
\end{array}
\) 
}
\casecomment{We compute the \lrf $f(x_1,x_2,x_3)=x_2-x_1$.  }
\hline\hline
\casedesc{5}{The update is affine linear with integer coefficients, and $\states$ is \emph{$N$-dimensional}. In this case we compute the integer hull of $\states$.}
\caseex{
  \( 
\begin{array}{l}
\while~( -x_1 + x_2 \le 0, -2x_1-x_2 \le -1 ) ~\wdo~  x_1' = x_1, x_2'=x_2-2x_1+1 
\end{array}
\)
}
\casecomment{Computing the integer hull of $\states$ adds $x_1\ge 1$. Then we 
compute the \lrf $f(x_1,x_2)=2x_1+x_2-1$. }
\hline\hline
\casedesc{6}{The update is affine linear with integer coefficients, and  $\states$ can be partitioned into independent sets where each is either a \emph{cone}, \emph{totally unimodular}, or \emph{$N$-dimensional}. In the case of \emph{$N$-dimensional} we compute its integer hull. }
\caseex{
  \( 
\begin{array}{l}
\while~( -x_1 + x_2 \le 0, -2x_1-x_2 \ge -1, x3 \le 1 ) ~\wdo~ \\
~~~~~ x_1' = x_1, x_2'=x_2-2x_1+x_3, x_3' = x_3 
\end{array}
\)
}
\casecomment{$\states$ is partitioned into $T_1=\{-x_1+x_2 \le 0 ,~\  -2x_1-x_2 \le -1\}$ and $T_2=\{x_3\le 1\}$. $T_1$ is \emph{$N$-dimensional} and $T_2$ is \emph{totally unimodular}. Computing the integer hull of $T_1$ adds $x_1\ge 1$. Then we compute the \lrf $f(x_1,x_2,x_3)=2x_1+x_2-1$. }
\hline\hline
\casedesc{7}{$\transitions$ can be partitioned into independent sets that are covered by cases (1)-(6).}
\caseex{
\( 
\begin{array}{l}
\while~(4x_1 \ge 1, x_2\ge 1)~\wdo~ 5x_1' \le 2x_1+1, 5x_1' \ge 2x_1-3, x_2'=x_2+1
\end{array}
\)
}
\casecomment{$\transitions$ is partitioned into  $T_1=\{{4x_1 \ge 1}, \ {5x_1' \le 2x_1+1}, \ {5x_1' \ge 2x_1-3}\}$ and $T_2=\{{x_2\ge 1}, {x_2'=x_2+1}\}$, which are covered by cases (2) and (4). The integer hull of $T_1$ is as in case (2). Then we compute the \lrf $f(x_1,x_2)=x_1-1$. 
}
\hline\hline
\casedesc{8}{An \mlc loop where each path is covered by cases (1)-(7).}
\hline
\end{tabular}
\end{center}

\caption{Summary of special PTIME cases of \linrfz : (1)-(7) summarize the
  special cases of sections~\ref{sec:special:intpoly}
  and~\ref{sec:special:bnv} for \slc loops; (8) summarizes the special
  cases of Section~\ref{sec:special:mlc} for \mlc loops.
  Recall that: $\transitions$ is the set of constraints that define the
  loop; $\states$ is the set of constraints that define the loop
  condition; and $N$-dimensional means at most $N$ variables for a fixed $N$ (above we have $N=2$).  }
\label{fig:special:recap}

\end{figure}
 
We conclude our discussion on the special PTIME cases for \linrfz with
a summary table (Figure~\ref{fig:special:recap}), that
briefly describes each case and illustrates it with an example.

\section{The Lexicographic-Linear Ranking Problem}
\label{sec:llinrf}

In this section we turn to the problems of finding a Lexicographic-Linear Ranking Function (\llrf),
or determining if one exist (as defined in Section~\ref{sec:prelim:llrf}).
We study the
complexity of both \llinrfz and \llinrfq and develop corresponding
complete algorithms for synthesizing \llrfs (moreover, \llrfs of smallest dimension).

In Section~\ref{sec:llinrf:int:alg} we consider the \llinrfz problem,
and develop a synthesis algorithm which has exponential-time
complexity in general, and polynomial-time complexity for the special
cases of Section~\ref{sec:ptime}. We also provide sufficient and
necessary conditions for the existence of a \llrf which imply the
completeness of our algorithm. These conditions are used in
Section~\ref{sec:llinrf:int:complexity} to show that
\llinrfz is coNP-complete.

In Section~\ref{sec:llinrf:rat} we consider the \llinrfq problem.  We
observe that applying the algorithm of
Section~\ref{sec:llinrf:int:alg}, which is complete for the integer
case, does not result in general in a \llrf for a rational loop, but just
what we call a \emph{weak} \llrf. This is a \llrf as in
Definition~\ref{def:lexlinearrf} but changing~\eqref{eq:llrf3} to
$\diff{\rho}(\vec{x}'')>0$.
It is not immediate that a weak ranking function even implies
termination, since $\diff{\rho}(\vec{x}'')$ can be arbitrarily close to
zero. However, we prove that it does, and in fact such a weak ranking function can
be converted to a \llrf. This provides a complete 
polynomial-time algorithm for \llinrfq (which is also optimal with respect to the dimension).

In the rest of this section we assume an input \mlc loop specified by
the transition polyhedra $\transitions_1,\cdots,\transitions_k$, where
each $\transitions_i$ is given as a system of inequalities $A\vec{x}''\le \vec{c}_i''$.
Since we handle \mlc loops, our results apply to \slc loops as a special case; we would like to point out, however,
that the coNP-hardness already applies to \slc loops (Section~\ref{sec:llinrf:int:complexity}), and that some interesting examples which demonstrate
the advantage of \llrfs over \lrfs use just \slc loops (e.g., Example~\ref{ex:kinds-of-lrfs-1} on Page~\pageref{ex:kinds-of-lrfs-1}).

\subsection{A Complete Algorithm for \llinrfz}
\label{sec:llinrf:int:alg}

The basic building blocks for our \llrfs are non-trivial quasi-\lrfs.
These are similar to \lrfs, except that
$\diff{\rho}(\vec{x}'')>0$ is not required to hold for all transitions,
but rather for at least one.

\bdfn
\label{def:quasirf}
We say that an affine linear function $\rho$ is a quasi-\lrf for $T \subseteq \rats^{2n}$
if for every
$\vec{x}''\in T$ the following holds:
\begin{align}
 \rho(\vec{x}) \ge 0 \label{eq:qlrf1}\\
 \diff{\rho}(\vec{x}'') \ge 0 \label{eq:qlrf2}
\end{align}
We say that it is \emph{non-trivial} if, in addition,
inequality~\eqref{eq:qlrf2} is strict, i.e., $\diff{\rho}(\vec{x}'') >
0$, for at least one $\vec{x}'' \in T$.
\edfn

We say that $\rho$ is a quasi-\lrf for a rational (respectively integer) loop if it is a quasi-\lrf
for its transition polyhedra (respectively, their integer points).

\bexm
\label{ex:qlrfs:1}
Consider the \slc loop~\eqref{eq:llrf:loop:1} of
Example~\ref{ex:kinds-of-lrfs-1}: 
$\rho_1(x_1,x_2,x_3)=x_2$ is a non-trivial quasi-\lrf;
$\rho_2(x_1,x_2,x_3)=x_1$ is not because $\diff{\rho_2}(\vec{x}'') \ge
0$ does not hold for all transitions; and $\rho_3(x_1,x_2,x_3)=x_3$ is
not because $\rho_3(\vec{x})<0$ for $\vec{x}=(2,1,-1)$.
Now consider the \mlc loop~\eqref{intro:ex:llrf} of Section~\ref{sec:introduction}:
$\rho_4(x_1,x_2)=x_1$ is a non-trivial quasi-\lrf for both paths of
this loop; and $\rho_5(x_1,x_2)=x_2$ is not quasi-\lrf since
$\diff{\rho_5}(\vec{x}'') \geq 0$ does not hold for all transitions,
e.g., it fails for $\vec{x}''=(2,2,1,3)$. Note that $\rho_5$ is a 
quasi-\lrf for the second path, but this is not enough.
\eexm

Note that when dealing with integer points, we can safely assume that whenever the function decreases in a transition,
it decreases at least by 1.  In fact, this holds for all affine functions with integer coefficients, and a function with non-integral
rational coefficients can always be scaled up to have integer ones.

Our \llrf synthesis algorithm is based on repeatedly finding
non-trivial quasi-\lrfs, and therefore we first focus on developing a
complete algorithm for synthesizing non-trivial quasi-\lrfs. The next
lemma explains how to represent the space $\poly{S}$ of all quasi-\lrfs,
afterwards, we explain how to pick a non-trivial one, if possible, from
this space.

\blem
\label{lem:qlrf-space}
Given $\transitions_1,\ldots,\transitions_k$, it is possible to build,
in polynomial time, a set of inequalities $\poly{S}$ whose solutions
define the coefficient vectors of all quasi-\lrfs for the
corresponding transitions
$\transitions_1\cup\cdots\cup\transitions_k$.
\elem

\bprf
Consider the constraints built by the Podelski-Rybalchenko procedure
of Theorem~\ref{th:pr04}, and change~\eqref{eq:pr:5} to
$\vect{\prcoeffb} \cdot\vec{c}'' \le 0$. Then, these constraints
describe the set of all quasi-\lrfs for $\transitions$, rather than
\lrfs. 
Using the construction of Theorem~\ref{thm:linrfq:mlc}, with this
change, we get a polyhedron $\poly{S}$ of dimension
$n'=n+1+\sum_{i=1}^k 2m_i$ where $m_i$ is the number of inequalities
in $\transitions_i$. 
Assume the first $n+1$ components correspond to the coefficients
$(\rfcoeff_0,\vect{\rfcoeff})$ (and the rest correspond to $\vect{\mu}$ and $\vect{\eta}$), then any point
$\pointinS{} \in \poly{S}$ defines a
quasi-\lrf $\rho(\vec{x})=\vect{\rfcoeff}\cdot\vec{x}+\rfcoeff_0$ for
$\transitions_1\cup\cdots\cup\transitions_k$.
\eprf

The next lemma explains how to pick a non-trivial quasi-\lrf $\rho$,
if any, from $\poly{S}$. Moreover, it shows how to pick one such that
$\diff{\rho}$ is strict for as many transitions as possible, i.e.,
there is no other  quasi-\lrf $\rho'$, and valid transition $\vec{x}''$, such that
$\diff{\rho'}(\vec{x}'')>0$ and $\diff{\rho}(\vec{x}'')=0$. We refer
to such non-trivial quasi-\lrfs as \emph{optimal}. The importance of
this optimal choice is in that it leads to an algorithm that
synthesizes \llrfs of minimal dimension.

\blem
\label{lem:qlrfalg}
There is a polynomial-time algorithm that finds a non-trivial
quasi-\lrf $\rho$, if there is any, for
$\transitions_1,\ldots,\transitions_k$; moreover, for any quasi-\lrf $\rho'$, and valid transition $\vec{x}''$, 
${\diff{\rho}(\vec{x}'')=0} \Rightarrow {\diff{\rho'}(\vec{x}'')=0}$.
\elem

\bprf
The algorithm follows the following steps:
\begin{enumerate}[(a)]
\item Construct a polyhedron $\poly{S}$ of all quasi-\lrfs as in
  Lemma~\ref{lem:qlrf-space};
\item If $\poly{S}=\emptyset$ return \textsc{None}, otherwise, pick
  $\pointinS{}$ in the \emph{relative
    interior}%
\footnote{For definitions related to faces of polyhedra, and the relative interior, see Section~\ref{sec:prelim:polyhedra}.}
 of $\poly{S}$;
\item If $\max\set{\vect{\rfcoeff}\cdot(\vec{x}-\vec{x}') \mid
    \vec{x}'' \in \transitions_i}> 0$, for some $1 \le i \le k$,
  return $\rho(\vec{x})=\vect{\rfcoeff}\cdot\vec{x}+\rfcoeff_0$,
  otherwise return \textsc{None}.
\end{enumerate}
When the above algorithm returns $\rho\neq\textsc{None}$,  it is a
non-trivial quasi-\lrf since it is a quasi-\lrf, and the last step
guarantees the existence of at least one $\vec{x}''$ for which
$\diff{\rho}(\vec{x}'')>0$.
To show completeness of the above algorithm and optimality of $\rho$,
it is enough to show that for any $\pointinS{'} \in \poly{S}$ and
$\vec{z}''\in \transitions_1\cup\cdots\cup\transitions_k$, we have
${\vect{\rfcoeff}\cdot(\vec{z}-\vec{z}')=0}
\Rightarrow
{\vect{\rfcoeff}'\cdot(\vec{z}-\vec{z}')=0}$.

So, assume that ${\vect{\rfcoeff}\cdot(\vec{z}-\vec{z}')=0}$.
Define the hyperplane $\poly{H}=\set{
  (\alpha_0,\vect{\alpha},\vect{\beta},\vect{\gamma}) \in \rats^{n'} \mid
  \vect\alpha\cdot (\vec z - \vec z') = 0 }$ where $\vect{\alpha}$ is
a vector of dimension $n$, and $n'$ is the dimension of $\poly{S}$.
By assumption, $\pointinS{} \in
\poly{S}\cap\poly{H}$.  Note that $\poly{S}\cap\poly{H}$ is a face of $\poly{S}$. If 
it equals to $\poly{S}$, then $\pointinS{'} \in\poly{H}$ and our claim holds. Otherwise,
it is a proper face of $\poly{S}$. Since $\pointinS{}$ was chosen from the
relative interior of $\poly{S}$, we have ${\vect{\rfcoeff}\cdot(\vec{z}-\vec{z}') > 0}$, and again our claim holds.

To justify the polynomial-time complexity note that the first step is
polynomial by Lemma~\ref{lem:qlrf-space}; the second step can be done
in polynomial time~\cite[Cor.~14.1\textbf{g}, p.~185]{Schrijver86}; and
the third is also polynomial since it consists of solving at most $k$
\lp problems over the rationals.
\eprf

Next we observe that finding a non-trivial quasi-\lrf for
$\intpoly{\transitions_1}\cup\cdots\cup\intpoly{\transitions_k}$,
i.e., over the integers, can be done by finding one for the
corresponding integer hulls.

\blem
\label{lem:qlrfint}
Function $\rho$ a is non-trivial quasi-\lrf for
$\intpoly{\transitions_1}\cup\cdots\cup\intpoly{\transitions_k}$ if
and only if it is a non-trivial quasi-\lrf for
$\inthull{\transitions_1}\cup\cdots\cup\inthull{\transitions_k}$.
\elem

\bprf
($\Rightarrow$) Suppose $\rho$ is a non-trivial quasi-\lrf for
$\intpoly{\transitions_1}\cup\cdots\cup\intpoly{\transitions_k}$. Then,
since $\intpoly{\transitions_i} \subseteq \inthull{\transitions_i}$,
there is an integer point
$\vec{x}''\in\inthull{\transitions_1}\cup\cdots\cup\inthull{\transitions_k}$
for which $\diff{\rho}(\vec{x}'')>0$.
It remains to show that for any
$\vec{x}''\in\inthull{\transitions_1}\cup\cdots\cup\inthull{\transitions_k}$
we have $\rho(\vec{x})\ge 0$ and $\diff{\rho}(\vec{x}'') \geq 0$. This
follows from the fact that, by definition of integer polyhedra, any
$\vec{x}''\in\inthull{\transitions_i}$ is a convex combination of some
points from $\intpoly{\transitions_i}$.
($\Leftarrow$) Suppose $\rho$ is a non-trivial quasi-\lrf for
$\inthull{\transitions_1}\cup\cdots\cup\inthull{\transitions_k}$. Then,
for any
$\vec{x}''\in\intpoly{\transitions_1}\cup\cdots\cup\intpoly{\transitions_k}$
we have $\rho(\vec{x})\ge 0$ and $\diff{\rho}(\vec{x}'') \geq 0$. It
remains to show that there is
$\vec{x}''\in\intpoly{\transitions_1}\cup\cdots\cup\intpoly{\transitions_k}$
for which $\diff{\rho}(\vec{x}'')>0$.
Let $\vec{x}''\in\inthull{\transitions_i}$ be a point for which
$\diff{\rho}(\vec{x}'')>0$, then, since $\vec{x}''$ is a convex
combination of some integer points from $\intpoly{\transitions_i}$,
there must be an integer point $\vec{z}''\in\intpoly{\transitions_i}$
for which $\diff{\rho}(\vec{z}'')>0$.
\eprf

%
%

\begin{algorithm}[t]
\caption{Synthesizing Lexicographical Linear Ranking Functions}
\label{alg:llrfsyn}
\DontPrintSemicolon
\LinesNumberedHidden
\SetKwFunction{procsyn}{LLRFSYN}
\SetKwFunction{procsynint}{LLRFint}
\procsynint{$\tuple{\poly{Q}_1,\ldots,\poly{Q}_k}$}\;
\KwIn{MLC loop defined by the polyhedra $\poly{Q}_1,\ldots,\poly{Q}_k$}
\KwOut{A \llrf for $\intpoly{Q_1},\ldots,\intpoly{Q_k}$, if exists, otherwise $\nollrf$ }
\Begin{
\setcounter{AlgoLine}{0} 	
\ShowLn Compute the integer hulls $\inthull{\transitions_1},\ldots,\inthull{\transitions_k}$ \; \label{alg:step:1}
\ShowLn \Return $\procsyn(\tuple{\inthull{\transitions_1},\ldots,\inthull{\transitions_k}})$.
}
\BlankLine 
\procsyn{$\tuple{\poly{P}_1,\ldots,\poly{P}_k}$}\;
\KwIn{MLC loop defined by the polyhedra $\poly{P}_1,\ldots,\poly{P}_k$}
\KwOut{A \llrf for $\poly{P}_1,\ldots,\poly{P}_k$, if exists, otherwise $\nollrf$ }
\Begin{
\setcounter{AlgoLine}{0} 	
\ShowLn  \lIf{$\tuple{\poly{P}_1,\ldots,\poly{P}_k}$ are all empty}{\Return \textbf{nil}}\;\label{alg:terminates}
\ShowLn  \ElseIf{ $\poly{P}_1\cup\cdots\cup\poly{P}_k$ has a non-trivial quasi-\lrf $\rho$}{
    \label{alg:qlrf}
\ShowLn    $\forall 1\le i \le k\ .\ \poly{P}_i' := \mathcal{P}_i \wedge \diff{\rho}(\vec{x}'')=0$\;\label{alg:constraint}
\ShowLn    $\llrfsym \leftarrow \procsyn(\tuple{\poly{P}'_1,\ldots,\poly{P}'_k})$\;\label{alg:rec}
\ShowLn    \lIf{$\llrfsym\neq\nollrf$}{\Return  $\rho{::}\tau$} \lElse{\Return $\nollrf$}
  }
\ShowLn  \lElse{\Return $\nollrf$}
}
\end{algorithm}
 
Now we are in a position for describing our algorithm for synthesizing
a \llrf, shown as the procedure $\procsynint$ in 
  Algorithm~\ref{alg:llrfsyn}. It either
  returns a \llrf $\llrfsym$ or $\nollrf$ if none exists.
Let us first explain the recursive procedure $\procsyn$.
It builds the \llrf component by component, or more precisely, by finding a suitable first
component and calling itself recursively to find the rest.
At Line~\ref{alg:qlrf} it finds a non-trivial quasi-\lrf $\rho$ for
the transitions $\poly{P_1}\cup\cdots\cup\poly{P_k}$. 
Assuming (as is always safe to do) that the coefficients returned are integer, 
this $\rho$ ranks all transitions for which $\diff{\rho}(\vec{x}'') \geq 1$, while for other 
transitions,
$\diff{\rho}(\vec{x}'')=0$. The set of these transitions is computed at Line~\ref{alg:constraint},
and at Line~\ref{alg:rec} $\procsyn$ is recursively called in order to
find a \llrf $\tau$ for them. If it finds one, then it returns
$\rho{::}\llrfsym$ as a \llrf for
$\poly{P_1}\cup\cdots\cup\poly{P_k}$.
The recursion stops when all transitions are ranked
(Line~\ref{alg:terminates}), or when there is no non-trivial
quasi-\lrf for the current set of transitions (Line \ref{alg:constraint}).
An important property of this algorithm is that when calling
$\procsyn$ with integral polyhedra, then the polyhedra passed to the recursive call 
are also integral.
This allows us to rely on
Lemmas~\ref{lem:qlrfalg} and \ref{lem:qlrfint}, which entail
the completeness of the overall algorithm. This also explains why it suffices 
to compute the integer hulls once, at Line~\ref{alg:step:1} of Procedure $\procsynint$.

\bexm
\label{ex:llralg:int:1}
Let us demonstrate the algorithm on the \slc
loop~\eqref{eq:llrf:loop:1} of Example~\ref{ex:kinds-of-lrfs-1}, which
is defined by 
\[
\transitions=\{ x_1 \ge 0,~ x_2 \ge 0,~ x_3 \ge -x_1,~ x_2'= x_2-x_1,~ x_3'= x_3+x_1-2\}.
\]
First note that in this
case $\inthull{\transitions}=\transitions$ and thus we can skip
Line~\ref{alg:step:1} of Procedure $\procsynint$.
\procsyn is first called with $\transitions$, and then, at
Line~\ref{alg:qlrf} it finds the non-trivial quasi-\lrf
$\rho_1(x_1,x_2,x_3)=x_2$ for $\transitions$, at
Line~\ref{alg:constraint} it sets $\poly{P'_1}$ to $\transitions\land
x_2-x'_2=0$, and at Line~\ref{alg:rec} \procsyn is  called
recursively with this $\poly{P'_1}$.
Then, at Line~\ref{alg:qlrf} it finds the non-trivial quasi-\lrf
$\rho_2(x_1,x_2,x_3)=x_3$ for $\transitions\land x_2-x'_2=0$, at
Line~\ref{alg:constraint} it sets $\poly{P'_1}$ to $\transitions\land
x_2-x'_2=0\land x_3-x'_3=0$ which is an empty polyhedron, and at
Line~\ref{alg:rec} \procsyn is called recursively with an empty
polyhedron.
Then, the check at Line~\ref{alg:terminates} succeeds and it returns
\textbf{nil}. Thus, the final returned value is $\tuple{x_2,x_3}$
which is a \llrf for $\intpoly{\transitions_1}$.
Now suppose that we remove $x_3 \ge -x_1$ from $\transitions$, and
note that we still have $\inthull{\transitions}=\transitions$.
Calling \procsyn with this modified $\transitions$ would proceeds as
above, however, it would fail to find a non-trivial quasi-\lrf for
$\transitions\land x_2-x'_2=0$ and thus it returns \nollrf. Indeed, in
this case $\intpoly{\transitions}$ does not have a \llrf since the
loop is non-terminating.
\eexm

Before formally proving soundness and completeness of
Algorithm~\ref{alg:llrfsyn}, we state a fundamental observation that
we will rely on.

\bobs 
\label{obs:face} 
Let $\transitions$ be a transition polyhedron.
If $\rho$ is a quasi-\lrf for $\transitions$, then the
points where $\diff{\rho}(\vec{x}'')=0$ holds, if any, form a
face of $\transitions$.
\eobs

\bprf
If there is $\vec{x}''\in\transitions$ such that
$\diff{\rho}(\vec{x}'')=0$, then $\min\{ \diff{\rho}(\vec{x}'')\mid
\vec{x}'' \in \transitions\}=0$. According to the definition of a
face, the intersection of the hyperplane $\{ \vec{x}''\in \rats^{2n}
\mid \diff{\rho}(\vec{x}'') = 0\}$ with $\transitions$ is a 
face of $\transitions$.
\eprf

Note that the statement that $\rho$ is non-trivial is equivalent to stating that the face, above, is proper.

\blem
\label{lem:llrfzalg}
If
$\procsynint(\tuple{\transitions_1,\ldots,\transitions_k})$
returns $\llrfsym$ different from $\nollrf$, then $\llrfsym$ is a
\llrf for $\intpoly{\transitions_1},\ldots,\intpoly{\transitions_k}$.
\elem

\bprf
We show that when $\poly{P}_1,\ldots,\poly{P}_k$ are integral, and
$\procsyn(\tuple{\poly{P}_1,\ldots,\poly{P}_k})$ returns
$\llrfsym\neq\nollrf$, then $\llrfsym$ is a \llrf for
$\intpoly{\poly{P}_1},\ldots,\intpoly{\poly{\poly{P}_k}}$.  The
conclusion of the lemma then follows because $\procsynint$
calls $\procsyn$ with the integer polyhedra
$\inthull{\transitions_1},\ldots,\inthull{\transitions_k}$.
The proof is by induction on $\sum\dim(\poly{P}_i)$.

\paragraph{Base-case} The base-case is when $\sum\dim(\poly{P}_i) =
-k$, i.e., all $\poly{P}_i$ are empty. In such case the algorithm
returns \textbf{nil}, which is trivially correct since there are no
transitions.

\paragraph{Induction hypothesis} If $\sum\dim(\poly{P}_i) < j$, each
$\poly{P}_i$ is integral, and
$\procsyn(\tuple{\poly{P}_1,\ldots,\poly{P}_k})$ returns $\llrfsym$, then
$\llrfsym$ is a \llrf for
$\intpoly{\poly{P}_1},\ldots,\intpoly{\poly{P}_k}$.

\paragraph{Induction step} Assume $\sum\dim(\poly{P}_i)=j$, and that
$\procsyn(\tuple{\poly{P}_1,\ldots,\poly{P}_k})$ returns
$\rho{::}\llrfsym$. Namely, at Line~\ref{alg:qlrf} it finds $\rho$,
and $\llrfsym\neq\nollrf$ is the result of
$\procsyn(\tuple{\poly{P}_1',\ldots,\poly{P}_k'})$ at
Line~\ref{alg:rec}. We show that $\rho{::}\llrfsym$ is a \llrf for
$\intpoly{\poly{P}_1},\ldots,\intpoly{\poly{P}_k}$.
First note the following:
\begin{enumerate}

\item\label{llrfalg:arg:1} Each $\poly{P}_i'$ is integral. This is
  because it is either empty, or a face of $\poly{P}_i$ (by Lemma~\ref{obs:face}), and all faces of an integral polyhedron are integral.
\item\label{llrfalg:arg:2} $\sum\dim(\poly{P}'_i) <
  \sum\dim(\poly{P}_i) = j$. This is because
   \begin{inparaenum}[\upshape(\itshape i\upshape)]
   \item $\forall 1\le i\le k \ .\ \dim(\poly{P}'_i) \leq
     \dim(\poly{P}_i)$; and
   \item there is $\vec{x}''\in\poly{P}_i$, for some $i$, such that
     $\diff{\rho}(\vec{x}'') > 0$, and thus $\poly{P}_i'$ is either
     empty or a proper face of $\poly{P}_i$ (by Lemma~\ref{obs:face}),
     in both cases $\dim(\poly{P}'_i) < \dim(\poly{P}_i)$.
   \end{inparaenum}
 \item\label{llrfalg:arg:3} 
We may assume that the function $\rho$ has been scaled, if necessary, so that
  for any
   $\vec{x''}\in\intpoly{\poly{P}_1}\cup\cdots\cup\intpoly{\poly{P}_k}$,
   either $\diff{\rho}(\vec{x}'') = 0$ and
   $\vec{x''}\in\intpoly{\poly{P}_1'}\cup\cdots\cup\intpoly{\poly{P}_k'}$,
   or $\diff{\rho}(\vec{x}'') \geq    1$.
\end{enumerate}
Using (\ref{llrfalg:arg:1},\ref{llrfalg:arg:2}), we apply the
induction hypothesis and conclude that $\llrfsym$ is a \llrf for
$\intpoly{\poly{P}'_1},\ldots,\intpoly{\poly{P}'_k}$. Using
\eqref{llrfalg:arg:3} we conclude that $\rho{::}\llrfsym$ is still a
\llrf for $\intpoly{\poly{P}'_1},\ldots,\intpoly{\poly{P}'_k}$, and
that $\rho$ ranks all transitions of
$\intpoly{\poly{P}_1}\cup\cdots\cup\intpoly{\poly{P}_k}$ that are not
in $\intpoly{\poly{P}_1'}\cup\cdots\cup\intpoly{\poly{P}_k'}$. Thus,
$\rho{::}\llrfsym$ is a \llrf for
$\intpoly{\poly{P}_1},\ldots,\intpoly{\poly{P}_k}$.
\eprf

Lemma~\ref{lem:llrfzalg} proves that 
Algorithm~\ref{alg:llrfsyn} is a sound procedure for \llinrfz. 
In Theorem~\ref{th:llrfz-completeness} below we combine this with a completeness proof. First, we give
sufficient and
necessary conditions for the existence of a \llrf for
$\intpoly{\transitions_1}, \ldots, \intpoly{\transitions_k}$.

\bobs
\label{obs:LLRF-only-if}
If there is a \llrf for $\intpoly{\transitions_1},\ldots,
\intpoly{\transitions_k}$, then every set of transitions $T \subseteq
\intpoly{\transitions_1} \cup \cdots \cup \intpoly{\transitions_k}$
has a non-trivial quasi-\lrf.
\eobs

\bprf
Let $\llrfsym=\tuple{\rho_1,\ldots,\rho_d}$ be a \llrf for
$\intpoly{\transitions_1},\ldots,\intpoly{\transitions_k}$, and $T$ be
a set of transitions.  Define $I=\{ i \mid \vec{x}''\in T \mbox{ is
  ranked by } \rho_i\}$, and let $j=\min(I)$. Then, from Definition~\ref{def:lexlinearrf}, it is easy to verify that $\rho_j$ is
a non-trivial quasi-\lrf for $T$.
\eprf

\bobs
\label{obs:LLRF-if}
If every set of transitions $T \subseteq \intpoly{\transitions_1} \cup
\cdots \cup \intpoly{\transitions_k}$ has a non-trivial quasi-\lrf,
then there is a \llrf for $\intpoly{\transitions_1},\ldots,
\intpoly{\transitions_k}$.
\eobs

\bprf 
In such case, Algorithm~\ref{alg:llrfsyn},  will
find a \llrf. 
This is because in every call to $\procsyn$, $\poly{P_1},\ldots,\poly{P_k}$ are integral,
and thus, by Lemmas~\ref{lem:qlrfalg}
and~\ref{lem:qlrfint} the check at Line~\ref{alg:qlrf} of $\procsyn$ is complete.
Moreover, the algorithm terminates since $\sum\dim(\poly{P}_i)$
decreases in each recursive call and has a lower bound $-k$.
\eprf

\bthm
\label{th:llrfz-completeness}
Algorithm~\ref{alg:llrfsyn} is sound and complete for
\llinrfz. Moreover, when it finds a \llrf, it finds one of a minimal
dimension.
\ethm

\bprf
If the algorithm returns $\llrfsym=\tuple{\rho_1,\ldots,\rho_d}$, then, by
Lemma~\ref{lem:llrfzalg}, it is a \llrf. If it is returns
$\nollrf$, then it has found a subset of integer points (at
Line~\ref{alg:qlrf} of $\procsyn$) that does not have a non-trivial quasi-\lrf. In
this case, by Observation~\ref{obs:LLRF-only-if}, there is no \llrf. Thus,
soundness and completeness have been established.

The minimality of the dimension stems from the fact that our algorithm
is greedy, i.e., in each step finds (by Lemma~\ref{lem:qlrfalg})
a \llrf that ranks as many transitions as possible.
Assume there is another \llrf
$\llrfsym'=\tuple{\rho'_1,\ldots,\rho'_{d'}}$. We show by
induction that the set of transitions that are not ranked by
$\tuple{\rho_1,\ldots,\rho_i}$, call it $\poly{U}_i$, is contained in the set of transitions not ranked by
$\tuple{\rho_1',\ldots,\rho'_i}$, call them $\poly{U}'_i$.  
Observe that since $\procsyn$ returns immediately if the input polyhedra are empty, we must have
$\poly{U}_i \ne \emptyset$ for $i\le d$.
It follows that $\poly{U}'_{i}\ne \emptyset$ for $i\le d$, hence $d' \ge d$.

The claim holds by definition for $i=0$ since
$\poly{U}_0=\poly{U}'_0=\intpoly{\transitions_1}\cup\cdots
\cup\intpoly{\transitions_1}$.
Assume $\poly{U}_i \subseteq \poly{U}'_i$ for some $0 \le i<d'$, we show
that $\poly{U}_{i+1} \subseteq \poly{U}'_{i+1}$. 
Since $\poly{U}_i \subseteq \poly{U}'_i$ then $\rho'_{i+1}$ is a
quasi-\lrf for $\poly{U}_i$, and since $\rho_{i+1}$ is optimal for
$\poly{U}_i$, by Lemma~\ref{lem:qlrfalg}, it cannot be that
$\rho'_{i+1}$ ranks a transition from $\poly{U}_i$ that is not ranked
by $\rho_{i+1}$, thus $\poly{U}_{i+1} \subseteq \poly{U}'_{i+1}$.
\eprf

The next corollary bounds the dimension of the \llrf inferred by
$\procsyn$ in terms of $n$, the number of variables in the loop.

\bcor
\label{cor:llrf-dim}
If $\procsyn$ returns
$\llrfsym=\tuple{\rho_1,\ldots,\rho_d}$, then $d \leq n$.
\ecor

\bprf
Let $\vect{\rfcoeff}_i$ be the coefficients of $\rho_i$ (i.e., we
ignore the constant $\rfcoeff_0$); for $1\le i \le d$. 
We claim that the vectors $\vect{\rfcoeff}_i$ must be linearly independent.
Assume the contrary; let $i$ be the first index such that 
$\vect{\rfcoeff}_i=c_1\cdot\vect{\rfcoeff}_1+\cdots+c_{i-1}\cdot\vect{\rfcoeff}_{i-1}$.
Pick a transition $\vec{x}''$ that is ranked by $\rho_i$, i.e.,
$\diff{\rho_i}(\vec{x}'')=\vect{\rfcoeff}_i\cdot(\vec{x}-\vec{x}') > 0$ and
$\forall 1 \le j < i\ .\
\diff{\rho_j}(\vec{x}'')=\vect{\rfcoeff}_j\cdot(\vec{x}-\vec{x}')=0$. Then
\begin{equation}
\label{eq:linindp}
\diff{\rho_i}(\vec{x}'')
=\vect{\rfcoeff}_i\cdot(\vec{x}-\vec{x}')
=(\sum_{j=1}^{i-1}c_j\cdot \vect{\rfcoeff}_j)\cdot(\vec{x}-\vec{x}')
=\sum_{j=1}^{i-1}c_j\cdot \vect{\rfcoeff}_j\cdot(\vec{x}-\vec{x}')
= 0
\end{equation}
which contradicts the assumption that $\diff{\rho_i}(\vec{x}'') > 0$.
Now since each $\vect{\rfcoeff}_i$ is a vector in $\rats^{n}$, linear independence implies
$d\leq n$. 
\eprf

The above Lemma provides the best bound possible for \mlc loops. To
see this, consider the \mlc loop~\eqref{intro:ex:llrf} of Section~\ref{sec:introduction}, for which
$n=2$, and note that it has a \llrf with $d=2$, namely
$\tuple{x_1,x_2}$, but no \llrf with $d=1$ (since it does not have a
\lrf). This can easily be extended to provide an example for any $n$.

Next, we argue that Procedure~$\procsyn$ can be implemented in
polynomial time. Note that this does not mean that \llinrfz is
PTIME-decidable since Algorithm~\ref{alg:llrfsyn}
has to compute the integer hulls first, which may take exponential time. 
However, this does mean that in certain special cases, \llinrfz is
PTIME-decidable.

\blem
\label{lem:alg:ptime}
Procedure~$\procsyn$ can be implemented in polynomial time.
\elem

\bprf 
First note that by Corollary~\ref{cor:llrf-dim} the recursion depth is
bounded by $n+1$, and that lines~\ref{alg:terminates}
and~\ref{alg:qlrf} can be performed in polynomial time in the bit-size
of (the current) $\poly{P_1},\ldots,\poly{P_n}$. However, we cannot
immediately conclude that the overall runtime is polynomial since as
recursion progresses, the procedure operates on polyhedra obtained by
adding additional constraints (at Line~\ref{alg:constraint}), that
could get bigger and bigger in their bit-size.
Thus, to complete the proof, we need to ensure that the bit-size of
$\poly{P_1},\ldots,\poly{P_n}$, at any stage of the recursion, is
polynomial in the bit-size of the original ones. Next we show how
Line~\ref{alg:constraint} can be implemented to ensure this,
exploiting the fact that when
$\poly{P_i}\land \diff{\rho}(\vec{x}'')=0$ is not empty, it is
a face of $\poly{P_i}$.

Recall that any face of $\poly{P_i}$ can be obtained by changing some
of its inequalities to equalities.  Hence, instead of adding
$\diff{\rho}(\vec{x}'')=0$ to $\poly{P_i}$ at
Line~\ref{alg:constraint}, we can identify those inequalities of
$\poly{P_i}$ that should be turned into equalities to get
$\poly{P_i}\land \diff{\rho}(\vec{x}'')=0$. Changing these
inequalities to equalities ensures that the bit-size of $\poly{P_i}$,
at any stage of the recursion, is at most twice its original bit-size.
Finding these inequalities can be done as follows: for each inequality
$\vec{a}\cdot\vec{x} \leq b$ of $\poly{P}_i$, we check if $\poly{P}_i
\land \diff{\rho}(\vec{x}'')=0 \Rightarrow \vec{a}\cdot\vec{x} \geq b$
holds, if so, then this inequality should be turned to equality. This
check can be done in polynomial time since it is an \lp problem and
the bit-size of $\rho$ is polynomial in the bit-size of
$\poly{P_1},\ldots,\poly{P_n}$.
\eprf

The above lemma implies that, as for \linrfz, if it is guaranteed
that the transition polyhedra are integral, or their integer hull can
be computed in polynomial time, then the \llinrfz problem can be
solved in polynomial time.

\bthm
\label{th:llrfz-ptime}
The \llinrfz problem for \mlc loops is PTIME-decidable if each path
corresponds to one of the special cases discussed in
sections~\ref{sec:special:intpoly} and \ref{sec:special:bnv}.
\ethm

\bprf
For those special cases either we do not compute the integer hulls
since they are already integral, or we compute them in polynomial
time. Then Algorithm~\ref{alg:llrfsyn} becomes polynomial-time since
Line~\ref{alg:step:1} of \procsynint can be done in polynomial time,
and \procsyn is polynomial according to Lemma~\ref{lem:alg:ptime}.
\eprf

It may be worthwhile to point out that even if we do not have a PTIME-decidable case,
we can always apply Procedure~$\procsyn$ to the given polyhedra---if it produces a \llrf,
we have a sound result in polynomial time.

\subsection{Complexity of \llinrfz}
\label{sec:llinrf:int:complexity}

In this section we show that the \llinrfz problem, in the general
case, is coNP-complete.
First, coNP-hardness follows from the coNP-hardness of \linrfz as in
Theorem~\ref{th:conp-hardness}. This is because the construction in
Theorem~\ref{th:conp-hardness} either produces a loop that has a \lrf
(which is also a \llrf) or else it is non-terminating (so it does not
have any kind of ranking function).
For the inclusion in coNP, we show that the complement problem, i.e.,
the nonexistence of a \llrf, has a polynomially checkable witness.

\bcor 
\label{cor:nollrf}
There is no \llrf for
$\intpoly{\transitions_1},\ldots,\intpoly{\transitions_k}$, if and
only if there is $T \subseteq
\intpoly{\transitions_1}\cup\cdots\cup\intpoly{\transitions_k}$ for
which there is no non-trivial quasi-\lrf.  
\ecor

\bprf
Immediate from Observations  \ref{obs:LLRF-only-if} and~\ref{obs:LLRF-if}.
\eprf

The above observation suggests that such $T$ can be used as a witness,
however, $T$ might include infinite number of transitions,
and thus it does not immediately meet our needs (polynomially
checkable witness).

\bexm
\label{ex:llrf:infT}
We show a case in which $T$ must consist of  infinitely many points. Let
$\transitions=\{ x' \le x-1 \}$ and take an arbitrary finite $T
\subseteq \transitions$. Now define $\lambda_0=\min\{ x \mid (x,x')
\in T \}$, then $\rho(x)=x - \lambda_0$ is a non-trivial quasi-\lrf
(actually \lrf) for $T$ and thus $T$ does not prove that there is no quasi-\lrf for $\transitions$. Any set of transitions 
out of $\transitions$
that does not have a quasi-\lrf must be infinite.
\eexm

To overcome this finiteness problem, we use notions similar to the
witness and h-witness that we have used for the case of \linrfz. In
particular, we show that the existence of $T$ as in
Corollary~\ref{cor:nollrf} can be witnessed by finite sets
$X\subseteq
\intpoly{\transitions_1}\cup\cdots\cup\intpoly{\transitions_k}$ and
$Y\subseteq
\intpoly{\recess{\transitions_1}}\cup\cdots\cup\intpoly{\recess{\transitions_k}}$,
whose bit-size is bounded polynomially in the bit-size of the input.

\bdfn
\label{def:nollrf:witness}
Let $X=X_1\cup\cdots\cup X_k$ and $Y=Y_1\cup\cdots\cup Y_k$, such that
\begin{inparaenum}[\upshape(\itshape i\upshape)]
\item\label{def:nollrf:witness:1} $X_i \subseteq \intpoly{\transitions_i}$; 
\item\label{def:nollrf:witness:2} $Y_i\subseteq \intpoly{\recess{\transitions_i}}$; and
\item\label{def:nollrf:witness:3} $Y_i\neq\emptyset \Rightarrow X_i\neq\emptyset$.
\end{inparaenum}
We say that $X$ and $Y$ form a witness against the existence of a
\llrf for $\intpoly{\transitions_1},\ldots,\intpoly{\transitions_k}$,
if the following set of linear constraints, denoted by
$\nollrfwitness{X}{Y}$, has no solution
\begin{subequations}
\begin{align}
\vect{\rfcoeff}\cdont \vec{x} + {\rfcoeff}_0 \ge 0  
   &~~~ \mbox{ for all } \vec{x}'' \in X   \label{eq:noLLRF:1}\\
\vect{\rfcoeff}\cdont\vec{y} \ge 0 
&~~~ \mbox{ for all } \vec{y}'' \in Y \label{eq:noLLRF:2}\\
\vect{\rfcoeff}\cdot(\vec{x} - \vec{x}')  \ge 0  
&~~~ \mbox{ for all } \vec{x}'' \in X \label{eq:noLLRF:3}\\
\vect{\rfcoeff}\cdot (\vec{y} - \vec{y}')  \ge 0 
&~~~ \mbox{ for all } \vec{y}'' \in Y \label{eq:noLLRF:4}\\
\sum_{\vec{x}'' \in X} \vect{\rfcoeff}\cdot (\vec{x} - \vec{x}') \,+&\sum_{\vec{y}'' \in Y} \vect{\rfcoeff}\cdot(\vec{y} - \vec{y}')  \ge 1  \label{eq:noLLRF:5}
\end{align}
\end{subequations}
\edfn

\blem 
\label{lem:nollrf:witness-1}
Let $X=X_1\cup\cdots\cup X_k$ and $Y=Y_1\cup\cdots\cup Y_k$ be as in
Definition~\ref{def:nollrf:witness}. Then there is $T \subseteq
\intpoly{\transitions_1}\cup\cdots\cup\intpoly{\transitions_k}$ that
has no non-trivial quasi-\lrf.
\elem

\bprf
We construct such $T$.
First note that for $\vec{x}''\in X_i$ and $\vec{y}''\in Y_i$, the
point $\vec{x}''+a\vec{y}''$, for any integer $a\ge 0$, is a
transition in $\intpoly{\transitions_i}$.
Now define 
\[
T = \{ \vec{x}''+a\vec{y}'' \mid \vec{x}''\in X_i,
\vec{y}''\in Y_i, \mbox { integer $a\ge 0$ } \}\; .
\] 
Clearly $T \subseteq
\intpoly{\transitions_1}\cup\cdots\cup\intpoly{\transitions_k}$.  We
claim that $T$ has no non-trivial quasi-\lrf.
Assume the contrary, i.e., there is
$\rfp{\rfcoeff_0}{\vect{\rfcoeff}}\in\rats^{n+1}$ such that
$\rho(\vec{x})= \vect{\rfcoeff}\cdont\vec{x}+\rfcoeff_0$ is a
non-trivial quasi-\lrf for $T$. We show that
$\rfp{c\rfcoeff_0}{c\vect{\rfcoeff}}$, for \emph{some}
$c>0$, is a solution of $\nollrfwitness{X}{Y}$, which contradicts the
assumption that $X$ and $Y$ form a witness as in
Definition~\ref{def:nollrf:witness}.

We first show that~(\ref{eq:noLLRF:1}--\ref{eq:noLLRF:4}) of
$\nollrfwitness{X}{Y}$ hold for
$\rfp{c\rfcoeff_0}{c\vect{\rfcoeff}}$ with \emph{any} $c>0$.
Pick arbitrary $\vec{x}''\in X_i$ and $\vec{y}''\in Y_i$.  
Since $\rho$ is a non-trivial quasi-\lrf for $T$, inequalities~(\ref{eq:qlrf1},\ref{eq:qlrf2}) on Page~\pageref{eq:qlrf1}
must hold for 
$\vec{x}''+a\vec{y}'' =
\tr{\vec{x}+a\vec{y}}{\vec{x}'+a\vec{y}'} \in T$. Namely,
the following must hold for any integer $a \ge 0$
\begin{align}
\rho(\vec{x}+a\vec{y})
{=}& \vect{\rfcoeff}\cdont(\vec{x}+a\vec{y})+\rfcoeff_0
{=} \vect{\rfcoeff}\cdont\vec{x}+\rfcoeff_0 + a\vect{\rfcoeff}\cdont\vec{y} \geq 0 \label{eq:llrfwitness:1}\\
\diff{\rho}(\vec{x}''+a\vec{y}'')
{=}& \vect{\rfcoeff}\cdont(\vec{x}+a\vec{y})-\vect{\rfcoeff}\cdont(\vec{x}'+a\vec{y}')
  {=}\vect{\rfcoeff}\cdont(\vec{x}-\vec{x}')+a\vect{\rfcoeff}\cdont(\vec{y}-\vec{y}') \geq 0\label{eq:llrfwitness:2}
\end{align}
This implies  
\begin{enumerate}[\upshape(\itshape i\upshape)]
\item\label{eq:llrf-if-1} $\vect{\rfcoeff}\cdont\vec{x}+\rfcoeff_0 \geq 0$,
otherwise~\eqref{eq:llrfwitness:1} is false for $a=0$;
\item\label{eq:llrf-if-2} $\vect{\rfcoeff}\cdont\vec{y}\geq 0$,
otherwise~\eqref{eq:llrfwitness:1} is false for $a>
-(\vect{\rfcoeff}\cdont\vec{x}+\rfcoeff_0)/(\vect{\rfcoeff}\cdont\vec{y})$; 
\item\label{eq:llrf-if-3} $\vect{\rfcoeff}\cdot(\vec{x}-\vec{x}') \geq 0$,
otherwise~\eqref{eq:llrfwitness:2} is false for $a=0$; and
\item\label{eq:llrf-if-4} $\vect{\rfcoeff}\cdot(\vec{y}-\vec{y}') \geq 0$,
otherwise~\eqref{eq:llrfwitness:2} is false for $a>
-\vect{\rfcoeff}\cdot(\vec{x}-\vec{x}')/\vect{\rfcoeff}\cdot(\vec{y}-\vec{y}')$.
\end{enumerate}
Note that the inequalities in~(\ref{eq:llrf-if-1}--\ref{eq:llrf-if-4})
above are those used in~(\ref{eq:noLLRF:1}--\ref{eq:noLLRF:4}). Hence  (\ref{eq:noLLRF:1}--\ref{eq:noLLRF:4}) hold for
$\rfp{\rfcoeff_0}{\vect{\rfcoeff}}$, and clearly, also 
for $\rfp{c\rfcoeff_0}{c\vect{\rfcoeff}}$ with any $c>0$.

Now we show that~\eqref{eq:noLLRF:5} of $\nollrfwitness{X}{Y}$ holds
for $\rfp{c\rfcoeff_0}{c\vect{\rfcoeff}}$, for \emph{some}
$c>0$. 
Since $\rho$ is a non-trivial quasi-\lrf, then,
inequality~\eqref{eq:qlrf2} must be strict for at least one
$\vec{x}''+a\vec{y}'' =
\tr{\vec{x}\phantom{'}+a\vec{y}}{\vec{x}'+a\vec{y}'} \in T$, i.e.,
$\diff{\rho}(\vec{x}''+a\vec{y}'') =\vect{\rfcoeff}\cdot(\vec{x}-\vec{x}') +
a\vect{\rfcoeff}\cdot(\vec{y}-\vec{y}') > 0$.
This means that either $\vect{\rfcoeff}\cdot(\vec{x}-\vec{x}')>0$ or
$\vect{\rfcoeff}\cdot(\vec{y}-\vec{y}')>0$ must hold. Taking $c>0$
large enough, we have $c\vect{\rfcoeff}\cdot(\vec{x}-\vec{x}')
\ge 1$ or $c\vect{\rfcoeff}\cdot(\vec{y}-\vec{y}') \geq 1$. Thus,
inequality~\eqref{eq:noLLRF:5} holds for
$\rfp{c\rfcoeff_0}{c\vect{\rfcoeff}}$.
Since~(\ref{eq:noLLRF:1}--\ref{eq:noLLRF:4}) also hold for this
$\rfp{c\rfcoeff_0}{c\vect{\rfcoeff}}$,  it is a solution
of $\nollrfwitness{X}{Y}$.
\eprf

\blem 
\label{lem:llrf:witness:2}
If there is $T \subseteq
\intpoly{\transitions_1}\cup\cdots\cup\intpoly{\transitions_k}$ that
has no non-trivial quasi-\lrf, then there are finite sets $X=X_1\cup\cdots\cup
X_k$ and $Y=Y_1\cup\cdots\cup Y_k$, fulfilling the conditions of
Definition~\ref{def:nollrf:witness}.
\elem

\bprf
Let $\vec{x}''$ be an arbitrary member of $T$. 
Let $\transitions \in \set{\transitions_1,\dots,\transitions_k}$ so that $\vec{x}''\in \transitions$, and
consider the generator representation
 $$\inthull{\transitions} = \convhull\{\vec x_1'',\dots,\vec x_{m}''\} + \cone\{\vec
y_1'',\dots,\vec y_{t}''\} \,.$$
Using these representation, we have
$\vec{x}''=\sum_{i=1}^m a_i \vec{x}_i'' + \sum_{j=1}^t b_j \vec{y}_j''$ for some rationals $a_i,b_j \ge 0$,
and $\sum_i a_i = 1$.
 We let $\mathit{ver}(\vec{x}'')$ be the set of all  vertices
$\vec{x}_i''$ with $a_i>0$ and $\mathit{rays}(\vec{x}'')$ be the set
of all rays $\vec{y}_j''$ with $b_j>0$.

For $\ell=1,\dots,k$,
define $X_\ell=\cup \{ \mathit{ver}(\vec{x}'') \mid \vec{x}''\in
T\cap\intpoly{\transitions_\ell} \}$ and $Y_\ell =\cup \{
\mathit{rays}(\vec{x}'') \mid \vec{x}''\in
T\cap\intpoly{\transitions_\ell} \}$.
Next we show that $X=X_1\cup\dots\cup X_k$ and $Y=Y_1\cup\dots\cup
Y_k$ form a witness as in Definition~\ref{def:nollrf:witness}.

Conditions~(\ref{def:nollrf:witness:1},\ref{def:nollrf:witness:2}) of
Definition~\ref{def:nollrf:witness} hold by construction, and
Condition~\eqref{def:nollrf:witness:3} holds 
because $\sum_i a_i = 1$.
  What is left to show is that
$\nollrfwitness{X}{Y}$ has no solution.
Assume the contrary, i.e., $\nollrfwitness{X}{Y}$ has a solution
$\rfp{\rfcoeff_0}{\vect{\rfcoeff}}\in\rats^{n+1}$. 
We claim that then $\rho(\vec{x})=
\vect{\rfcoeff}\cdont\vec{x}+\rfcoeff_0$ is a non-trivial quasi-\lrf
for $T$, which contradicts the lemma's assumption.
Pick an arbitrary $\vec{x}''\in T$ and write it, using the
corresponding $X_\ell$ and $Y_\ell$, as $\vec{x}''=\sum_{i=1}^m
a_i\vec{x}''_i + \sum_{j=1}^t b_j \vec{y}''_j$ where $a_i,b_j >
0$ and $\sum_{i=1}^m a_i = 1$.
Since~(\ref{eq:noLLRF:1},\ref{eq:noLLRF:3}) hold for each
$\vec{x}''_i\in X$ and~(\ref{eq:noLLRF:2},\ref{eq:noLLRF:4}) hold for
each $\vec{y}''_j\in Y$, we have
\begin{align*}
\rho(\vec{x}) 
=& \vect{\rfcoeff}\cdot(\sum_{i=1}^m  a_i\vec{x}_i+\sum_{j=1}^t b_j\vec{y}_j)+\rfcoeff_0 \\
=& \sum_{i=1}^m  a_i\cdot(\vect{\rfcoeff}\cdont\vec{x}_i+\rfcoeff_0)+\sum_{j=1}^t b_j\vect{\rfcoeff}\cdont
\vec{y}_j \ge 0\\[1.5ex]
\diff{\rho}(\vec{x}'')
=&\vect{\rfcoeff}\cdot(\sum_{i=1}^m  a_i\vec{x}_i+\sum_{j=1}^t b_j \vec{y}_j) - 
\vect{\rfcoeff}\cdot(\sum_{i=1}^m  a_i\vec{x}'_i+\sum_{j=1}^t b_j \vec{y}'_j)\\
=&\sum_{i=1}^m  a_i\vect{\rfcoeff}\cdot(\vec{x}_i-\vec{x}_i')+ 
\sum_{j=1}^t  b_j\vect{\rfcoeff}\cdot(\vec{y}_j-\vec{y}_j') \geq 0
\end{align*}
Thus, $\rho$ satisfies~(\ref{eq:qlrf1},\ref{eq:qlrf2}) for any
$\vec{x}''\in T$.
Now since~\eqref{eq:noLLRF:5} holds, there must be $\vec{x}_i''\in X$ or
$\vec{y}_j''\in Y$ for which $\vect{\rfcoeff}\cdot(\vec{x}_i-\vec{x}'_i)>0$ or
$\vect{\rfcoeff}\cdot(\vec{y}_j-\vec{y}'_j)>0$. %
Now note that since $X$ and $Y$ were constructed from the vertices and
rays of the transitions in $T$, these $\vec{x}_i''$ or $\vec{y}_j''$
must correspond to some $\vec{x}''\in T$, and thus it must be the case
that $\diff{\rho}(\vec{x}'')>0$ for this specific $\vec{x}''$, i.e.,
inequality~\eqref{eq:qlrf2} is strict for $\vec{x}''$.
\eprf

\bexm
\label{ex:llrf:witness:1}
For $\transitions=\{ x' \leq x-1\}$ of Example~\ref{ex:llrf:infT}, we
claim that $X=\{(0,-1)\}$ and $Y=\{(1,1),(-1,-1)\}$ form a witness as
in Definition~\ref{def:nollrf:witness}. It is easy to check that
$X$ and $Y$ satisfy
conditions~(\ref{def:nollrf:witness:1}--\ref{def:nollrf:witness:3}).
Then, $\nollrfwitness{X}{Y}$ is the set of inequalities
$\{ {\rfcoeff_0 \geq 0},~ {\rfcoeff_1 \geq 0},~ {-\rfcoeff_1 \geq 0},~ {\rfcoeff_1\geq 1 }\}$ which has no solution.
\eexm

\bexm
\label{ex:llrf:witness:2}
Consider an \mlc loop represented by 
\[
\begin{array}{rlll}
\transitions_1 & = \{ x_1 \geq 0, &  x_2\ge 0, &  x_1' = x_1-1 \} \\
\transitions_2 & = \{ x_1 \geq 0, &  x_2\ge 0, & x_2' = x_2-1 \}
\end{array}
\]
and let 
\[
\begin{array}{rlrl}
X_1 &= \{ (0,0,-1,~0) \},& Y_1 & = \{(0,0,0,1)\}, \\
X_2 &= \{ (0,0,~0,-1) \},& Y_2 & = \{(0,0,1,0)\}.
\end{array}
\]
We claim that these sets form a witness as
in Definition~\ref{def:nollrf:witness}.
It is easy to check that they satisfy
conditions~(\ref{def:nollrf:witness:1}--\ref{def:nollrf:witness:3}) of
Definition~\ref{def:nollrf:witness}. Substituting these points in \eqref{eq:noLLRF:5}
gives $0 \geq 1$,
so clearly (\ref{eq:noLLRF:1}--\ref{eq:noLLRF:5}) are unsatisfiable.
\eexm

The next lemma concerns the bit-size of the witness. 

\blem
\label{lem:nollrf:witness-size}
If there is a finite witness for the nonexistence of \llrf for
$\intpoly{\transitions_1},\ldots,\intpoly{\transitions_k}$, then there
is one defined by $X=X_1\cup\cdots\cup X_k$ and $Y=Y_1\cup\cdots\cup
Y_k$ such that $\sum_{i=1}^k |X_i|+|Y_i| \le 6n+2$; and its bit-size
is polynomial in the bit-size of
${\transitions_1},\ldots,{\transitions_k}$.
\elem

\bprf
Consider the witness constructed in Lemma~\ref{lem:llrf:witness:2},
and recall that $\Phi_1=\nollrfwitness{X}{Y}$ has no solution.
Let $Z$ be any maximal linearly-independent subset of $X\cup Y$.
Clearly, $|Z|\le 2n$.
Let $\Phi_2$ be the formula obtained from $\Phi_1$ by
replacing~\eqref{eq:noLLRF:5} with
\begin{equation}
\label{eq:zxzy}
\sum_{\vec{z}'' \in Z} \vect{\rfcoeff}\cdot (\vec{z} - \vec{z}') \ge 1
\end{equation}
We claim that $\Phi_2$ has no solution. To see this, take arbitrary
$\rfp{\rfcoeff_0}{\vect{\rfcoeff}} \in \rats^{2n}$, we know it is not
a solution of $\Phi_1$. If this is because one of the inequalities
in~(\ref{eq:noLLRF:1}-\ref{eq:noLLRF:4}) is false, then it is clearly
not a solution of $\Phi_2$ since it includes all such inequalities.
If all inequalities in~(\ref{eq:noLLRF:1}-\ref{eq:noLLRF:4}) are true,
then \eqref{eq:noLLRF:5} must be false. Since all terms in the sum are %
non-negative, they must all be zero, that is,
$\vect{\rfcoeff}\cdot (\vec{z} - \vec{z}')=0$ for any $\vec{z}''\in X\cup Y$. Otherwise, $\rfp{c\rfcoeff_0}{\vect{c\rfcoeff}}$ for $c \geq 1$ large enough would be a solution of $\Phi_1$.
 Thus, inequality~\eqref{eq:zxzy} is false. 

A corollary of Farkas' Lemma~\cite[p.~94]{Schrijver86} states that: if
a set of inequalities over $\rats^d$ has no solution, there is a
subset of at most $d+1$ inequalities that has no solution.
Let $\Phi_3$ be such a subset of $\Phi_2$, it has at most $n+2$
inequalities (since $\Phi_2$ is over $\rats^{n+1}$).
Note that $\Phi_3$ must include inequality~\eqref{eq:zxzy}, otherwise
it is trivially satisfiable.
Let $X' = X_1'\cup \ldots \cup X'_k \subseteq X \mbox{ and } Y'=Y_1'
\cup \ldots \cup Y'_k\subseteq Y$ be the points involved in the
inequalities of $\Phi_3$ (including \eqref{eq:zxzy}),
then $\sum_{i=1}^k |X'_i|+|Y'_i| \leq n+1+2n = 3n+1$.
To get a witness as per Definition~\ref{def:nollrf:witness},
if, for any $i\le k$, $Y'_i \neq \emptyset$ and $X'_i=\emptyset$, we
include an arbitrary point $\vec{x}''\in X_i$ to $X_i'$.
This can at most double the size of these sets, i.e., $\sum_{i=1}^k
|X'_i|+|Y'_i| \leq 6n+2$ (or $\sum_{i=1}^k |X'_i|+|Y'_i| \leq 3n+1+k$
when $k<3n+1$).

We claim that $\tuple{X',Y'}$ is a witness that fulfills the
conditions of Definition~\ref{def:nollrf:witness}.
It satisfies
conditions~(\ref{def:nollrf:witness:1}-\ref{def:nollrf:witness:3}) by
construction. Next, we show that $\Phi_4=\nollrfwitness{X'}{Y'}$ has no
solution.
Take arbitrary $\rfp{\rfcoeff_0}{\vect{\rfcoeff}} \in \rats^{n+1}$, we
know it is not a solution for $\Phi_2$.  If it is because one of the
inequalities in~(\ref{eq:noLLRF:1}-\ref{eq:noLLRF:4}) is false, then
it is clearly not a solution of $\Phi_4$ since it includes all such
inequalities.
If all inequalities in~(\ref{eq:noLLRF:1}-\ref{eq:noLLRF:4}) are true,
then~\eqref{eq:zxzy} must be false, and then we must have
$\vect{\rfcoeff}\cdot (\vec{z} - \vec{z}')=0$ for any $\vec{z}''\in Z$.
Now since any $\vec{z}''\in X' \cup Y'$ is a linear combination of points from
$Z$,  $\vect{\rfcoeff}\cdot (\vec{x} - \vec{x}')=0$ for
any $\vec{x}''\in X'$ and $\vect{\rfcoeff}\cdot(\vec{y} - \vec{y}')=0$
for any $\vec{y}''\in Y'$. Thus, inequality~\eqref{eq:noLLRF:5} of
$\Phi_4$ is false.

Finally, we show that the bit-size of the witness is polynomial in the
bit-size of the input.
Recall that the points of $X'$ and $Y'$ come from the generator
representations of
$\inthull{\transitions_1},\ldots,\inthull{\transitions_k}$, and that
there is a generator representation for each
$\inthull{\transitions_i}$ in which each vertex/ray can fit in
$\vtxsize{\inthull{\transitions_i}}$ bits.
Thus, the bit-size of $X'$ and $Y'$ is bounded by
$(6n+2)\cdot\max_i\vtxsize{\inthull{\transitions_i}}$.
By Theorem~\ref{thm:PIsize}, since the dimension of each
$\transitions_i$ is $2n$,
\[
(6n+2)\cdot\max_i\vtxsize{\inthull{\transitions_i}} \le
(6n+2)\cdot (6\cdot (2n)^3\cdot\max_i\fctsize{\transitions_i}) \le 
(288n^4 + 96n^3)\cdot\max_i\bitsize{\transitions_i}
\]
which is polynomial in the bit-size of the input.
\eprf

\bthm
\label{th:llrf:conp-inc}
$\llinrfz \in \mathrm{coNP}$ for \mlc loops.
\ethm

\bprf 
We show that the complement of \linrfz has a polynomially checkable
witness. The witness is a listing of sets of integer points
$X=X_1\cup\cdots\cup X_k$ and $Y=Y_1\cup\cdots\cup Y_k$ of at most
$6n+2$ elements and has a polynomial bit-size (specifically, a
bit-size bounded as in Lemma~\ref{lem:nollrf:witness-size}).
Verifying a witness consists of the following steps:

\paragraph{Step 1} Verify that each $\vec{x}''\in X_i$ is in
  $\intpoly{\transitions_i}$, which can be done by verifying  $A_i''
  \vec{x}'' \le \vec{c}_i''$; and that each $\vec{y}''\in Y_i$ is in
  $\intpoly{\recess{\transitions}}$, which can be done by verifying
   $A_i'' \vec{y}'' \le 0$. This is done in polynomial time.
   Note that according to Lemma~\ref{lem:nollrf:witness-1} it is not
   necessary to check that $X$ and $Y$ come from a particular
   generator representation.

\paragraph{Step 2} Verify that $\nollrfwitness{X}{Y}$ has no solutions,
which can be done in polynomial time since it is an \lp problem over
$\rats^{n+1}$.  
\eprf

\subsection{Lexicographic Ranking Functions over the Rationals}
\label{sec:llinrf:rat}

In this section we address the \llinrfq problem. In particular, we show
that Procedure $\procsyn$, when applied to the input polyhedra
$\transitions_1,\ldots,\transitions_k$ instead of their integer hulls,
can be used to decide the existence of a \llrf for
$\transitions_1,\ldots,\transitions_k$.
However, in such case, the returned value
$\llrfsym=\tuple{\rho_1,\ldots,\rho_d}$ of the algorithm does not fit
in the class of \llrfs as in Definition~\ref{def:lexlinearrf}.
We define a new class of \llrfs that captures such functions, and
prove that it is actually equivalent to that of
Definition~\ref{def:lexlinearrf} as far as the existence of a \llrf is
concerned.

First recall that in Section~\ref{sec:prelim:llrf} we discussed the
possibility of replacing inequality $\diff{\rho_i}(\vec{x}'')\geq 1$ by
$\diff{\rho_i}(\vec{x}'')\geq \delta_i$ in condition~\eqref{eq:llrf3}
of Definition~\ref{def:lexlinearrf}.
With this change, $\llrfsym=\tuple{\rho_1,\dots,\rho_d}$ is a
\llrf if and only if there are positive $\delta_1,\ldots,\delta_d$
such that, for any $\vec{x}''\in
\transitions_1\cup\cdots\cup\transitions_k$ there exists $i$ for which
the following hold
\begin{alignat}{ 2 }
 \forall j < i \ .\  && \diff{\rho_j}(\vec{x}'') &\ge 0  \label{eq:rat:llrf1d}\\
 \forall j \le i \ .\ && \rho_j(\vec{x}) &\ge 0  \label{eq:rat:llrf2d}\\
                      && \diff{\rho_i}(\vec{x}'') &\geq \delta_i  \label{eq:rat:llrf3d} 
\end{alignat}
This is equivalent to Definition~\ref{def:lexlinearrf}, as far as the
existence of a \llrf is concerned, since $c\llrfsym$, for any $c >
\min(\vec\delta_i)^{-1}$, is a corresponding \llrf as in
Definition~\ref{def:lexlinearrf}.
In the rest of this section, for the sake of simplifying the formal
presentation, we use this notion of \llrfs.

Let us start by explaining why the returned value of
Procedure $\procsyn$, in the rational case, does not fit in the
above class of \llrfs.
For this, let us consider a non-trivial quasi-\lrf $\rho$ synthesized
at Line~\ref{alg:qlrf}.
In the integer case, all integer transitions of
$\poly{P_1},\ldots,\poly{P_k}$ that do not pass to
$\poly{P'_1},\ldots,\poly{P'_k}$ are ranked by this $\rho$. This is
because $\diff{\rho}(\vec{x}'') \geq 1$ for all such transitions (see
the proof of Lemma~\ref{lem:llrfzalg}, point~\eqref{llrfalg:arg:3}).
This, however, is not true when considering rational transitions.
 In this case, all transitions that do not pass to
$\poly{P'_1},\ldots,\poly{P'_k}$ satisfy $\diff{\rho}(\vec{x}'')>0$,
but it is not guaranteed that $\diff{\rho}(\vec{x}'')$ has a minimum
$\delta$ over this set of transitions.
For example, take $\poly{P_1}=\{ x \ge 0, x=2x' \}$ and $\rho(x)=x$,
then $\poly{P_1'}=\{x=0,x'=0\}$. The transitions that do not pass to
$\poly{P_1'}$ are those specified by the non-closed polyhedron $\{x >
0, x=2x'\}$, in which $\diff\rho$ does not have a positive lower bound.
This leads us to introduce \emph{weak} \llrfs. 

\bdfn
\label{def:llrfw}
We say that $\llrfsym=\tuple{\rho_1,\dots,\rho_d}$ is a \emph{weak}
  \llrf for $\transitions_1\cup\cdots\cup\transitions_k$, if and only
if for any $\vec{x}''\in \transitions_1\cup\cdots\cup\transitions_k$
there exists $i$ for which
(\ref{eq:rat:llrf1d},\ref{eq:rat:llrf2d}) hold, as well as 
\begin{align}
    & \diff\rho_i(\vec{x}'')  > 0  \label{eq:rat:llrf3w}  
\end{align}
(which replaces \eqref{eq:rat:llrf3d}).
\edfn

While any \llrf is a also
weak \llrf, the converse is more subtle. Over the integers,
the existence of a weak \llrf implies the existence
of a \llrf (since $\diff\rho_i(\vec{x}'')  > 0$ means $\diff\rho_i(\vec{x}'')  \ge 1$ when the coefficients and state variables are integer). Over the rationals,
such an implication is not immediate. Moreover, even whether a weak ranking function implies termination
is unclear, as infinitely descending sequences of positive
rationals exist.

\bexm
\label{ex:wllrf}
Consider the following \mlc loop
\begin{equation}
\begin{array}{rcllll@{}l@{}l}
\mlcwhile: 
& &\{ x_1 \geq 0, &   &  &  x_1' = x_1-1 \} && \\
&\vee& \{ x_1 \geq 0, &  x_2 - x_1\geq 0, & & x_1' = x_1, &  x_2' = x_2-1 \} & \\
&\vee& \{ x_1 \geq 0, & x_2 - x_1\geq 0, &  x_3\geq 0, &  x_1' \le \frac{1}{2}x_1, &  x_2' = x_2, &  x_3' = x_3-1 \}
\end{array}
\label{eq:llrf:loop:3}
\end{equation}
Applying Procedure $\procsyn$ to the corresponding transition
polyhedra $\transitions_1,\transitions_2,\transitions_3$ possibly returns
$\llrfsym=\tuple{x_1,x_2-x_1,x_3}$.
It is easy to see that it is a weak \llrf over the rationals, and, consequently,
it is a \llrf over the integers. To see why it is not a \llrf over the
rationals, assume the first component of $\tau$ decreases by at least
$\delta_1>0$. All transitions for which $x_1-x_1'<\delta_1$ are not
ranked by this component  and thus should be ranked by
either the second or the third.
Let us take $\vec{x}''\in\transitions_3$ such that
$\vec{x}=(\delta_1,1,1)$ and $\vec{x}'=(\frac{1}{2}\delta_1,1,0)$.
This transition is not ranked by the first component since
$\diff{\rho_1}(\vec{x}'')=\frac{1}{2}\delta_1<\delta_1$, and it is not
ranked by the second or the third since
$\diff{\rho_2}(\vec{x}'')=-\frac{1}{2}\delta_1 < 0$.
Nonetheless, this loop is terminating over the rationals and has a \llrf, and later
we show how to obtain it.
\eexm

Over the rationals, Procedure $\procsyn$ is sound and complete
for synthesizing weak \llrfs. Moreover, as in the integer case, it
synthesizes one with minimal dimension. 

\blem
\label{lem:llrfqalg}\label{th:llrfq-completeness}
Procedure $\procsyn$, when applied to $\transitions_1, \cdots,
\transitions_k$, is sound and complete for the existence of a weak
\llrf for $\transitions_1,\ldots,\transitions_k$. Moreover,
if $\procsyn(\tuple{\transitions_1,\ldots,\transitions_k})$ returns
$\llrfsym$ different from $\nollrf$, then $\llrfsym$ is a weak \llrf of minimal dimension
for $\transitions_1,\ldots,\transitions_k$.
\elem

\bprf
Suppose that $\procsyn(\tuple{\transitions_1,\ldots,\transitions_k})$ returns $\tau$. Then, as
in the proof of Lemma~\ref{lem:llrfzalg}, we can show that $\tau$ is a weak \llrf.  We prefer not to repeat
the  whole proof but just indicate the difference, which boils down to
drop points~\eqref{llrfalg:arg:1} and \eqref{llrfalg:arg:3} regarding the integrality of
corresponding polyhedra and a non-zero decrease being at least 1.

This gives soundness; for completeness,
the proof is as that of Theorem~\ref{th:llrfz-completeness}. In fact,
the sufficient and necessary condition for the existence of a
\llrf, stated in Observations~\ref{obs:LLRF-only-if}
and~\ref{obs:LLRF-if}, is a condition for existence of a weak \llrf when
applied to $\transitions_1,\ldots,\transitions_k$.

The minimality follows from the same consideration as in the proof of
Theorem~\ref{th:llrfz-completeness}.  
\eprf

In the rest of this section we show how one can construct a \llrf for
$\transitions_1,\ldots,\transitions_k$ from a weak \llrf.
This implies soundness and completeness of Procedure $\procsyn$ as a decision
procedure for \llinrfq, and its usage for synthesis of \llrfs.
To simplify notation, we shall consider the polyhedra $\transitions_1,\ldots,\transitions_k$ to be fixed
up to the completion of the proof.

Here is a brief outline of the construction. The first step, culminating in Lemma~\ref{lem:rat:consolidate}, shows how to
transform the \llrf $\tuple{\rho_1,\ldots,\rho_d}$ into another one $\tuple{f_1,\ldots,f_d}$, where each $f_i$ will be a linear
combination of $\rho_1,\ldots,\rho_i$, so that if component $i$ is used for ranking some transition of one of the transition polyhedron $\transitions_\ell$, we
will be ensured that $f_i$ is non-increasing over all of this $\transitions_\ell$ (even over transitions that are already ranked by
a previous component). Consequently, in Lemmas~\ref{lem:rat:llrf} and \ref{thm:rat:llrf}, we show how thanks to this property,
the ranking-function components can be ``nudged'' so that the weak 
\llrf becomes a proper one.

\bdfn
Let $\llrfsym=\tuple{\rho_1,\ldots,\rho_d}$ be a weak \llrf for
$\transitions_1,\ldots,\transitions_k$. The \emph{ranking chain} for $\tau$ is the $(d+1)$-tuple
of sets, $U_1,\dots,U_{d+1}$, defined by $U_1=\transitions_1\cup\cdots\cup\transitions_k$, and
$U_{i+1}=U_i\wedge (\diff{\rho_{i}}(\vec{x}'')=0)$. 
\edfn
Observe that
\begin{equation*}
 \transitions_1\cup\cdots\cup\transitions_k = U_1 \supseteq U_2 \supseteq \dots \supseteq U_d \supseteq U_{d+1}=\emptyset.
 \end{equation*}
 It is easy to see that if for some $j$, $U_j = U_{j+1}$, it is possible to omit $\rho_j$ from $\tau$ without
 any harm. We say that $\tau$ is \emph{irredundant} if
 \begin{equation}
 \label{eq:irredundant}
 \transitions_1\cup\cdots\cup\transitions_k = U_1 \supset U_2 \supset \dots \supset U_d \supset U_{d+1}=\emptyset.
 \end{equation}
 
 \bobs
A weak \llrf computed by Procedure $\procsyn$ is irredundant. In fact,
$U_i$ is the union  $\poly{P}_1\cup\cdots\cup\poly{P}_k$ of the arguments to the $i$-th
recursive call.
\eobs

\newcounter{eq:rat:Unonegative}
\setcounter{eq:rat:Unonegative}{\theequation}
\addtocounter{eq:rat:Unonegative}{1}
\newcounter{eq:rat:Ustict}
\setcounter{eq:rat:Ustict}{\theequation}
\addtocounter{eq:rat:Ustict}{2}
\newcounter{eq:rat:Uirredundant}
\setcounter{eq:rat:Uirredundant}{\theequation}
\addtocounter{eq:rat:Uirredundant}{3}
By the definition of a weak \llrf, and the definition of $U_1,\dots,U_{d+1}$, the following
properties clearly follow:

\begin{indexedequations}{i}
\begin{alignat}{2}
\forall \vec{x}'' \in U_i \ .\ && \rho_i(\vec{x})&\ge 0, \label{eq:rat:Unonegative}\\
\forall \vec{x}'' \in U_i \setminus U_{i+1}\ .\ && \diff{\rho_i}(\vec{x}'') &> 0, \label{eq:rat:Ustict}\\
\forall \vec{x}'' \in U_{i+1}\ .\ && \diff{\rho_i}(\vec{x}'') &= 0  \,. \label{eq:rat:Uirredundant}
\end{alignat}
\end{indexedequations}

\noindent 
Note that each $U_i$ is a finite union of closed polyhedra, obtained
by intersecting $U_1$ with some hyperplanes.
For $1\le i\le d$, let $J_{i} = \{ j \mid \transitions_j \cap U_i \ne
\emptyset \}$, and let $\closure{i} = \bigcup_{ j\in J_i }
\transitions_j $. This means that if $U_i$ includes a point from
$\transitions_j$, then $\closure{i}$ includes all points of
$\transitions_j$. Note that $\closure{i} \supseteq \closure{i+1}$.
The next lemma shows that one can construct, for each $U_i$, a function $f_i$
such that the domain on which~\myeqref{eq:rat:Ustict}{i} holds is extended to
$\closure{i} \setminus U_{i+1}$. These functions are later used in
constructing a \llrf for $\transitions_1,\ldots,\transitions_k$.

\blem
\label{lem:rat:consolidate}
Given an irredundant weak \llrf, $\tau$, and its ranking chain $\{U_i\}$,
one can construct, for each $1\le i\le d$, an affine function $f_i : \rats^{n} \to\rats$ such that
\newcounter{eq:rat:consolidate.1}
\setcounter{eq:rat:consolidate.1}{\theequation}
\addtocounter{eq:rat:consolidate.1}{1}
\newcounter{eq:rat:consolidate.2}
\setcounter{eq:rat:consolidate.2}{\theequation}
\addtocounter{eq:rat:consolidate.2}{2}
\newcounter{eq:rat:consolidate.3}
\setcounter{eq:rat:consolidate.3}{\theequation}
\addtocounter{eq:rat:consolidate.3}{3}
\begin{indexedequations}{i}
\begin{alignat}{2}
\forall \vec{x}''\in U_{i}\ .\ && f_i(\vec{x}) \ge \rho_i(\vec{x}) &\ge 0 \label{eq:rat:consolidate.1} \\
\forall \vec{x}''\in \closure{i}\setminus U_{i+1}\ . && \diff{f_i}(\vec{x}'') &> 0 \label{eq:rat:consolidate.2} \\
\forall \vec{x}''\in U_{i+1}\ . && \diff{f_i}(\vec{x}'') &= 0  \,. \label{eq:rat:consolidate.3}  
\end{alignat}
\end{indexedequations}
\elem

\bprf 
The proof proceeds by induction.

\paragraph{Base-case} For the base-case we take $i=1$, and define
$f_1(\vec{x}) = \rho_1(\vec{x})$.
Since $\closure{1}=U_1$,
(\myref{eq:rat:consolidate.1}{1}--\myref{eq:rat:consolidate.3}{1})
hold (they are equivalent
to~(\myref{eq:rat:Unonegative}{1}--\myref{eq:rat:Uirredundant}{1}) in
this case).

\paragraph{Induction hypothesis} Let $1\le i< d$, and assume that
$f_1,\dots, f_{i}$ have been defined. In particular,  $f_i$
satisfies
(\myref{eq:rat:consolidate.1}{i}--\myref{eq:rat:consolidate.3}{i}). Only
$f_i$ is used in the induction step below.

\paragraph{Induction step} We show that
$f_{i+1}(\vec{x})=\rho_{i+1}(\vec{x})+(\xi+1)\cdont
f_i(\vec{x})$, for some $\xi\geq 0$,
satisfies~(\myref{eq:rat:consolidate.1}{i+1}--\myref{eq:rat:consolidate.3}{i+1}).
Most of the proof deals with finding $\xi$ and constructing some
related properties.
Consider $\vec{x}''\in \closure{i+1}$. If $\vec{x}''\in U_{i+1}$ then
by~\myeqref{eq:rat:consolidate.3}{i} we have $\diff{f_i}(\vec{x}'') =
0$, and if $\vec{x}''\not\in U_{i+1}$ then
$\vec{x}''\in\closure{i+1}\setminus U_{i+1} \subseteq
\closure{i}\setminus U_{i+1}$ and by~\myeqref{eq:rat:consolidate.2}{i}
we have $\diff{f_i}(\vec{x}'')>0$.
This means that the conjunction $\vec{x}'' \in \closure{i+1} \land
\diff{f_i}(\vec{x}'') \le 0$ refers only to the points of $U_{i+1}$,
and such points,
by~(\myref{eq:rat:Ustict}{i+1},\myref{eq:rat:Uirredundant}{i+1}),
satisfy $\diff{\rho_{i+1}}(\vec{x}'') \ge 0$. Thus, we get
\begin{equation}
\label{eq:rat:consolidate:farkas.0}
\vec{x}'' \in \closure{i+1} \land \diff{f_i}(\vec{x}'')\le 0 \Rightarrow  \diff{\rho_{i+1}}(\vec{x}'') \ge 0\, .
\end{equation}
Take $j\in J_{i+1}$, since $\transitions_j \subseteq \closure{i+1}$,
\eqref{eq:rat:consolidate:farkas.0} still holds when replacing
$\closure{i+1}$ by $\transitions_j$
\begin{equation}
\label{eq:rat:consolidate:farkas.1}
\vec{x}'' \in \transitions_{j} \land  \diff{f_i}(\vec{x}'') \le 0 \Rightarrow \diff{\rho_{i+1}}(\vec{x}'') \ge 0\, .
\end{equation}
Note that~\eqref{eq:rat:consolidate:farkas.1} has a non-vacant
antecedent since $U_{i+1} \cap \transitions_j\neq\emptyset$ by
definition of $J_{i+1}$, this allows using Farkas' lemma below.
Let $\rho_{i+1}(\vec{x}) = \vect{a}\cdont \vec{x} + a_0$ and
$f_i(\vec{x}) = \vect{b}\cdont \vec{x} + b_{0}$, where
$\vect{a}$ and $\vect{b}$ are row vectors of $n$ elements each.
Recall that $\transitions_{j}$ is given as a system of inequalities
$A_j''\vec{x}'' \le \vec{c}''_j$, where $A_j''$ is a matrix of dimension
$m \times 2n$.
Using these representations for $\rho_{i+1}$, $f_i$, and
$\transitions_{j}$ we can present~\eqref{eq:rat:consolidate:farkas.1} as follows:
\[
\begin{array}{rl}
 \begin{pmatrix} 
 \multicolumn{2}{c}{A_j''} \\
  ~\vect{b}, & -\vect{b} 
 \end{pmatrix}
 \cdot \vec{x}'' &
 \le 
 \begin{pmatrix}
 \vec{c}''_j \\ 
 0 
 \end{pmatrix} \\[2ex]
\hline
\rule{0pt}{12pt}
\tev{-\vect{a}}{\vect{a}}\,\cdot \vec{x}'' &\le 0
\end{array}
\]
Farkas' Lemma guarantees the existence of a vector
$\vect{\mu}_j=(\mu_{j1},\ldots,\mu_{jm}) \ge 0$,
and a scalar $\xi_j \ge 0$, such that
\begin{align}
-\vect{\mu}_j\cdot A''_j + \xi_j\cdot \tev{ -\vect{b}}{ \vect{b} }
& = \tev{\vect{a}}{ -\vect{a}} ,
 \label{eq:rat:f1}\\
\vect{\mu}_j\cdot \vec{c}''_j &\le 0. \label{eq:rat:f2}\\
\intertext{This means that}
\label{eq:rat:f1alt}
\tev{\vect{a} + \xi_j\cdont \vect{b}} { -(\vect{a} + \xi_j\cdont \vect{b})} & =
-\vect{\mu}_j\cdot A_j'' \,.
\end{align}
Now since the entries of $\vect{\mu}_j$ are non-negative,
from $A_j''\vec{x}'' \leq
\vec{c}''_j$ we get $\vect{\mu}_j\cdot A_j''\vec{x}'' \leq
\vect{\mu}_j\cdot \vec{c}''_j \leq 0$. By \eqref{eq:rat:f1alt},
\[
-\vect{\mu}_j\cdot A_j''\vec{x}''=  \tev{\vect{a} + \xi_j\cdont \vect{b}}%
{-(\vect{a} + \xi_j\cdont \vect{b})}\cdot\vec{x}'' = (\vect{a} +
\xi_j\cdont \vect{b})\cdot(\vec{x} - \vec{x}'),
\]
so we get
\begin{equation}
\label{eq:rat:farkas.2}
\forall \vec{x}''\in\transitions_j\ .\ (\vect{a} + \xi_j\cdont
\vect{b})\cdot(\vec{x} - \vec{x}') \ge 0.
\end{equation}
Note that $\xi_j\cdont \vect{b}\cdot(\vec{x} -
\vec{x}')=\xi_j\cdont\diff{f_i}(\vec{x}'')$, and that
by~(\myref{eq:rat:consolidate.2}{i},\myref{eq:rat:consolidate.3}{i})
we have $\diff{f_i}(\vec{x}'') \geq 0$ over $\closure{i}$, and thus
over $\transitions_j\subseteq\closure{i+1}\subseteq\closure{i}$.
This means that~\eqref{eq:rat:farkas.2} still holds when replacing
$\xi_j$ by any $\xi \ge \xi_j$. Now define $\xi = \max\set{\xi_j \mid
  j\in J_{i+1}}$, then~\eqref{eq:rat:farkas.2} holds for any $j\in
J_{i+1}$ and this $\xi$. Since $\closure{i+1} = \bigcup_{ j\in J_{i+1}
} \transitions_j$, we get
\begin{equation}
\label{eq:rat:farkas.3}
\forall \vec{x}'' \in \closure{i+1} \ .\ (\vect{a} + \xi\cdont \vect{b})\cdot(\vec{x} - \vec{x}') \ge 0 \,. 
\end{equation}
Now we show that $f_{i+1}(\vec{x}) = \rho_{i+1}(\vec{x}) + (\xi+1)\cdont f_i(\vec{x})$ satisfies
(\myref{eq:rat:consolidate.1}{i+1}--\myref{eq:rat:consolidate.3}{i+1}).

\begin{itemize}
\item[(\myref{eq:rat:consolidate.1}{i+1})]
  By~\myeqref{eq:rat:consolidate.1}{i} we know that $f_i(\vec{x})\ge
  0$ over $U_{i} \supset U_{i+1}$, and
  by~\myeqref{eq:rat:Unonegative}{i+1} we know that~$\rho_{i+1}(\vec{x})
  \ge 0$ over $U_{i+1}$.  Thus, for any $\vec{x}''\in U_{i+1}$ we have
  $f_{i+1}(\vec{x}) = \rho_{i+1}(\vec{x})+ (\xi + 1)\cdont
  f_i(\vec{x}) \geq \rho_{i+1}(\vec{x}) \ge 0$.

\smallskip
\item[(\myref{eq:rat:consolidate.2}{i+1})] 
  Pick an arbitrary $\vec{x}''\in\closure{i+1}\setminus U_{i+2}$, and
  consider the two complementary cases $\vec{x}''\in U_{i+1} \setminus
  U_{i+2}$ and $\vec{x}'' \not\in U_{i+1} \setminus U_{i+2}$:

  \smallskip
  \begin{enumerate}[\upshape(\itshape a\upshape)]
  \item If $\vec{x}''\in U_{i+1} \setminus U_{i+2} \subseteq U_{i+1}$, then
    by~\myeqref{eq:rat:consolidate.3}{i} we get $\diff{f_{i}}(\vec{x}'')=0$
    and by~(\myref{eq:rat:Ustict}{i+1}) we get
    $\diff{\rho_{i+1}}(\vec{x}'')>0$. Thus,
    $\diff{f_{i+1}}(\vec{x}'')=\diff{\rho_{i+1}}(\vec{x}'')+(\xi+1)\cdont\diff{f_{i}}(\vec{x}'')=\diff{\rho_{i+1}}(\vec{x}'')>0$;

  \smallskip
\item If $\vec{x}''\not\in U_{i+1}\setminus U_{i+2}$, then
  $\vec{x}''\in(\closure{i+1}\setminus U_{i+1})\setminus U_{i+2} =
  \closure{i+1}\setminus U_{i+1}$.
  Write $\diff{f_{i+1}}(\vec{x}'')$ as $(\vect{a} + \xi\cdont
  \vect{b})\cdot(\vec{x} - \vec{x}') + \diff{f_i}(\vec{x}'')$.
  On one hand $\vec{x}''\in\closure{i+1}\setminus U_{i+1}
  \subseteq \closure{i+1}$ so by~\eqref{eq:rat:farkas.3} we get
  $(\vect{a} + \xi\cdont \vect{b})\cdot(\vec{x} - \vec{x}')\ge 0$, and
  on the other hand $\vec{x}''\in\closure{i+1}\setminus U_{i+1} \subseteq
  \closure{i}\setminus U_{i+1}$ so by~\myeqref{eq:rat:consolidate.2}{i} we
  get $\diff{f_{i}}(\vec{x}'') > 0$. Thus
  $\diff{f_{i+1}}(\vec{x}'')=(\vect{a} + \xi\cdont
  \vect{b})\cdot(\vec{x} - \vec{x}') + \diff{f_i}(\vec{x}'') \geq \diff{f_{i}}(\vec{x}'')>0$.
\end{enumerate}

\smallskip
\item[(\myref{eq:rat:consolidate.3}{i+1})]
  Pick an arbitrary $\vec{x}''\in U_{i+2}$.
  By~(\myref{eq:rat:Uirredundant}{i+1}) we have
  $\diff{\rho_{i+1}}(\vec{x}'')=0$, and
  by~\myeqref{eq:rat:consolidate.3}{i}, since $U_{i+2}\subset U_{i+1}$, we
  have $\diff{f_i}(\vec{x}'') = 0$. Thus, $$\diff{f_{i+1}}(\vec{x}'') =
  \diff{\rho_{i+1}}(\vec{x}'')+
  (\xi+1)\cdont\diff{f_{i}}(\vec{x}'') = 0+(\xi+1)\cdont 0 = 0 \,.$$
\end{itemize}
This completes the proof.
\eprf

\bexm
\label{ex:wllrf:consolidate}
We compute $f_1,f_2$ and $f_3$ for the weak \lrf
$\tau=\tuple{x_1,x_2-x_1,x_3}$ of Example~\ref{ex:wllrf}. So we have
\[
\begin{array}{lll}
\rho_1(x_1,x_2,x_3)=x_1, &\quad \rho_2(x_1,x_2,x_3)=x_2-x_1, &\quad \rho_3(x_1,x_2,x_3)=x_3.\\
\end{array}
\]
We let $A''_i\vec{x}'' \le \vec{c}''_i$, for
$1 \le i \le 3$, be the constraint representations of the
transition polyhedra.
\begin{itemize}
\item[($f_1$)] We set
  $f_1(x_1,x_2,x_3)=\rho_1(x_1,x_2,x_3)=x_1$, as in the base-case of
  the induction. %
\item[($f_2$)] We have $\closure{2}=\transitions_2\cup\transitions_3$,
  thus we solve~(\ref{eq:rat:f1},\ref{eq:rat:f2}) twice, once with
  $A''_2\vec{x}'' \le \vec{c}''_2$ and once with $A''_3\vec{x}'' \le
  \vec{c}''_3$. In both cases 
\[
\begin{array}{ll}
(\vect{a},-\vect{a})=(-1,1,0,1,-1,0), &\qquad
  (-\vect{b},\vect{b})=(-1,0,0,1,0,0). 
\end{array}
\]
We get $\xi_1=0$ and
  $\xi_2=1$, and thus we take $\xi=1$. Then we define
\[
  f_2(x_1,x_2,x_3)=\rho_2(x_1,x_2,x_3)+(\xi + 1)\cdont
  f_1(x_1,x_2,x_3) = x_2+x_1.
\]
\item[($f_3$)] We have $\closure{3}=\transitions_3$, thus we
  solve~(\ref{eq:rat:f1},\ref{eq:rat:f2}) for $A''_3\vec{x}'' \le
  \vec{c}''_3$, $(\vect{a},-\vect{a})=(0,0,1,0,0,-1)$ and
  $(-\vect{b},\vect{b})=(-1,1,0,1,-1,0)$. We get $\xi=0$, and thus
\[
  f_3(x_1,x_2,x_3)=\rho_3(x_1,x_2,x_3)+ (\xi + 1)\cdont f_2(x_1,x_2,x_3)
  =x_3+x_2+x_1.
\]
\end{itemize}
\eexm

The procedure to construct $f_1,\ldots,f_d$ in
Lemma~\ref{lem:rat:consolidate} is not necessarily polynomial. This is
because the inference of $f_{i+1}$ depends on $f_i$, in particular we
add the constraint $\diff{f_i}(\vec{x}'') \le 0$ to $\transitions_j$
before using Farkas' lemma to find $\xi_j$. This means that the
bit-size of the problem can grow exponentially when repeating this
process $n$ times, since $\xi_j$ (which becomes part of $f_{i+1}$) is
of bit-size polynomial in the bit-size of the corresponding \lp
problem.
In Lemma~\ref{lem:rat:consolidate:fix} below, we describe an
alternative procedure to compute $f_1,\ldots,f_d$ in polynomial time%
\footnote{Lemma~\ref{lem:rat:consolidate:fix} does not appear in the
  journal version of this technical report~\cite{BG14}.}.
The construction in the proof of Lemma~\ref{lem:rat:consolidate} is
still needed to guarantee that there are $f_1,\ldots,f_d$ of a
particular form that we will seek using a polynomial time procedure
(we may obtain different $f_1,\ldots,f_d$, but they are of a
particular form that is enough for the statement of
Lemma~\ref{lem:rat:consolidate} to hold).

\blem
\label{lem:rat:consolidate:fix}
It is possible to construct $f_1,\ldots,f_d$ that
satisfy~(\ref{eq:rat:consolidate.1}--\ref{eq:rat:consolidate.3}) in
polynomial time.
\elem

\bprf First recall that in the proof of
Lemma~\ref{lem:rat:consolidate} we have
$f_{i+1}(\vec{x})=\rho_{i+1}(\vec{x})+\xi\cdot
f_{i}(\vec{x})+f_{i}(\vec{x})$.
We claim
that~(\myref{eq:rat:consolidate.1}{i+1}--\myref{eq:rat:consolidate.3}{i+1}) still
hold if we replace $\xi\cdot f_{i}$ by any $g_{i}$ such that
\begin{align}
\forall \vec{x}''  \in U_{i+1} \ .\ g_{i}(\vec{x}) & \ge 0\,, \label{lem:rat:consolidate:fix:1}\\
\forall \vec{x}''  \in U_{i+1} \ .\ \diff{g_{i}}(\vec{x}) & = 0\,, \label{lem:rat:consolidate:fix:2}\\
\forall \vec{x}'' \in \closure{i+1} \ .\ \diff{\rho_{i+1}}(\vec{x}'')+\diff{g_{i}}(\vec{x}'') & \ge 0 \,.\label{lem:rat:consolidate:fix:3} 
\end{align}
Next we show that this new definition of $f_{i+1}(\vec{x})$ satisfies
(\myref{eq:rat:consolidate.1}{i+1}--\myref{eq:rat:consolidate.3}{i+1}),
following the same steps (by induction) as in the proof of
Lemma~\ref{lem:rat:consolidate}. We let the base-case be $f_1=\rho_1$,
and assume that the statement holds for $f_i$, then the justification
for $f_{i+1}$ is as follows:

\begin{itemize}
\item[(\myref{eq:rat:consolidate.1}{i+1})]
  By~\myeqref{eq:rat:consolidate.1}{i} we know that
  $f_i(\vec{x})\ge 0$ over $U_{i} \supset U_{i+1}$,
  by~\eqref{lem:rat:consolidate:fix:1} we know that
  $g_{i}(\vec{x}) \ge 0$ over $U_{i+1}$, and
  by~\myeqref{eq:rat:Unonegative}{i+1} we know
  that~$\rho_{i+1}(\vec{x}) \ge 0$ over $U_{i+1}$.  Thus, for any
  $\vec{x}''\in U_{i+1}$ we have
  $f_{i+1}(\vec{x}) = \rho_{i+1}(\vec{x})+ g_{i}(\vec{x})+
  f_i(\vec{x}) \ge \rho_{i+1}(\vec{x}) \ge 0$.

\smallskip
\item[(\myref{eq:rat:consolidate.2}{i+1})] 
  Pick an arbitrary $\vec{x}''\in\closure{i+1}\setminus U_{i+2}$, and
  consider the two complementary cases $\vec{x}''\in U_{i+1} \setminus
  U_{i+2}$ and $\vec{x}'' \not\in U_{i+1} \setminus U_{i+2}$:

  \smallskip
  \begin{enumerate}[\upshape(\itshape a\upshape)]
  \item If $\vec{x}''\in U_{i+1} \setminus U_{i+2} \subseteq U_{i+1}$, then
    by~\myeqref{eq:rat:consolidate.3}{i} we get $\diff{f_{i}}(\vec{x}'')=0$, by~\eqref{lem:rat:consolidate:fix:2} we know that
    $\diff{g_{i}}(\vec{x}'')=0$,
    and by~(\myref{eq:rat:Ustict}{i+1}) we get
    $\diff{\rho_{i+1}}(\vec{x}'')>0$. Thus,
    $\diff{f_{i+1}}(\vec{x}'')=\diff{\rho_{i+1}}(\vec{x}'')+\diff{g_i}(\vec{x}'')+\diff{f_{i}}(\vec{x}'')=\diff{\rho_{i+1}}(\vec{x}'')>0$;

  \smallskip
\item If $\vec{x}''\not\in U_{i+1}\setminus U_{i+2}$, then
  $\vec{x}''\in(\closure{i+1}\setminus U_{i+1})\setminus U_{i+2} =
  \closure{i+1}\setminus U_{i+1}$.
  On one hand $\vec{x}''\in\closure{i+1}\setminus U_{i+1}
  \subseteq \closure{i+1}$ so by~\eqref{lem:rat:consolidate:fix:3} we get
  $\diff{\rho_{i+1}}(\vec{x}'')+ \diff{g_i}(\vec{x}'') \ge 0$, and
  on the other hand $\vec{x}''\in\closure{i+1}\setminus U_{i+1} \subseteq
  \closure{i}\setminus U_{i+1}$ so by~\myeqref{eq:rat:consolidate.2}{i} we
  get $\diff{f_{i}}(\vec{x}'') > 0$. Thus
  $\diff{f_{i+1}}(\vec{x}'')= \diff{\rho_{i+1}}(\vec{x}'')+ \diff{g_i}(\vec{x}'')+\diff{f_i}(\vec{x}'') \geq \diff{f_{i}}(\vec{x}'')>0$.
\end{enumerate}

\smallskip
\item[(\myref{eq:rat:consolidate.3}{i+1})]
  Pick an arbitrary $\vec{x}''\in U_{i+2}$.
  By~(\myref{eq:rat:Uirredundant}{i+1}) we have
  $\diff{\rho_{i+1}}(\vec{x}'')=0$,
  by~\eqref{lem:rat:consolidate:fix:2} we have
  $\diff{g_i}(\vec{x}'')=0$ over $U_{i+1}\supset U_{i+2}$, and
  by~\myeqref{eq:rat:consolidate.3}{i} we have
  $\diff{f_i}(\vec{x}'') = 0$ over $U_{i+1}\supset U_{i+2}$. Thus,
  $\diff{f_{i+1}}(\vec{x}'') =
  \diff{\rho_{i+1}}(\vec{x}'')+\diff{g_i}(\vec{x})+\diff{f_{i}}(\vec{x}'')
  = 0+0+0 = 0 \,.$
\end{itemize}
This completes the proof
that~(\myref{eq:rat:consolidate.1}{i+1}--\myref{eq:rat:consolidate.3}{i+1}
hold for $f_{i+1}$ when replacing $\xi\cdot f_{i}$ by any $g_{i}$ that
satisfies~(\ref{lem:rat:consolidate:fix:1}--\ref{lem:rat:consolidate:fix:3}).
Note that it is enough to infer $g_1,\ldots,g_{d-1}$ that
satisfy~(\ref{lem:rat:consolidate:fix:1}--\ref{lem:rat:consolidate:fix:3}),
which by Lemma~\ref{lem:rat:consolidate} we know that they exist ($g_i$
is $\xi \cdot f_i$ that was constructed there), and then define
$f_{i+1}(\vec{x})=\rho_{i+1}(\vec{x})+g_i(\vec{x})+f_i(\vec{x})$ where
$f_1=\rho_1$.
Inferring $g_i$ can be done as follows. Since
$U_{i+1}=\bigcup_{\ell=1}^k (\transitions_\ell \wedge
\diff{\rho_1}(\vec{x}'')= 0\wedge\cdots\wedge\diff{\rho_i}(\vec{x}'')=
0)$,
and by the definition of a weak \llrf, for each $\transitions_\ell$ we have
\begin{equation}
\transitions_\ell \wedge \diff{\rho_1}(\vec{x}'')\le 0\wedge\cdots\wedge\diff{\rho_i}(\vec{x}'')\le 0 \to \diff{\rho_{i+1}}(\vec{x}'') \ge 0
\end{equation}
When the left-hand side is not empty, we can use Farkas lemma (in a
similar way to what we have done in the proof of
Lemma~\ref{lem:rat:consolidate}) to find $\mu_1,\ldots,\mu_i$ such
that
\begin{equation}
\label{fix:farkas}
\transitions_\ell \to \diff{\rho_{i+1}}(\vec{x}'') + \sum_{j=1}^i \mu_j\cdot\diff{\rho_j}(\vec{x}'') \ge 0
\end{equation}
Let $g_i(\vec{x}) = \sum_{j=1}^i \mu_j\cdot\rho_j(\vec{x}'')$ and note that:
\begin{itemize}
\item \eqref{lem:rat:consolidate:fix:1} holds since
  by~\myeqref{eq:rat:Unonegative}{j}, for any $1 \le j \le i$, we have
  $\rho_j(\vec{x}) \ge 0$ for any
  $\vec{x}''\in U_{j} \supset U_{i+1}$. This means that
  $g_i(\vec{x}) \ge 0$ over $U_{i+1}$ since $g_i$ is a non-negative
  combination of $\rho_1,\ldots,\rho_i$.
\item \eqref{lem:rat:consolidate:fix:2}
holds since
  by~\myeqref{eq:rat:Uirredundant}{j}, for any $1 \le j \le i$, we have
  $\diff{\rho}_j(\vec{x}) = 0$ for any
  $\vec{x}''\in U_{j+1} \supseteq U_{i+1}$. This means that
  $\diff{g}_i(\vec{x}) = 0$ over $U_{i+1}$ since $g_i$ is a non-negative combination of
  $\rho_1,\ldots,\rho_i$.
\item \eqref{lem:rat:consolidate:fix:3} holds for the part of
  \closure{i+1} that corresponds to $\transitions_\ell$
  by~\eqref{fix:farkas}. To make it hold for all $\closure{i+1}$, we
  have to find $\mu_1,\ldots,\mu_i$ that work for all
  $\transitions_\ell$, and by Lemma~\ref{lem:rat:consolidate} we know
  that they exists (those that define $\xi\cdot f_i$). To compute
  them, we can solve~\eqref{fix:farkas} simultaneously for all
  $\transitions_\ell$ using several instances of Farkas' lemma that use
  the same $\mu_1,\ldots,\mu_i$.
\end{itemize}
Computing each $g_i$ can be done in polynomial time, since it is based
on solving an \lp problem of bit-size polynomial in the bit-size of the
loop $\transitions_1,\ldots,\transitions_k$:
\begin{inparaenum}[\upshape(\itshape i\upshape)]
\item the number of variables and constraints is polynomial in the
  number of paths, variables, and constraints of the loop; and
\item each constraint uses coefficients that appear in the loop or in
  $\rho_1,\ldots,\rho_d$, and by Lemma~\ref{lem:alg:ptime} the
  coefficients of each $\rho_i$ are of bit-size polynomial in the
  bit-size of the loop.
\end{inparaenum}
This completes the proof.
\eprf

Now we show how to use $f_1,\ldots,f_d$ of 
Lemma~\ref{lem:rat:consolidate:fix} in order to construct a \llrf for
$\transitions_1,\ldots,\transitions_k$. We first state an auxiliary
definition.

\bdfn
For affine functions $\rho_1,\dots,\rho_j:\rats^n \to \rats$, and
positive constants $\delta_1,\dots,\delta_j$,  define
$\ranked(\tuple{\rho_1,\dots,\rho_j},\tuple{\delta_1,\ldots,\delta_j})$
to be the set of $\vec{x}'' \in \rats^{2n}$ for which there is an 
$1 \le i\le j$ satisfying  (\ref{eq:rat:llrf1d}--\ref{eq:rat:llrf3d}).
We say that such transitions $\vec{x}''$ are ranked by $\tuple{\rho_1,\dots,\rho_j}$ (with $\delta_1,\ldots,\delta_j$),
or, to name the position, that they are ranked by $\rho_i$ in $\ranked(\tuple{\rho_1,\dots,\rho_j},\tuple{\delta_1,\ldots,\delta_j})$.
\edfn

In the next lemma we construct a \llrf $\llrfsym_\ell$ that ranks all
transitions of $\transitions_\ell$, for each $1 \le \ell \le k$. 
Afterwards, we show how $\llrfsym_1,\ldots,\llrfsym_k$ are combined
into a \llrf $\llrfsym$ for $\transitions_1,\ldots,\transitions_k$.

\blem
\label{lem:rat:llrf}
Let $1 \le d' \le d$ be the largest $d'$ such that $U_{d'} \cap
\transitions_\ell \neq \emptyset$ for a given $\transitions_\ell$. Then, $\llrfsym_\ell=\tuple{\rho_1',\ldots,\rho_{d'}'}$, where
$\rho_i'=f_i+i-1$, is a \llrf for $\transitions_\ell$.
\elem

\bprf
\newcommand{\fp}[2]{{\ensuremath{f_{#2}^{[#1]}}}}
For $1\le i\le d'$,
let $X_i=U_i\cap \transitions_\ell$. Note that
$X_1,\ldots,X_{d'}$ are closed polyhedra,
$\transitions_\ell=X_1\supseteq\ldots\supseteq X_{d'}\neq\emptyset$,
and $X_{d'}\cap U_{d'+1}=\emptyset$.
We find $\delta_1,\ldots,\delta_{d'}$ such that
\begin{equation}
\label{eq:rat:ranked}
\ranked(\tuple{\rho'_1, \dots, \rho'_{d'}},\tuple{\delta_1,\dots,\delta_{d'}}) \supseteq X_1 \,.
\end{equation}
This implies the lemma's statement since $X_1=\transitions_\ell$.  The
proof is by induction, where we start from $i=d'$ and proceed
backwards. In the $i$-th step we find $\delta_i$ such that

\newcounter{eq:rat:ranked:ind}
\setcounter{eq:rat:ranked:ind}{\theequation}
\addtocounter{eq:rat:ranked:ind}{1}
\begin{indexedequations}{i}
\begin{equation}
R_i \eqdef 
\ranked(\tuple{\fp{i}{i}, \dots, \fp{i}{d'}},\tuple{i\cdont\delta_i, i\cdont\delta_{i+1},\dots, i\cdont\delta_{d'}}) \supseteq X_i  \:,
\label{eq:rat:ranked:ind}
\end{equation}
\end{indexedequations}

\noindent 
where $\fp{i}{j}=f_j+j-i$. Then, for $i=1$ we get~\eqref{eq:rat:ranked}.
First note that $\diff{\fp{i_1}{j}}=\diff{\fp{i_2}{j}}=\diff{f_j}$ for any
$1 \le i_1 < i_2 \le d'$, this relation is fundamental to our proof.
The intuition behind the offset $j-i$ in $\fp{i}{j}$ is
explained below, at the beginning of the induction step.

\paragraph{Base-case} 
We take $i=d'$, then $\fp{d'}{d'}=f_{d'}$ and thus
$R_{d'}=\ranked(\tuple{f_{d'}},\tuple{d'\cdont\delta_{d'}})$.
Since $X_{d'}\subseteq U_{d'}$ and $X_{d'}\cap U_{d'+1}=\emptyset$,
then, for any $\vec{x}''\in X_{d'}$, by~\myeqref{eq:rat:consolidate.1}{d'}
we have $f_{d'}(\vec{x}) \ge 0$ and by~\myeqref{eq:rat:consolidate.2}{d'} we
have $\diff{f_{d'}}(\vec{x}'')>0$.
Now since $X_{d'}$ is a closed polyhedron and $\diff{f_{d'}}$ is
positive over $X_{d'}$, $\diff{f_{d'}}$ must have a minimum
$\mu>0$ in $X_{d'}$. Define $\delta_{d'}=\frac{\mu}{d'}$, then
$\diff{f_{d'}}(\vec{x}'') \geq \mu = d'\cdont\delta_{d'}$. Thus,
$X_{d'} \subseteq R_{d'}$.

\paragraph{Induction hypothesis} $X_{i+1}\subseteq R_{i+1}$.

\paragraph{Induction step} We find a value for $\delta_i$, and show
that $X_{i}\subseteq R_{i}$. Note that $R_{i}$ uses the same
$\delta_{i+1},\ldots,\delta_{d'}$ as $R_{i+1}$.

Let us first intuitively explain how the induction step is carried
out.
We first split $X_i$ into two sets, $C_i$ and $X_i\setminus C_i$, and
then show that each transition in $X_i\setminus C_i$ is ranked by
$\fp{i}{j}$ for some $j>i$, and that each transition in $C_i$ is
ranked by $\fp{i}{i}$.
To construct $C_i$, we simply start by considering the set of
transitions that violate the \llrf conditions
(\ref{eq:rat:llrf1d}-\ref{eq:rat:llrf3d}) for all components
$j>i$. This set is not closed, and, in order close it, we include also
transitions that are on the ``edge'' (simply by turning strict
inequalities to non-strict ones). Being closed is fundamental for a
later step in the proof.
Going back to the definition of $\fp{j}{i}$, the reason for which we
use the offset $j-i$ (so it becomes larger as $i$ becomes smaller)
can be explained as moving the transitions of $R_{i+1}$ away
from some ``edge''.
Next we define $C_i$, and then prove the desired properties of
$X_i\setminus C_i$ and $C_i$.

\medskip

Recall that $C_i$ should be a superset of the transitions that
are not ranked by any component $i\le j\le d'$ in $R_{i}$.
Note that for any $i\le j\le d'$,
by~(\myref{eq:rat:consolidate.2}{j},\myref{eq:rat:consolidate.3}{j})
we have $\diff{\fp{i}{j}}(\vec{x}'')=\diff{f_j} \geq 0$ for any
$\vec{x}'' \in X_i$, thus it is not possible to
violate~\eqref{eq:rat:llrf1d}. This means that if $\vec{x}''$ is not
ranked by some $i<j\le d'$ in $R_{i}$, then one of the following must
hold:
\begin{itemize}
\item $\diff{\fp{i}{j}}(\vec{x}'') < i\cdont\delta_j$ for any $i< j
  \leq d'$, to violate (\ref{eq:rat:llrf3d}); or
\item if there is $i< j' \le d'$ for which
  $\diff{\fp{i}{j'}}(\vec{x}'') \geq i\cdont\delta_{j'}$, assuming it
  is the smallest $j'$, then there must be $l \leq j'$ for which
  $\fp{i}{l}(\vec{x}) < 0$, to violate (\ref{eq:rat:llrf2d}).
\end{itemize}
The set of transitions that satisfy either of the above conditions is
not necessarily closed --- due to the use of strict inequalities. To
obtain a closed set, we simply turn $<$ to $\leq$, and define $C_i$
to be the set of all transitions $\vec{x}''\in X_i$ for which one
of the following holds
\begin{align}
\forall i < j \le d' \ .\  &\diff{\fp{i}{j}}(\vec{x}'') \leq i\cdont\delta_j\; , \label{eq:rat:Ci:1}\\
\exists l\geq i \ .\ &(\forall i < j < l \ .\ \diff{\fp{i}{j}}(\vec{x}'') \leq i\cdont\delta_j) \land \fp{i}{l}(\vec{x}) \leq 0\; . \label{eq:rat:Ci:2}
\end{align}
 Thus $C_i$ is
closed, and consists of a finite union of closed
polyhedra.
Note that the role of $i\cdont\delta_j$ is similar to the offset in
$\fp{j}{i}$, it moves the transitions of $R_{i+1}$ away from some
``edge'' (since $R_{i+1}$ uses $(i+1)\cdot\delta_j$ while $R_{i}$ uses
$i\cdot\delta_j$).

\medskip

We now prove that each transition in $X_i\setminus C_i$ is ranked by
$\fp{i}{j}$, for some $i < j \le d'$, in $R_i$.
Pick an arbitrary transition $\vec{x}''\in X_i \setminus C_i$, we show
that it is ranked by $\fp{i}{j}$ in $R_i$, for some $j > i$.
To see this, note the following:

\begin{itemize}
\item Since $\vec{x}'' \not\in C_i$,  it
  violates~\eqref{eq:rat:Ci:1} and~\eqref{eq:rat:Ci:2}.
  To violate~\eqref{eq:rat:Ci:1}, there must be $i < j \le d'$ for
  which 
\begin{equation}
\label{eq:desc.j}
\diff{\fp{i}{j}}(\vec{x}'') > i\cdont\delta_j \,.
\end{equation}
  Take minimal such $j$, then, for any $i< j' < j$, we have
  $\diff{\fp{i}{j'}}(\vec{x}'') \le i\cdont\delta_{j'}$.
  This means that the first conjunct of~\eqref{eq:rat:Ci:2} is not
  violated by $\vec{x}''$ for any $i<l\le j$, and thus, to
  violate~\eqref{eq:rat:Ci:2}, the second conjunct
  $\fp{i}{l}(\vec{x})\le 0$ must be violated, that is:
\begin{equation}
\label{eq:positive.l}
  \forall i < l \le j\ .\ \fp{i}{l}(\vec{x})> 0 \,.
\end{equation}
\item 
Let $i \le l \le d'$.
Since  $X_l = U_l \cap \transitions_\ell$ is not empty,  $\transitions_\ell\subseteq\closure{l}$.
  This means that $\vec{x}''\in\closure{l}$,
  and thus
  by~(\myref{eq:rat:consolidate.2}{l},\myref{eq:rat:consolidate.3}{l})
  we have 
\begin{equation}
\label{eq:desc.l}
\diff{\fp{i}{l}}(\vec{x}'')=\diff{f_l}(\vec{x}'') \ge 0 \,.
\end{equation}  %
  Moreover, since $\vec{x}''\in X_i \subseteq U_i$,   by~\myeqref{eq:rat:consolidate.1}{i} we have
\begin{equation}
\label{eq:positive.i}
  \fp{i}{i}(\vec{x})=f_i(\vec{x}) \ge 0
\end{equation}
\end{itemize}
Inequalities (\ref{eq:desc.j}--\ref{eq:positive.i}) show that $\vec{x}''$ is ranked by
$\fp{i}{j}$ in $R_i$.

\medskip

Now we show that the transitions of $C_i$ are ranked by $\fp{i}{i}$ in
$R_i$, for some $\delta_i$. If $C_i=\emptyset$ then we simply take
$\delta_i=\delta_{i+1}$, and clearly $X_i\subseteq R_i$ (since the
transitions of $X_i\setminus C_i$ are ranked as we have seen above
independently from $\delta_i$).
Assume $C_i\neq\emptyset$. 
We first claim that $C_i \cap X_{i+1} = \emptyset$. To see this, take
$\vec{x}'' \in X_{i+1}$, by the induction hypothesis we have
$X_{i+1}\subseteq R_{i+1}$ and thus there must be $\fp{i+1}{j}$, for
some $i < j \le d'$, that ranks $\vec{x}''$, thus:
\begin{itemize}
  \item $\diff{\fp{i}{j}}(\vec{x}'')=\diff{\fp{i+1}{j}}(\vec{x}'') \ge
  (i+1)\cdont\delta_j > i\cdont\delta_j$, so~\eqref{eq:rat:Ci:1} is
  violated;
\item $\fp{i+1}{l}(\vec{x}) \ge 0$ for any $i< l \le j$, and thus
  $\fp{i}{l}(\vec{x})=\fp{i+1}{l}(\vec{x})+1\geq 1$. This means
  that~\eqref{eq:rat:Ci:2} cannot be true for any $i < l \le j$, it
  also cannot be true for any $j < l \le d'$ since
  $\diff{\fp{i}{j}}(\vec{x}'') > i\cdont\delta_j$ as we have seen in
  the previous point.
\end{itemize}
Now since $C_i\cap X_{i+1}=\emptyset$ and $C_i \subseteq X_i$ we get
$C_i \subseteq X_i \setminus X_{i+1}$.
We also know that $X_i \setminus X_{i+1} \subseteq
\closure{i}\setminus U_{i+1}$ by definition, and that
by~\myeqref{eq:rat:consolidate.2}{i} we have $\diff{f_i}(\vec{x}'')>0$
throughout $\closure{i}\setminus U_{i+1}$.
This means that $\diff{f_i}(\vec{x}'')>0$ throughout $C_i$ as
well. Now since $C_i$ is a finite union of closed polyhedra,
$\diff{f_i}(\vec{x}'')$ must have a minimum $\mu>0$.
Define $\delta_i=\frac{\mu}{i}$ then
$\fp{i}{i}(\vec{x}'')=f_i(\vec{x})'' \geq \mu = i\cdont \frac{\mu}{i}$.
Moreover, by~\myeqref{eq:rat:consolidate.1}{i} we have $f_i(\vec{x})
\ge 0$ and thus $\fp{i}{i}(\vec{x})=f_i(\vec{x}) \ge 0$. This proves
that $\vec{x}''\in C_i$ is ranked by $\fp{i}{i}$ in $R_i$.
\eprf

\blem
\label{thm:rat:llrf}
$\llrfsym = \tuple{\rho_1',\ldots,\rho_d'}$, where $\rho_j'=f_j+j-1$, is a \llrf
for $\transitions_1,\ldots,\transitions_k$. Moreover, it has a minimal
dimension, at most $n$.
\elem

\bprf 
That $\llrfsym$ is a \llrf
follows immediately from Lemma~\ref{lem:rat:llrf}, because the
transitions of each $\transitions_\ell$ are ranked in $\llrfsym_\ell$,
and each $\llrfsym_\ell$ is a prefix of $\llrfsym$. 
The minimality of the
dimension follows from that of the weak \llrf: if there were a shorter \llrf, since every \llrf is a weak \llrf,
it would contradict Lemma~\ref{th:llrfq-completeness}.
\eprf

\bexm 
\label{ex:wllrr:llrf}
Consider again the weak \llrf of Example~\ref{ex:wllrf}, and
$f_1=x_1$, $f_2=x_2+x_1$ and $f_3=x_3+x_2+x_1$ that we have computed
in Example~\ref{ex:wllrf:consolidate}. The corresponding \llrf is
$\llrfsym=\tuple{x_1,x_1+x_2+1,x_1+x_2+x_3+2}$, with $\delta_1=1$,
$\delta_2=\frac{1}{2}$ and $\delta_3=\frac{1}{3}$.
\eexm

\bthm
\label{thm:rat:ptime}
\llinrfq is PTIME-decidable.
\ethm

\bprf
Procedure $\procsyn$, which has polynomial-time complexity by
Lemma~\ref{lem:alg:ptime}, is complete for the existence of a weak
\llrf. If no weak \llrf exists then no \llrf exists either, and by
Lemma~\ref{thm:rat:llrf}, if one exists then there is a \llrf.
\eprf

Note that if only termination is of interest, then there is no reason
to actually perform the construction of
Lemmas~\ref{lem:rat:consolidate:fix} and~\ref{lem:rat:llrf}, it suffices
to check the existence of a weak \llrf.
Ranking functions are also used to bound the number of iterations of
loops, as discussed in the next subsection. 
In this context, an explicit upper bound is desirable, so we may need to
carry out the construction of Lemmas~\ref{lem:rat:consolidate:fix}
and~\ref{lem:rat:llrf}, which is polynomial.

\subsection{Lexicographic Ranking Functions and Iteration Bounds}

 \citeN{ADFG:2010} showed how lexicographic ranking functions can be used to bound the
 number of steps in a program---in our restricted form of programs this is just the number of iterations
 of the loop. What is sought is a symbolic bound, as an expression in terms of the input variables.
\lrfs clearly provide linear bounds, and \llrfs provide
polynomial bounds when each component of the \llrf has a linear upper bound (derived using a linear-invariant generator).
Clearly, this bound is at most
the product of the bounds on the individual components, and hence a polynomial of degree given by the dimension of the
\llrf (this motivates the interest in \llrfs of minimal dimension).
 In the next theorem we show that, in
fact, for \slc loops we can always find a piecewise linear bound (this observation applies
whether one is interested in ranking all rational points or just integer ones).
Note that \citeN{ADFG:2010} proved that an \slc loop has a \llrf if and
only if it has a \lrf, and thus has a linear bound on the number of
iterations. However, our definition of \llrf captures some \slc loops that do not have a \lrf,
as seen in Example~\ref{ex:kinds-of-lrfs-1}.

\bthm
\label{thm:loop-bound}
Let $\transitions$ be the transition polyhedron of an \slc loop,
$\tuple{\rho_1,\ldots,\rho_{d}}$ a (weak) \llrf inferred by
Procedure $\procsyn$, and
$\llrfsym=\tuple{\rho_1',\ldots,\rho_{d'}'}$ a \llrf as constructed in
Lemma~\ref{lem:rat:llrf} with corresponding
$\delta_1,\ldots,\delta_{d'}$.
Given an input $\vec{x}\in\rats^n$, let $j$ be the minimum $1 \le j
\le d'$ such that $\rho_j'(\vec{x}) < 0$, or $j=d'$ if no one exists,
then $\sum_{i=1}^{j-1}(\lfloor \rho_i'(\vec{x})/\delta_i\rfloor+1)$ is
an upper bound on the number of iterations of $\transitions$ when
starting from $\vec{x}$.
\ethm

\bprf
By Lemma~\ref{lem:rat:consolidate}, any $\vec{z}''\in\transitions$
satisfies $\diff{\rho_i'}(\vec{z}'') \ge 0$; for any $1 \le i \le d'$,
which means that once the $i$-th component of $\llrfsym$ become
negative, it is then disabled and cannot rank any transition anymore
(since it remains negative).
In addition, when a transition is ranked by the $i$-th component,
$\diff{\rho_i'}(\vec{x}'') \geq \delta_i$ which, together with the
above argument, means that the $i$-th component of $\llrfsym$ can rank
at most $\lfloor \rho_i'(\vec{x})/\delta_i\rfloor+1$ transitions
before it becomes negative.
Now since every transition in the execution trace must be ranked by
some component $\rho'_i$ of $\llrfsym$, and $i$ cannot be $\ge
j$ since such components are disabled right from the beginning,
we get the
upper bound $\sum_{i=1}^{j-1}(\lfloor
\rho_i'(\vec{x})/\delta_i\rfloor+1)$.
\eprf

\noindent Remarks:
\begin{enumerate}
\item
If we are only interested in an upper bound up to a constant
factor, we can avoid the construction of
Lemmas~\ref{lem:rat:consolidate} and~\ref{lem:rat:llrf} because
$\sum_{i=1}^{j-1}(\lfloor \rho_i'(\vec{x})/\delta_i\rfloor+1)$ is
$O(\sum_{i=1}^{d}\max(0, \rho_i(\vec{x}))$.
\item
The theorem is easily extended to conclude that the
piecewise linear bound is also valid
for \mlc loops, when $\rho_d$ ranks at least one transition from each
$\transitions_i$, that is, $U_d\cap\transitions_i\neq\emptyset$ for all
$1 \le k \le d$.
\end{enumerate}

One of the interesting parts of \cite{ADFG:2010} is the way they compute an iteration bound 
which is sometimes better than the product of the bounds on the \llrf components. The idea:
 Since $\rho$ always decreases, the number of steps is bounded by the number of distinct values it
 takes throughout the computation.
 Let $\states$ be the polyhedron which circumscribes the state space (in our case, the loop condition);
$\rho(\states)$ is a $d$-dimensional polyhedron, and, assuming that the program computes over integers,
the number of steps is bounded by
the number of integer points in this polyhedron, i.e., $|\intpoly{\rho(\states)}|$.
Alias et al.~estimate this number using techniques related to Ehrhart polynomials,  as implemented in the
PolyLib library~\cite{PolyLib:1993}. Such an approach can also be used with our class of functions, but it is an open problem how to get the
best results out of such computations. For example, is it possible to find a computation method that will always get
a piecewise linear bound in the situations described by the above theorem?

\section{Prototype Implementation}
\label{sec:implementation}

The different algorithms presented in this paper for synthesizing
\lrfs an \llrfs, both for the general cases and the special PTIME
cases, have been implemented. Our tool, \irankfinder, can be tried out
via \url{http://www.loopkiller.com/irankfinder}.  It receives as input
an \mlc loop in constraint representation, and allows applying
different algorithms for \linrfz, \linrfq, \llinrfz, or \llinrfq.
For \lrfs, the implementation includes the algorithms of
Theorems~\ref{thm:rf-synth} and~\ref{thm:linrfq:mlc}. By default it
uses the second one since the first one relies on the
generator representation of the transition polyhedron, which may take
exponential time to compute. For \llrfs it uses Algorithm~\ref{alg:llrfsyn}.

\begin{algorithm}[t]
\caption{Find a point in the relative interior}
\label{alg:rintpoint}
\DontPrintSemicolon
\LinesNumberedHidden
\SetKwFunction{procrint}{InteriorPoint}
\procrint{$\poly{S}$}\;
\KwIn{Space of quasi-\lrfs $\poly{S}$}
\KwOut{ A point $(\lambda_0,\lambda)$ in the relative interior }
\Begin{
\setcounter{AlgoLine}{0}
\ShowLn \For{$i=1\to n$}{
\ShowLn  $a \leftarrow \mbox{minimize } \rfcoeff_i \mbox{ wrt } \poly{S}$\;
\ShowLn  $b \leftarrow \mbox{maximize } \rfcoeff_i \mbox{ wrt } \poly{S}$\;
\ShowLn  \lIf{$a=b$}{ $c_i=a$ }\;
\ShowLn  \lElse{pick $c_i$ in the non-closed interval $(a,b)$, prioritizing $0$ and integers}\;\label{alg:rintpoint:0}
\ShowLn  $\poly{S} \leftarrow \poly{S}\land \{\rfcoeff_i=c_i\}$}
\ShowLn  $c_0 \leftarrow \mbox{minimize } \rfcoeff_0 \mbox{ wrt } \poly{S}$\;
\ShowLn \Return $\rfp{c_0}{\vect{c}}$
}
\end{algorithm}

Our algorithm for synthesizing non-trivial quasi-\lrfs, as described
in Lemma~\ref{lem:qlrfalg}, requires finding a point in the relative
interior of a polyhedron $\poly{S}$. Note that $\poly{S}$ is of dimension
$n'=n+1+\sum_{i=1}^k 2m_i$ and is defined by $m'= k(8n+2)+\sum_{i=1}^k
2m_i$ inequalities, where $m_i$ is the number of inequalities in
$\transitions_i$.
Existing algorithms for finding an interior point require solving at most
$n'$ or $m'$ \lp problems, and they have polynomial-time
complexity~\cite[Sec.~8.3]{fukuda13:lecture}. 
Now note that instead of finding a point in the relative interior of
$\poly{S}$, we could also project $\poly{S}$ onto $\vect{\rfcoeff}$,
and then find a point in the relative interior of the resulting polyhedron $\poly{S}_{|\vect\lambda}$. It is easy to see that 
Lemma~\ref{lem:qlrfalg} remains valid.
In our implementation we find such point without actually computing
$\poly{S}_{|\vect\lambda}$, by solving only $2n+1$ \lp problems.
The underlying procedure is depicted in Algorithm~\ref{alg:rintpoint},
it finds values for $\vect{\rfcoeff}$ iteratively as follows: in the
$i$-th iteration it computes the minimum and maximum values of
$\rfcoeff_i$ in $\poly{S}$, and then sets $\rfcoeff_i$ to a value that
lies between those extremes. Once all $\rfcoeff_i$ are computed, we
look for the minimum compatible value of $\rfcoeff_0$, and then
$\rfp{c_0}{\vect{c}}$ is the desired point.
We do not claim that the complexity of this algorithm is polynomial, since we
add $\rfcoeff_i=c$ to $\poly{S}$ in each iteration and thus the
bit-size might grow exponentially. However, we have experimentally
observed that it performs far better than an algorithm that finds a
point in the relative interior of $\poly{S}$.
Note that at Line~\ref{alg:rintpoint:0}, we prioritize $0$ over any
other coefficient, as a heuristic to obtain ``small'' ranking
functions. Moreover, we prioritize integer over fractional coefficients.
Both measures are intended to get more readable results, but we think they may
also improve time bounds inferred from our ranking functions.

Computing the integer hull of a polyhedron, in the case of \linrfz and \llinrfz, is
done by first decomposing  its set of inequalities into independent components,
and then computing the integer hull of each component separately.
Each set of inequalities is first matched against the PTIME cases of
sections~\ref{sec:special:intpoly}. %
If this matching fails,  the integer hull is computed using the 
algorithm described by~\citeN{DBLP:conf/aaim/CharlesHK09}. 
Note that this algorithm
supports only bounded polyhedra, the integer hull of an unbounded
polyhedron is computed by considering a corresponding bounded
one~\cite[Th.~16.1, p.~231]{Schrijver86}.
In addition, for octagonal relations, it gives the possibility of
computing the tight closure instead of the integer hull. As we have
seen in Section~\ref{sec:special:octagons}, when this option is used,
completeness of \linrfz is not guaranteed.

The Parma Polyhedra Library~\cite{BagnaraHZ08} is used for converting
between generator and constraints representations, solving (mixed) \lp
problems, etc.

\section{Related Work}
\label{sec:related-work}

There are several
works~\cite{DBLP:conf/pods/SohnG91,DBLP:conf/tacas/ColonS01,DBLP:conf/vmcai/PodelskiR04,DBLP:journals/tplp/MesnardS08,ADFG:2010}
that directly address the \linrfq problem for \slc or \mlc loops. In
all these works, the underlying techniques allow synthesizing \lrfs
and not only deciding if one exists.
The common observation to all these works is that synthesising \lrfs
can be done by inferring the implied inequalities of a given
polyhedron (the transition polyhedron of the loop), in particular
inequalities like conditions~\eqref{eq:lrf1} and \eqref{eq:lrf2} of
Definition~\ref{def:linearrf} that define a \lrf.
Regarding completeness, all these methods are complete for \linrfq but
not for \linrfz. They can also be used to approximate \linrfz
by  relaxing the loop such that its variables range over $\rats$
instead of $\ints$, thus sacrificing completeness.
All these methods have a corresponding PTIME algorithm.
Exceptions in this line of research are the work of~\citeN{DBLP:conf/concur/BradleyMS05} and~\citeN{CookKRW10} that directly address the \linrfz problem for \mlc loops. Below, we comment in more detail on each of these works.
\citeN{DBLP:conf/pods/SohnG91} considered \mlc
loops with variables ranging over \nats. These are abstractions of
loops from logic programs.
The loops were relaxed from \nats to \ratsp before seeking a
\lrf, however, this is not explicitly mentioned.
The main observation in this work is that the duality theorem of
\lp~\cite[p.~92]{Schrijver86} can be used to infer inequalities that are
implied by the transition polyhedron. The authors also mention that
this was observed before by~\citeN{DBLP:conf/pods/Lassez90} in
the context of solving {\sc CLP($\reals$)} queries.
Completeness was not addressed in this work, and the PTIME complexity
was mentioned but not formally addressed.
Later, \citeN{DBLP:journals/tplp/MesnardS08} formally proved that the
techniques of~\citeN{DBLP:conf/pods/SohnG91} provide a complete PTIME
method for \linrfq, also for the case of \mlc loops. They pointed out
the incompleteness for \linrfz.

Probably the most popular work on the synthesis of \lrfs is the one
of~\citeN{DBLP:conf/vmcai/PodelskiR04}. They also observed the need for
deriving inequalities implied by the transition polyhedron, but
instead of using the duality theorem of \lp they used the affine form
of Farkas' lemma~\cite[p.~93]{Schrijver86}.
Completeness  was claimed, and the statement did not make it clear that
the method is complete for \linrfq but not for \linrfz.
This was clarified, however, in the PhD thesis of~\citeN{Rybalchenko04}.
One of the reasons for the impact of this work is its use in the Terminator tool~\cite{CPR06},
which demonstrated the use of \lrfs in termination analysis of complex, real-world programs.

\citeN{DBLP:journals/iandc/BagnaraMPZ12} proved that the
methods of~\citeN{DBLP:journals/tplp/MesnardS08} and
\citeN{DBLP:conf/vmcai/PodelskiR04} are actually equivalent, i.e.,
they compute the same set of \lrfs. They also showed that the method
of Podelski and Rybalchenko can, potentially, be more efficient since
it requires solving rational constraints systems with fewer variables
and constraints.

The earliest appearances of a solution based on Farkas' Lemma, that we
know of, are by~\citeN{DBLP:conf/tacas/ColonS01}, in the context of
termination analysis, and by~\citeN{Feautrier92.1}, in the context of
automatic parallelization of computations.
\citeN{DBLP:conf/tacas/ColonS01} did not claim that the problem can be
solved in polynomial time, and indeed their implementation seems to
have exponential complexity since they use generators and polars,
despite the similarity of the underlying theory to that
of~\citeN{DBLP:conf/vmcai/PodelskiR04}.
Completeness was claimed, however it was not
explicitly mentioned that the variables range over \rats and not \ints
(the programs in the examples used integer variables). 
In this work the input loop comes with an initial condition on the
input, which is used to infer a supporting invariant.

\citeN{Feautrier92.1} described scheduling of computations that can be
described by recursive equations. An abstraction to a form similar to
an \mlc loop allowed him to compute a so-called \emph{schedule}, which
is essentially a ranking function, but used backwards, since the
computations at the bottom of the recursion tree are to be completed
first. \citeN{Feautrier92.2} extends this work to lexicographic rankings; this work was
subsequently extended by~\citeN{ADFG:2010} to \llrf generation, as described below.

\citeN{CookKRW10} observed that the Farkas-lemma based solution is
complete for \linrfz when the input \mlc loop is specified by integer
polyhedra. They also mention that any polyhedron can be converted to
an integer one, and that this might increase its size
exponentially. Unlike our work, they do not address PTIME cases or the
complexity of \linrfz. In fact, the main issue in that work is the
synthesis of ranking functions for machine-level integers
(bit-victors).

\citeN{DBLP:conf/concur/BradleyMS05} directly addressed the \linrfz
problem for \mlc loops, and stated that the methods
of~\citeN{DBLP:conf/tacas/ColonS01}
and~\citeN{DBLP:conf/vmcai/PodelskiR04} are not complete for \linrfz.
Their technique is based on the observation that if there is a \lrf,
then there exists one in which each coefficient $\rfcoeff_i$ has a
value in the interval $[-1,1]$, and moreover with denominators that
are power of $2$.
Using this observation, they recursively search for the coefficients
starting from a region defined by a hyper-rectangle in which each
$\rfcoeff_i$ is in the interval $[-1,1]$.
Given a hyper-rectangle, the algorithm first checks if one
of its corners defines a \lrf, in which case it stops. Otherwise, the
region is either pruned (if it can be verified that it contains no
solution), or divided into smaller regions for recursive search.
Testing if a region should be pruned is done by checking the
satisfiability of a possibly exponential (in the number of variables)
number of Presburger formulas. %
The algorithm will find a \lrf if exists, but it might not
terminate if no \lrf exists. To make it practical, it is parametrized
by the search depth, thus sacrificing completeness.
It is interesting to note that the search-depth parameter in their
algorithm actually bounds the bit-size of the ranking function
coefficients. Our Corollary~\ref{cor:bitsize} shows that it is possible to
deterministically bound this depth, that turns their algorithm into
a complete one, though still exponential.
In addition to \lrfs, this technique is extended for inferring linear
invariants over $\ints$.

The interest of~\citeN{DBLP:conf/concur/BradleyMS05} was in \mlc loops
in which \emph{integer division by constants} is allowed.  It is
incorrect to replace integer division $x'=\frac{x}{c}$ by precise
division, but the operation can be simulated by two paths of linear
constraints: $\{x \ge 0, c\cdot x'+y=x, 0\le y\le c-1\}$ and $\{x \le
0, c\cdot x'-y=x, 0\le y\le c-1\}$.
This illustrates the usefulness of (multipath) linear-constraint loops.

\citeN{DBLP:conf/iclp/CodishLS05} studied the synthesis of \lrfs for
\slc loops with \emph{size-change} constraints (i.e., of the form $x_i
\ge x_j'+c$ where $c\in\{0,1\}$), and \emph{monotonicity} constraints
(i.e., of the form $X \ge Y+c$, where $X$ and $Y$ are variables or
primed variables, and $c\in\{0,1\}$). In both cases the variables
ranged over \nats.
For size-change constraints, they proved that the loop terminates if
and only if a \lrf exists, moreover, such function has the form $\sum
\lambda_i\cdot x_i$ with $\lambda_i\in\{0,1\}$.
For the case of monotonicity constraints, they proved that the loop
terminates if and only if a \lrf exists for the \emph{balanced}
version of the loop, and has the form $\sum \lambda_i\cdot x_i$ with
$\lambda_i\in\{0,\pm 1\}$.
Intuitively, a balanced loop includes the constraint $x_i' \ge x_j'+c$ if and only if it
includes $x_i \ge x_j+c$.  They showed how to balance the loop while
preserving its termination behavior.
Recently, \citeN{DBLP:conf/tacas/BozgaIK12} presented similar results
for \slc loops defined by octagonal relations, implying that \emph{termination}
is decidable (even PTIME) for such loops.

\citeN{DBLP:conf/vmcai/Cousot05} used Lagrangian relaxation for
inferring possibly non-linear ranking functions. In the linear case,
Lagrangian relaxation is similar to the affine form of Farkas' lemma.

The earliest work that we know, that addresses lexicographic-linear ranking functions,
is that of~\citeN{DBLP:conf/cav/ColonS02}. As in their previous work, they use LP methods based on the computation of polars.
The \llrf is not constructed explicitly but can be inferred from the results of the algorithm.
\citeN{DBLP:conf/cav/BradleyMS05}  employed a constraint-solving approach to
search for lexicographic-linear ranking functions, where a template solution is set up and linear programming
is used to find the unknown coefficients in the template. 
\citeN{DBLP:conf/icalp/BradleyMS05} also relaxed the notion of
ranking functions to functions that \emph{eventually}
decrease, while in another work~\cite{DBLP:conf/vmcai/BradleyMS05} they considered
\mlc loops with polynomial transitions and the synthesis of
lexicographic-polynomial ranking functions. All these works actually tackle an even more complex problem,
since they also search for \emph{supporting invariants}, based on the transition constraints and on given
preconditions. 
\citeN{harris2011alternation} demonstrate that it is advantageous, to a tool that is
based on a CEGAR loop, to search for \llrfs instead of constructing transition invariants
from \lrfs only as in the original Terminator tool. They use a simplified version of the template method of \citeN{DBLP:conf/cav/BradleyMS05}. %
Similar observations have been reported by
\citeN{DBLP:conf/tacas/CookSZ13},
\citeN{DBLP:conf/cav/BrockschmidtCF13}
and
\citeN{LarrazORR13}.

\citeN{ADFG:2010} again extended the Farkas-lemma based solution
for \linrfq to the construction of \llrfs.
Like~\citeN{DBLP:conf/cav/ColonS02}, they  do it for programs with an arbitrary control-flow
graph. Unlike the latter, they prove completeness of their procedure (which means completeness over the rationals),
and their algorithm is of polynomial time.
The goal of~\citeN{ADFG:2010} was
to use these functions to derive cost bounds (like a bound on the
worst-case number of transitions in terms of the initial state); this
bound is (when it can be found) a polynomial, whose degree is at most
the dimension of the (co-domain of the) lexicographic ranking
function. Their construction produces a function of minimum
dimension  (within their class of ranking functions, which is narrower than ours, as discussed
in Section~\ref{sec:prelim}).

Decidability and complexity of termination (in general, not
necessarily with \lrfs or \llrfs) of \slc and \mlc loops has been intensively
studied for different classes of constraints.
For \slc loops, \citeN{Tiwari:04} proved that the problem is decidable
when the update is affine linear and the variables range over
$\reals$. \citeN{Braverman06} proved that this holds also for \rats,
and for the homogeneous case it holds for \ints. Both considered
universal termination, i.e., for all input.
Also, in both cases they allow the use of strict inequalities in the condition.
\citeN{Ben-AmramGM:toplas2012} showed that 
the termination of \slc loops is 
undecidable if the use of a single
irrational coefficient is allowed, as well as for \mlc loops with at least two paths, and certain other variants.

For some specific forms of integer \mlc loops termination is decidable:
Extending previous work on Size-Change Termination~\cite{DBLP:conf/popl/LeeJB01},
\citeN{DBLP:journals/corr/abs-1105-6317} proved that termination is decidable (more precisely:
PSPACE-complete) for \mlc loops with
monotonicity constraints (as defined above).
\citeN{DBLP:conf/vmcai/BozzelliP12} further extended the result (still
PSPACE-complete) for Gap Constraints, which are constraints of the form
$X-Y\ge c$ where $c \in \nats$ and $X$ and $Y$ are variables or primed
variables. This is, clearly, an extension of monotonicity constraints, which in particular
allows for more precise representation of relations of variables to constants.
 \citeN{DBLP:journals/toplas/Ben-Amram08} proved
that for difference constraints over the integers, specifically
updates of the form $x_i-x_j' \ge c$ where $c\in\ints$, and guards $x_i \ge 0$,
the termination problem becomes undecidable. However for a subclass in
which each target (primed) variable might be constrained only once (in
each path of a multiple-path loop) the problem is PSPACE-complete.

Regarding ranking functions, \citeN{DBLP:journals/corr/abs-1105-6317} shows that every terminating
program of the considered form has a ranking function which is \emph{piecewise lexicographic}.
This is achieved by transforming the program (by splitting CFG nodes) into one that is guaranteed to have
a \llrf. Such a result is probably achievable for the gap constraints of~\citeN{DBLP:conf/vmcai/BozzelliP12}
as well.  However, it is unknown how to explicitly construct ranking functions for the difference constraints of~\citeN{DBLP:journals/toplas/Ben-Amram08}. 

\section{Concluding Remarks}
\label{sec:conclusions}

We have studied the Linear Ranking problem for \slc
and \mlc linear-constraint loops and observed the difference between
the \linrfq problem, where variables range over the rationals, and the \linrfz problem,
where variables only take integer values. In practice, the latter is more common, but
the complexity of the problem has not been studied before; the common approach
has been to relax the problem to the rationals, where complete,
polynomial-time decision procedures have been known.

We have confirmed that  $\linrfz$ is a harder problem, proving it to be coNP-complete.
On a positive note, this shows that there \emph{is} a complete solution, even if exponential-time.
We further showed that some special cases of
importance do have a PTIME solution.
The latter results arise from a proof that
for integer polyhedra,  $\linrfz$ and $\linrfq$ are
equivalent. Interestingly, this is not the case for termination in general. For
example, the transition polyhedron of the \slc loop 
``$while~ x \ge 0~do~ x' = 10-2x$''
is integral; the loop terminates when the variables range over \ints 
but does not terminate when they range over \rats, specifically for $x=\frac{10}{3}$.
Note that this loop does not have a \lrf over the integers.

We have obtained results similar to the above regarding the \llinrfz problem, the existence of lexicographic-linear
ranking functions.  Our polynomial-time algorithm for \llinrfq is also new, and extends the class of functions
that can be found by the previously known polynomial-time algorithm of~\citeN{ADFG:2010}. Our algorithm is optimal, in the sense that it synthesizes \llrfs with minimal dimension.

A more general notion of ranking function applies to an arbitrary control-flow graph with transitions
specified by source and target nodes as well as linear constraints on the values of variables. In this setting,
one seeks to associate a (possibly different)  lexicographic-linear (or linear) function $\llrfsym_\nu$ with each node $\nu$, so that on 
a transition from $\nu$ to $\nu'$ we should have $\llrfsym_{\nu}(\vec x) \succ_{lex} \llrfsym_{\nu'}(\vec x')$. Such functions can be found by \lp, a procedure complete over the rationals,
using a simple extension of the solution for the loops we have discussed~\cite{DBLP:journals/tplp/MesnardS08,ADFG:2010}. The considerations regarding the complexity of the corresponding problems over integers are essentially the same as those we have presented, and we preferred to use the simpler
model for clearer presentation.

In all examples that we have discussed in this paper, when a loop has
a \lrf over \ints but not over \rats, then the loop did not terminate
over \rats. This is, however, not the case in general. A counter-example can be
constructed by combining (i.e., executing simultaneously) the loop of
Example~\ref{ex:gen:witness1} and Loop~\eqref{eq:intro:loop1} of Section~\ref{sec:introduction}.

In the context of complexity (cost) analysis, there is a special interest in
\lrfs that decrease at least by $1$ in each iteration, since they
bound the number of iterations of a given loop.
In order to get tight bounds, even if $\transitions$ has a \lrf it
might be worthwhile to compute one for $\intpoly{\transitions}$.
To see this, let us add $4x_1\ge 3$ to the condition of Loop~\eqref{eq:intro:loop1}
in Section~\ref{sec:introduction}. Then, both $\transitions$ and
$\intpoly{\transitions}$ have \lrfs.
For $\intpoly{\transitions}$ the most tight one (under the requirement to decrease by at least $1$) is
$f_1(x_1,x_2)=x_1+x_2-1$, while for $\transitions$ it is
$f_2(x_1,x_2)=2x_1+2x_2-2$.  Hence, a better bound is obtained using $\intpoly{\transitions}$. The same observation
applies to loop parallelization: the functions' value gives the schedule's \emph{latency} (depth of the computation tree)
and a lower value is preferable.

In Section~\ref{sec:prelim:loops} we have discussed the differences
between our \llrfs and those of~\citeN{ADFG:2010}
and~\citeN{DBLP:conf/cav/BradleyMS05}. This raises the question of how
our results extend to these other definitions of \llrfs.
\citeN{ADFG:2010} already show that their algorithm is complete and
PTIME over the rationals, and it is easy to show that it is complete
over the integers when computing the integer hull first, in which case
our special PTIME case also apply. Over the integers, the decision
problem is clearly coNP-hard (using the same reduction of
Section~\ref{sec:conp-hardness}), and we conjuncture that it is in
coNP as well.
The algorithm of~\citeN{DBLP:conf/cav/BradleyMS05} is exponential over
the rationals, since they search also for supporting invariants
starting from a given preconditions.
If one is interested only in \llrfs which are valid for any input, we
conjuncture that it can be done in polynomial time, by iteratively
seeking functions that are similar to our quasi-\lrfs. Over the
integers, the corresponding decision problem is clearly coNP-hard
(using the same reduction of Section~\ref{sec:conp-hardness}), and we
conjuncture that it is in coNP as well.  The technical development of
the above conjunctures is left for future work.

In Section~\ref{sec:special:octagons} we have discussed the \linrfz
problem for loops specified by octagonal relations. We showed that it
is not possible to obtain a polynomial-time algorithm that is based on
computing the integer hull as in our special PTIME cases. The question
of whether this special case of \linrfz is in PTIME or not is still
open.

In this paper we have considered \lrfs and \llrfs which are valid for
\emph{any} initial input. However, loops often come with a
precondition that restricts the space of valid input. This is the
case, for example, of the counter-example ``lassos" generated by
approaches that are based on
CEGAR~\cite{CPR06,DBLP:conf/tacas/CookSZ13,DBLP:conf/cav/BrockschmidtCF13,harris2011alternation}.
The complexity classification of the corresponding decision problems,
both over rationals and integers, is still open. Recent
work~\cite{HeizmannHLP:ATVA2013,leike2013} provides partial answers
for the rational case.

A more general definition for \llrfs can by obtained by
requiring~\eqref{eq:llrf2} of Definition~\ref{def:lexlinearrf} to hold
only for $j=i$. This is similar to the definition
of~\citeN{DBLP:conf/cav/BradleyMS05}, however, it is more general since
it does not require a fixed association of ranking-function components
with the paths of the loop.
Additional generalizations  of linear ranking functions are
\emph{eventual ranking functions}~\cite{BagnaraM13PPDP}
and Polyranking functions~\cite{DBLP:conf/icalp/BradleyMS05}.
The complexity classification of the corresponding decision problems,
over the integers (and in the latter case, also over rationals), is still open.

Regarding the potential practical impact of our results, recent
work~\cite{DBLP:conf/tacas/CookSZ13,DBLP:conf/cav/BrockschmidtCF13}
argues that the performance of a Terminator-like~\cite{CPR06} tool can
be dramatically improved by the use of \llrfs, instead of disjunctive
well-founded relations~\cite{DBLP:conf/lics/PodelskiR04}.  This is
demonstrated by their experiments, despite of using an
exponential-time algorithm. 
While we have not implemented our methods in a complete tool, their
results indicate that using a polynomial-time \llrfs algorithm could
significantly improve such analyzers.
In addition, our special PTIME cases that are based on affine linear
updates are also appealing in practice, because loops (in real
programs) that operate on integer variables often have this
form. Thus, for such cases, one can trust the answer of the
polynomial-time algorithm over the rationals.

Our algorithm for computing \llrfs, similarly to
others~\cite{ADFG:2010,LarrazORR13}, is based on iteratively
eliminating transitions. When the algorithm fails to find a \llrf, it
is guaranteed that no infinite execution can involve any of the
eliminated transitions infinitely often. In other words, any infinite
execution must have a suffix that consists only of the remaining
transitions (the potentially non-terminating kernel). 
\citeN{GantyG13} show how this kernel can be used to infer
preconditions on the input that guarantee termination, however, their
technique is developed for a more general kind of termination witness,
namely disjunctive well-founded
relations~\cite{DBLP:conf/lics/PodelskiR04}. Exploiting this approach
in our setting might have practical advantages, since the performance
bottleneck in the algorithm of~\citeN{GantyG13} is the computation of
the potentially non-terminating kernel, which we can compute (or
approximate) in polynomial time.

Finally, a theoretical study does not capture all aspects of the relative merits of different types of termination
witnesses. In practice, first, the performance of algorithms is a more involved issue than just a complexity class;
e.g., some polynomial algorithms are better than others, and some super-polynomial algorithms are nonetheless
practical. In addition, considerations such as simplicity of the
termination witnesses, information provided for certifying the witness, etc., may be important, depending on the
application. Thus, we conclude that empirical studies and algorithm-engineering are still an important objective for
future research.
 
\iftr 
\bibliographystyle{plainnat}
\fi

\bibliography{./integer-loops}

\end{document}